\documentclass[twocolumn,prd,nofootinbib]{revtex4}

\usepackage{natbib}

\usepackage{graphicx}
\usepackage{amsmath}
\usepackage{amssymb}
\usepackage{eufrak}
\usepackage[ansinew]{inputenc}
\usepackage{hyperref}
\usepackage{slashed}

\begin{document}

\title{Intrinsic Quantum Mechanics.\\
Particle physics applications on U(3) and U(2).}

\author{Ole L. Trinhammer}
\affiliation{Technical University of Denmark\\
ole.trinhammer@fysik.dtu.dk}

\begin{abstract}
We suggest how quantum fields derive from quantum mechanics on intrinsic configuration spaces with the Lie groups U(3) and U(2) as key examples. Historically the intrinsic angular momentum, the spin, of the electron was first seen as a new degree of freedom in 1925 by Uhlenbeck and Goudsmit to explain atomic spectra in magnetic fields. 
Today intrinsic quantum mechanics seems to be able to connect the strong and electroweak interaction sectors of particle physics. Local gauge invariance in laboratory space corresponds to left-invariance in intrinsic configuration space. We derive the proton spin structure function and the proton magnetic moment as novel results of the general conception presented here. We hint at the origin of the electroweak mixing angle in up and down quark flavour generators. We show how to solve for baryon mass spectra by a Rayleigh-Ritz method with all integrals found analytically. We relate to existing and possibly upcoming experiments like LHCb, KATRIN, Project 8, PSI-MUSE and ILC to test our predictions for neutral pentaquarks, proton radius, precise Higgs mass, Higgs self-couplings, beta decay neutrino mass and dark energy to baryon matter ratio. We take intrinsic quantum mechanics to represent a step, not so much beyond the Standard Model of particle physics, but to represent a step behind the Standard Model.

\end{abstract}

\maketitle
\tableofcontents

\section{Introduction}

Historically the intrinsic angular momentum, the spin, of the electron was first seen as a new degree of freedom in 1925 by George  Uhlenbeck and Samuel Goudsmit to explain atomic spectra in magnetic fields \cite{UhlenbeckGoudsmitErsetzungZwang}. After Goudsmit had told about the newest development in spectroscopy, Uhlenbeck realized that the four quantum numbers used to explain the spectroscopy should not only be ascribed to the electron as Pauli had done \cite{UhlenbeckGoudsmitErsetzungZwang,PaisNielsBohrsTimesSpin} but that to each quantum number should be ascribed an independent {\it degree of freedom}. Uhlenbeck and Goudsmit writes it like this: "...To us yet another road seems open: Pauli does not fix himself on an imagined model. The 4 quantum numbers ascribed to the electron have lost their original Land{\'e} meaning. It now lies at hand to give the electron  with its 4 quantum numbers also 4 degrees of freedom. One can then e.g. give the quantum numbers the following meaning: $n$ and $k$ remain as hitherto the main and azimutal quantum numbers for the electron in its orbit. But $R$ will be ascribed to an eigenrotation of the electron." In relation to this eigenrotation, Uhlenbeck and Goudsmit further notices: "...The ratio between the magnetic moment of the electron to its mechanical must be twice as large for its eigenrotation as for its orbital movement." \footnote{Translated from German: "...Uns scheint noch ein anderer Weg offen: Pauli bindet sich nicht an eine Modellvorstellung. Die jedem Elektron zugeordneten 4 Quantenzahlen haben Ihr urspr\"ungliche Land{\'e}che Bedeutung verlohren. Es liegt vor der Hand, nun jedem Elektron mit seinem 4 Quantenzahlen auch 4 Freiheitsgrade zu geben. Man kann dann den Quantenzahlen z.B. volgende Bedeutung geben: $n$ und $k$ bleiben wie fr\"uher die Haupt- und azimutale Quantenzahl des Elektrons in seiner Bahn. $R$ aber wird man eine eigene Rotation des Elektrons zuordnen." In relation to this eigenrotation, Uhlenbeck and Goudsmit further notices: "Das Verh\"altnis des magnetischen Momentes des Elektrons zum mechanischen {mu\ss} f\"ur die Eigenrotation doppelt so {gro\ss} sein als f\"ur die Umlaufsbewegung." \cite{UhlenbeckGoudsmitErsetzungZwang}}

Spin - by its relation to magnetic moment - also explains \cite{WeinbergSternGerlach} the twofold deflection of a silver atom beam in an inhomogeneous magnetic field in the Stern-Gerlach experiment from 1922 \cite{GerlachSternExperiment, GerlachSternMagneticMoment}. We shall return to the question of spin in section \ref{sec:quantumNumbers} but first we want to develop the idea of intrinsic degrees of freedom in more general terms.

Today intrinsic quantum mechanics seems to be able to connect the strong and electroweak interaction sectors of particle physics. 
We have used intrinsic quantum mechanics to derive the electron to nucleon mass ratio and parton distributions for the up and down quark content of the proton \cite{TrinhammerEPL102}, baryon spectra, electroweak energy scale and Higgs mass \cite{TrinhammerNeutronProtonMMarXivWithAppendices25Jun2012, TrinhammerBohrStibiusHiggsPreprint, TrinhammerBohrStibiusEPS2015}. Further we have  predicted beta decay neutrino mass scenarios and Higgs self-couplings \cite{TrinhammerNeutrinoMassHiggsSelfcoupling}. The latter are at a slight variance with Standard Model expectations by a presence of the up-down quark mixing matrix element as a factor in the quartic Higgs self-coupling.

\begin{figure}
\begin{center}
\includegraphics[width=0.45\textwidth]{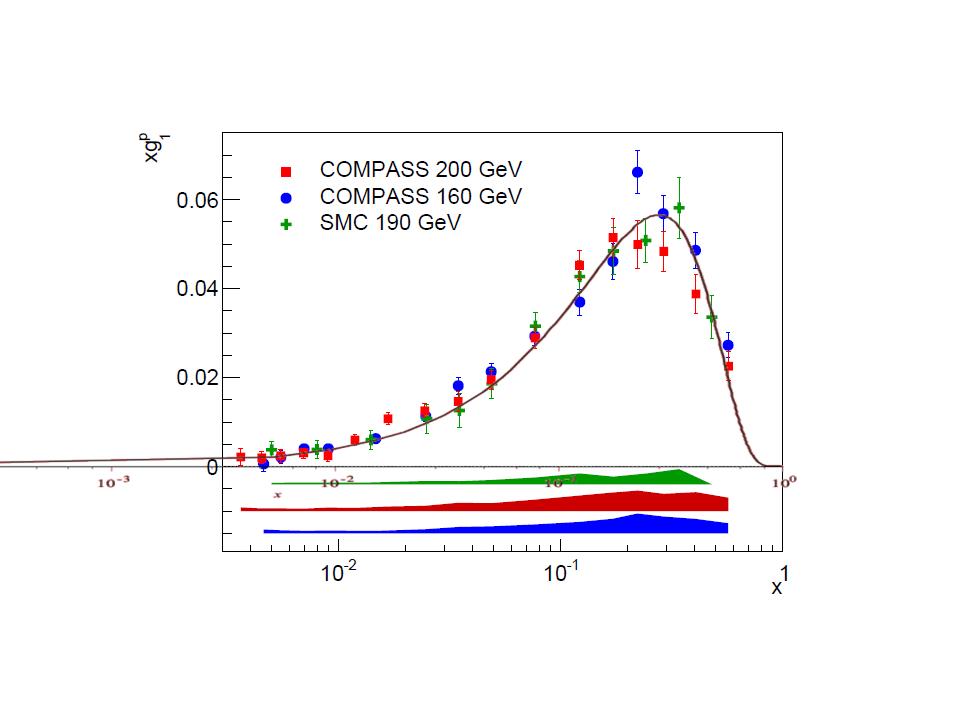}
\caption{Spin structure function for the proton as a function of the momentum fraction, $x$ of the scattering center. Data from the COMPASS Collaboration \cite{COMPASSspinStuctureFunctionProton} compared to a weighted sum of $T_{\rm u}$ and $T_{\rm d}$ distributions (\ref{eq:spinStructureFunc}) from an intrinsic protonic state (solid brown line) \cite{TrinhammerEPL102} overlaid on the COMPASS data. The state is an approximate protonic state and the weights are the squared charges $(\frac{2}{3})^2$ and $(-\frac{1}{3})^2$ of u and d quarks respectively.}
\label{fig:protonSpinStructureFunctionCOMPASS}
\end{center}
\end{figure}

\begin{figure}
\begin{center}
\includegraphics[width=0.45\textwidth]{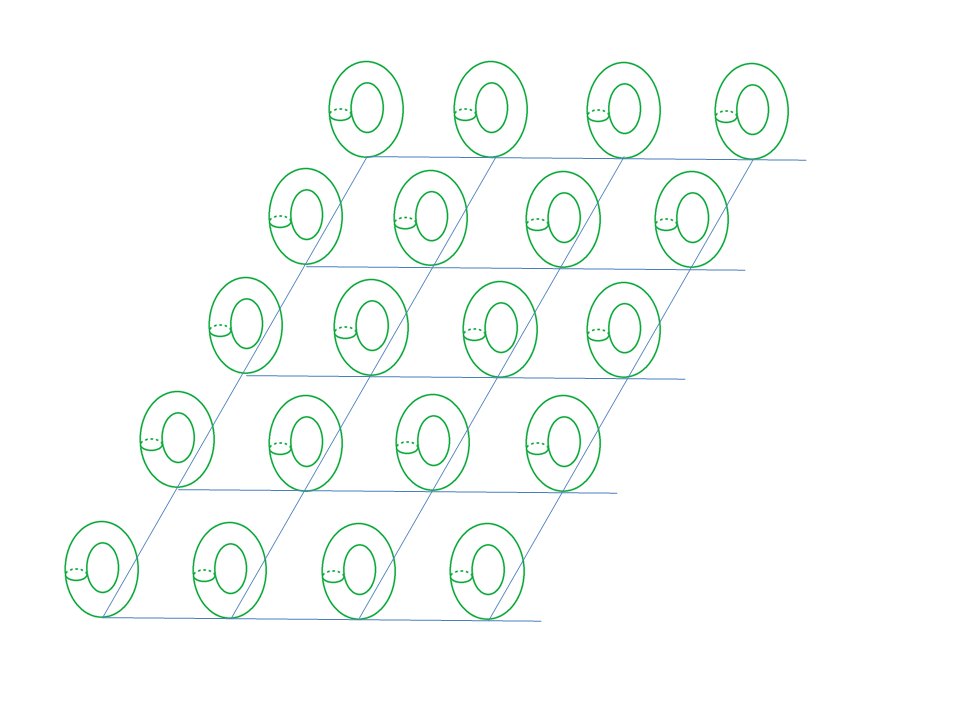}
\caption{The intrinsic space - shown as a torus - can be reached from any point in laboratory space - shown as the patterned floor. After Juan Maldacena \cite{MaldacenaIntrinsicDrawing}. Letting the floor pattern to be visible through the intrinsic toroidal space is to stress that the intrinsic space should not be considered as just "extra" spatial dimensions like in string theory. Rather the intrinsic space should be likened to a non-spatial, generalized spin space, i.e. no gravitational interaction in the intrinsic space.}
\label{fig:MaldacenaWithTori}
\end{center}
\end{figure}

The present work gives a more systematic presentation of the intrinsic point of view and presents a derivation of the proton spin structure function $g_1^{\rm p}$. Figure \ref{fig:protonSpinStructureFunctionCOMPASS} shows comparison with recent data from the COMPASS Collaboration \cite{COMPASSspinStuctureFunctionProton}. We also present a calculation of the proton magnetic moment and give considerations on the origin of the electroweak mixing angle $\theta_{\rm W}$ in up and down quark generators. The proton spin structure function, its magnetic dipole moment and the considerations on the electroweak mixing angle are new results from the intrinsic conception of dynamics in Lie group configuration spaces. It should be noted that the intrinsic space is not to be considered as extra spatial dimensions like in string theory. Rather it should be considered as a generalized spin space, see fig. \ref{fig:MaldacenaWithTori}. In other words, there is no gravitational interaction in intrinsic spaces.

\begin{figure}
\begin{center}
\includegraphics[width=0.45\textwidth]{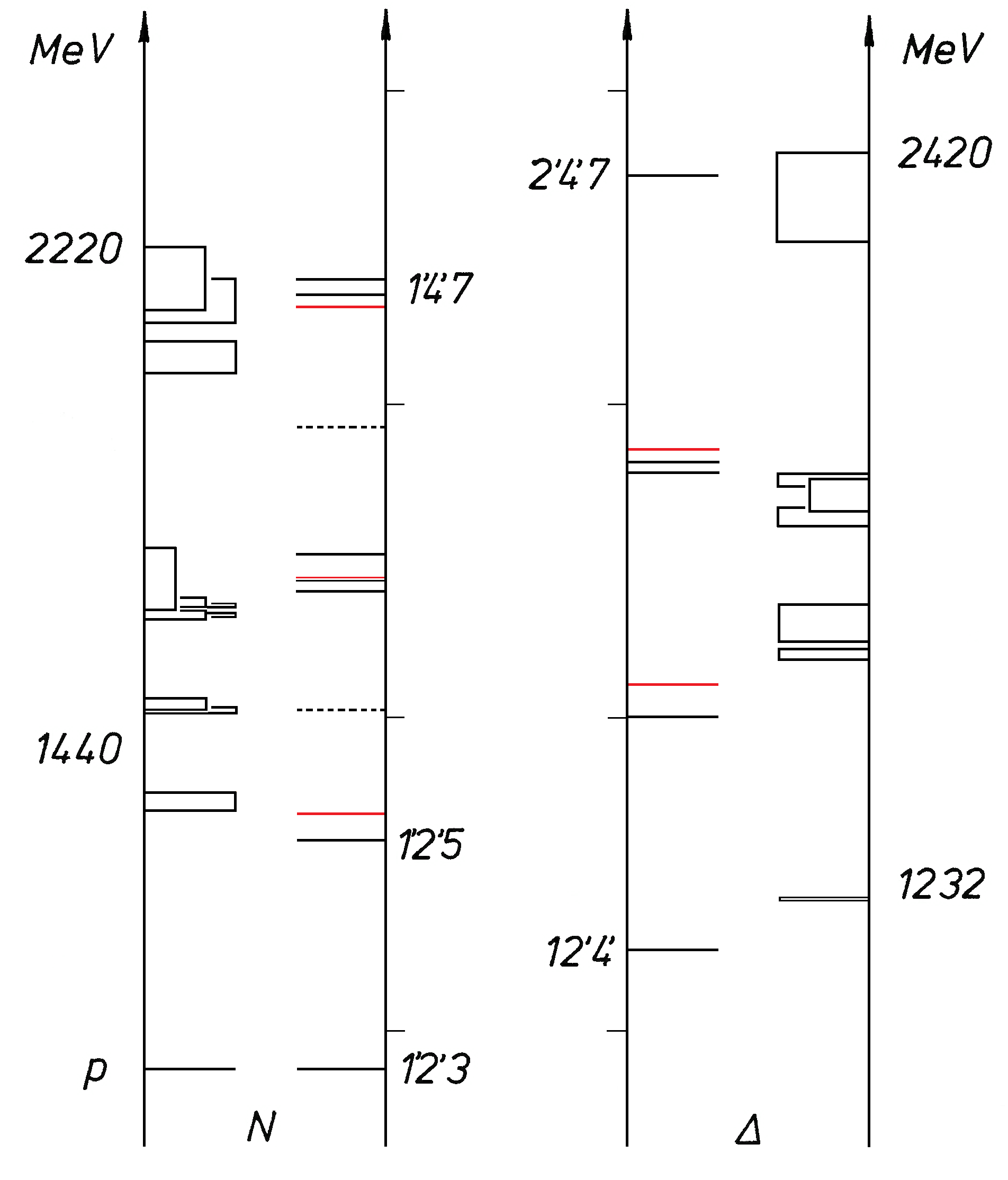}
\caption{Neutral flavour baryon spectra. The boxes represent baryons observed with certainty \cite{RPP2016}. The box widths represent the uncertainty in mass pole peak, not resonance widths, which are much larger. The lines are our approximate predictions  \cite{TrinhammerNeutronProtonMMarXivWithAppendices25Jun2012,TrinhammerBohrStibiusHiggsPreprint} based on Slater determinants constructed from solutions to a one-dimensional case (\ref{eq:oneDimSchroedinger}) as shown in fig. \ref{fig:oneDimWavefunctions}.}
\label{fig:NandDeltaSpectrum}
\end{center}
\end{figure}

The work is structured as follows. First we generalize the canonical commutation relations by use of differential forms and left-invariant coordinate fields on the intrinsic space. Then we show how a quantum field can be created when one reads off the intrinsic dynamics by use of exterior derivatives (momentum forms). We show that left-invariance in the intrinsic configuration space corresponds to local gauge invariance in laboratory space, provided the intrinsic space is unitary. In section \ref{sec:caseForU3} we give a specific example for $U(3)$. Since the intrinsic spaces we consider are compact Lie groups, the potentials in the Hamiltoniae are required to be periodic functions of the dynamical variables that parametrize the configuration variables. In section \ref{sec:intrinsicPotentialBlochPhase} we show how this opens for the introduction of Bloch phase factor degrees of freedom known from solid state physics. In sections \ref{sec:quantumNumbers} and \ref{sec:flavourIncolour} we describe the spectrum for the centrifugal term of the Laplacian and the intermingling of colour and flavour in the case of $U(3)$.  In section \ref{sec:neutralPentaquarks} we describe predictions of neutral pentaquarks. In section \ref{sec:protonSpinStructureFunction} we derive an approximate proton spin structure function. In section \ref{subsec:MixingAngleQuarkGenerators} we discuss the electroweak mixing angle in comparison with Standard Model descriptions. In section \ref{sec:RayleighRitz} we solve exactly for eigenvalues of a particular Hamiltonian on $U(3)$ which we used to describe baryon spectra of neutral electric charge and neutral flavour, i.e. neutral electric charge members of the $N$ and $\Delta$ spectrum. For electrically charged partners one needs to expand on a bases that exploits the Bloch phase degrees of freedom. In fig. \ref{fig:NandDeltaSpectrum} we show results from an approximate solution. We have only recently found a basis for charged states which can be integrated to exact analytical results and have not yet carried through all the integrals for the Rayleigh-Ritz method in this case. The main problem is the increasing number of terms, up to $144$ terms in one integral. One may fear that this would lead to a too slow processing when diagonalizing the Hamiltonian to get the resulting eigenvalues. But this is not the case. All integrals can be expressed as sums of Kronecker delta-like factors which are rapidly evaluated. The main problem is rather banal: to get all the signs of the different terms correct! The reader may wonder why we are not satisfied with solving the integrals numerically. The answer is twofold: Numerical solutions are exceedingly slow to carry through. Numerical solutions will therefore never be able to reach the accuracy we want. For instance we have found the neutron to proton mass shift by an approximate base used also for constructing fig. \ref{fig:NandDeltaSpectrum}. We find
\begin{equation}	\label{eq:mNmPapproxTheory}
 \frac{m_{\rm n}-m_{\rm p}}{m_{\rm p}}=0.13847(14)\%.
\end{equation}
This compares rather well with the value calculated from the neutron and proton masses which are known experimentally with eight significant digits \footnote{Respectively $m_{\rm n}c^2=939.565413(6)\ \rm MeV$ and $m_{\rm p}c^2=938.2720813(58)\ \rm MeV$ \cite{RPP2016}.}
\begin{equation}
 \frac{m_{\rm n}-m_{\rm p}}{m_{\rm p}}|_{\rm exp}=0.1378420(13)\%.
\end{equation}
The discrepancy is small but of principal importance which is why we need exact integrals in the Rayleigh-Ritz solution that we aim for. A solution that is more direct and potentially much more accurate than the otherwise successful (lattice) quantum field theory calculations from separately handled $QCD$ and $QED$ contributions within the Standard Model \cite{BorsanyiEtAlMnMp, HorsleyEtAlMnMp}.

\section{Scales overview}

Before we indulge into the detailed mathematics, an overview may be appropriate on how the different length scales come about in respectively the baryon sector, the electroweak sector and the neutrino sector\footnote{This section may be more intelligible after study of the more detailed sections to follow. Placed here anyhow to list the scales used in the different particle sectors.}. Our input shall be the electron mass $m_{\rm e}$ and the unit electric charge $e$ together with Planck's constant $h$ and the speed of light $c$ in vacuum. From the electron mass and electric charge we get a length scale, the classical electron radius $r_{\rm e}$ \cite{Heisenberg, LandauLifshitz} defined as the distance from the electron at which the classical electrostatic energy of a similar charge $e$ in the field of the electron equates the rest-energy of the electron $m_{\rm e}c^2$, that is 
\begin{equation}
 \frac{1}{4\pi\epsilon_0}\frac{e^2}{r_{\rm e}}\equiv m_{\rm e}c^2.
\end{equation}

We shall quantize our dynamics on intrinsic configuration spaces, where angular variables carry the dynamical degrees of freedom. We start out from the Lie group $U(3)$ which has three toroidal degrees of freedom parametrized by $\theta_j\in\mathbb{R},j=1,2,3$. These angles are projected to laboratory space by use of a length scale $a$ 
\begin{equation}	\label{eq:spaceprojectionPhysics}
 x_j=a\theta_j
\end{equation}
and the canonical quantization is generalized to
\begin{equation}
 [a\theta_i,\frac{-i\hbar}{a}\partial_j]=i\hbar\delta_{ij}
\end{equation}
where $\partial_j$ are derivatives on the configuration space. At the {\it origo} of the configuration space we have $\partial_j=\partial/\partial\theta_j$.

The Lie group $U(3)$ has nine generators which correspond to nine kinematical generators in laboratory space, namely $iT_j=\frac{\partial}{\partial\theta_j}$ corresponding to the three momentum operators together with six non-commuting operators $S_j$ and $M_j$ which take care about spin and flavour. We take these generators to generate excitations of the intrinsic degrees of freedom in high energy scattering experiments.

We can describe the baryon spectrum by a length scale $a$ defined as\footnote{In the neutron to proton decay one may heuristically think of the electron as a "peel-off" from the neutron leaving a "charge-scarred"  nucleon, the proton. Simultaneously we experience the creation of the electron and an anti-electron-neutrino. }
\begin{equation}
 \pi a=r_{\rm e}.
\end{equation}

\begin{figure} 
\begin{center}
\includegraphics[width=0.45\textwidth]{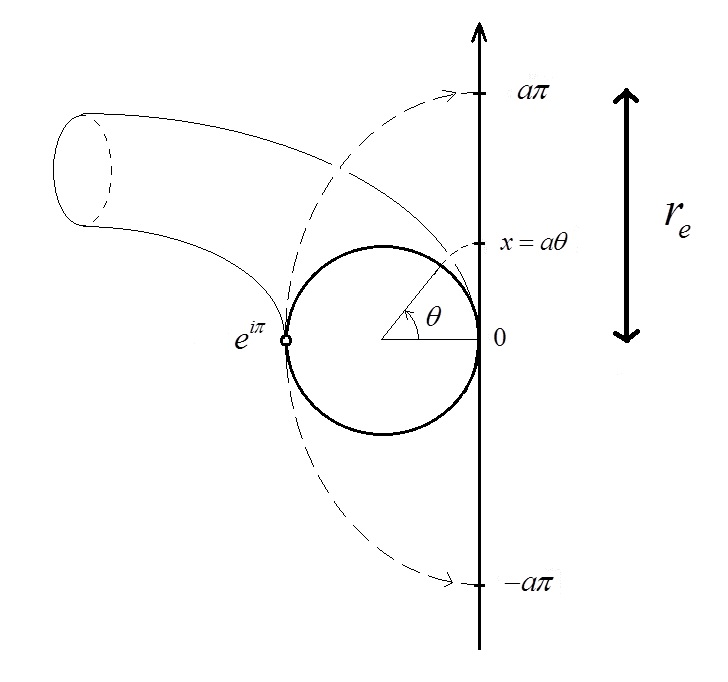}
\caption{Projection of the Lie group configuration space to the algebraic parameter space  \cite{TrinhammerEPL102} with the circle as a $U(1)$ toroidal dimension. The projection is a good description for going from intrinsic quantum mechanics to quark fields in space. The algebra approximates the group in the neighbourhood of the origo and the projection is scaled by the classical electron radius \cite{Heisenberg, LandauLifshitz} $r_{\rm e}$ as a measure for the extension of the charge "peel off" in the neutron decay. The scaling corresponds excellently to the measured value for the electron to neutron mass ratio \cite{TrinhammerEPL102}. Figure from ref. \cite{TrinhammerBohrStibiusHiggsPreprint}.}
\label{fig:ProjectionGroupAlgabra}
\end{center}
\end{figure}

This corresponds to the space projection in (\ref{eq:spaceprojectionPhysics}), illustrated in fig. \ref{fig:ProjectionGroupAlgabra} and yields an energy scale
\begin{equation}
 \Lambda=\frac{\hbar c}{a}
\end{equation}
With this energy scale we reproduce the baryon spectrum shown in fig. \ref{fig:NandDeltaSpectrum} from a Hamiltonian on the configuration space $U(3)$
\begin{equation}
 \Lambda\left[-\frac{1}{2}\Delta+\frac{1}{2}{\rm\ Tr}\chi^2\right]\Psi(u)={\cal E}\Psi(u).
\end{equation}
The configuration variable $u\in U(3)$ is parametrized by nine angular variables $\theta_j,\sigma_j,\mu_j$
\begin{equation}
 u=e^{i(\theta_jT_j+\sigma_jS_j+\mu_jM_j)}\equiv e^{i\chi}.
\end{equation}
The potential is periodic as a reflection of the compactness of the configuration space\footnote{Actually our specific choice for potential is only dependent on $\theta_j$ because the trace is invariant under conjugation $u\rightarrow v^{-1}uv$ where $u,v\in U(3)$. The same goes for a Wilson inspired potential.}.

The neutron to proton transformation in the baryon sector is undertaken by period doublings in the parametrization of the wavefunction $\Psi(u)$. The period doublings introduce Bloch phase factors known from periodic systems in solid state physics. To allow for the period doublings in the present context we invoke the Higgs mechanism. We let the Higgs field $\phi$ take up phase changes and let the leptonic sector take up spin structure and carry away released energy in the form of rest energy and kinetic energy together with that of the proton. We shape the Higgs potential by the intrinsic $U(3)$ potential and assume the exchange of one quantum of action between the strong and electroweak sectors. This yields
\begin{equation}	\label{eq:phi0PhysicsOverview}
 2\pi\Lambda=\alpha\varphi_0
\end{equation}
which determines the electroweak energy scale\footnote{Since we consider the neutron $\beta$ decay our $v$ is related to the standard model by $v_{\rm SM}=v\sqrt{V_{ud}}$, where $V_{ud}$ is the up-down quark mixing matrix element.} as $v/\sqrt{2}=\varphi_0$.

The period doublings in the baryonic sector have to come in pairs. This singles out the Lie group $U(2)$ as a representation space for the Higgs field and as a configuration space for the electron and the anti-electron-neutrino. We thus assume the electron and the neutrino to be ground states of
\begin{equation}	\label{eq:electronU2physicsOverview}
 \Lambda_{\rm e}\left[-\frac{1}{2}\Delta+\frac{1}{2}{\rm\ Tr}\chi^2\right]\Psi(u)={\cal E}\Psi(u), u\in U(2)
\end{equation}
and
\begin{equation}	\label{eq:neutrinoUphysicsOverview}
 \Lambda_\nu\left[-\frac{1}{2}\Delta+\frac{1}{2}{\rm\ Tr}\chi^2\right]\Psi(u)={\cal E}\Psi(u), u\in U(2)
\end{equation}
respectively. We have already set the electron mass as a basic input, so
\begin{equation}
 {\cal E}_{\rm e}=m_{\rm e}c^2\equiv {\rm E}_{\rm e}\Lambda_{\rm e}
\end{equation}
where ${\rm E}_{\rm e}\equiv {\cal E}_{\rm e}/\Lambda_{\rm e}$ is the dimensionless ground state eigenvalue of (\ref{eq:electronU2physicsOverview}).

\begin{figure}
\begin{center}
\includegraphics[width=0.45\textwidth]{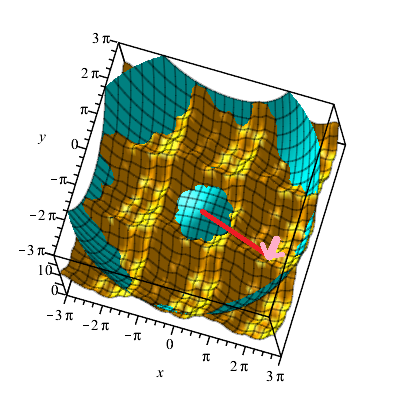}
\caption{The Higgs potential (cyan) as a wine bottle bottom on a periodically rippled egg tray (orange). The egg-tray structure is the periodic parametric potential scaled from the baryonic sector and the ripples are scaled from the leptonic sector. Both are active in the neutron decay where the neutron changes to a charged proton and a charge-compensating electron. The size of the ripples is grossly exaggerated for clarity (drawing for $\sin\zeta=1/3$ as opposed the physical case $\sin\zeta\approx1/1000$ in (\ref{eq:misalignmentPhysicsOverview})). The size of the Higgs field vacuum expectation value $\varphi_0$ in (\ref{eq:phi0PhysicsOverview}) is shown by the red line. The misalignment even means a slight rotation into the third toroidal coordinate. This is not shown in the figure. The free movement of the Goldstone bosons in the Higgs potential ditch is prohibited by the periodic potentials and the pion field is caught in the ripples leading to physical pion particles with masses determined by the vacuum misalignment. Figure and adapted caption from \cite{TrinhammerBohrIQM3researchGate}.}
\label{fig:HiggsFitEggTrayRipples}
\end{center}
\end{figure}

We expect the neutrino scale to follow from the exchange of one quantum of action between the electric potential $\phi_{\rm B}=\hbar c/a_\infty$ of the proton-electron system at a length scale given by the Bohr radius $a_\infty$ and with a neutral weak coupling in a slightly misaligned Higgs field vacuum with misalignment angle $\zeta$ given by
\begin{equation}	\label{eq:misalignmentPhysicsOverview}
 \sin\zeta=\frac{\Lambda_{\rm e}}{\Lambda},
\end{equation}
see fig. \ref{fig:HiggsFitEggTrayRipples}. We thus have
\begin{equation}
 2\pi\Lambda_\nu=\left(\frac{1}{2}\sqrt{g'^2+g^2}\right)^2\varphi_{\rm B}\sin\zeta
\end{equation}
from which a prediction for the neutrino mass can be made by the fact that (\ref{eq:electronU2physicsOverview}) and (\ref{eq:neutrinoUphysicsOverview}) share dimensionless eigenvalues $\rm E_\nu = E_e$ and therefore
\begin{equation}
 \frac{m_\nu}{m_{\rm e}}=\frac{\Lambda_\nu}{\Lambda_{\rm e}}.
\end{equation}

\section{Choice of coordinate system} \label{ch:coordinateSystem}

The description of a quantum state $\psi$ is based on configuration variables like position $\bf x$, spin $\bf S$, occupation number $n$ or e.g. phase angle $\theta$. For the parametrization of a configuration variable one needs a coordinate system where the coordinates become the parameters on which an operational theory can be formulated. \footnote{We use the word parameter as a continuous, dynamical variable - not an arbitrary, specific value as in "fitting parameter". A parameter in the present sense is a generalized coordinate but need not have the dimension of length. In stead, e.g. it could be an angular variable.}  A well known example is the Schr\"odinger equation for the hydrogen atom
\begin{equation}	\label{eq:SchroedingerHatom}
 \left[-\frac{\hbar^2}{2m}\left(\frac{\partial^2}{\partial x_1^2}+\frac{\partial^2}{\partial x_2^2}+\frac{\partial^2}{\partial x_3^2}\right)-\frac{e^2}{4\pi\epsilon_0}\frac{1}{r}\right]\psi({\bf x})={\cal E}\psi({\bf x})
\end{equation}
Here the configuration variable ${\bf x}=(x_1,x_2,x_3)$ parametrizes into three coordinates $x_1,x_2,x_3$, the values of which depend on the choice of the coordinate system. The first term in the Hamiltonian is usually called the kinetic term by the analogy $\hat{p}=-i\hbar {\bf \nabla}$ between the quantum momentum operator $\hat{\bf p}$ and the classical momentum ${\bf p}$ which enters the kinetic energy $T=\frac{p^2}{2m}$. The second term  is analogous to the potential energy in a classical Coulomb field with $r^2=x_1^2+x_2^2+x_3^2$ being the squared distance from a center charge $+e$ (the proton in the hydrogen atom) to a negative charge $-e$ (representing the electron of reduced mass $m$). Both terms are independent on the orientation of the coordinate system and we say that the potential has radial symmetry. This means that we would expect the state $\psi({\bf x})$ to be independent on coordinate rotations, i.e.
\begin{equation}
 \psi({\bf x}')=\psi({\bf x}), {\rm where}\ \ \ {\bf x}'={\cal R}{\bf x}
\end{equation}
with rotation matrix, e.g.
\begin{equation}
 {\cal R}=\left(\begin{matrix} \cos\beta & -\sin\beta & 0 \\
                          \sin\beta & \cos\beta & 0 \\
                          0 & 0 & 1 \end{matrix}\right)
\end{equation}
for a rotation through an angle $\beta$ about the $z$-axis. If however a direction in space is singled out by the presence of e.g. a magnetic field, one is able to observe the phenomenae of orbital and intrinsic angular momentum (spin), which influence the energy eigenvalues $\cal E$ in (\ref{eq:SchroedingerHatom}). But let us first take a look again at the interpretation $\hat{\bf p}=-i\hbar {\bf \nabla}$. It corresponds to a definition of the {\it momentum form} ${{\rm d}\psi}$ \cite{GuilleminPollack, HolgerBechNielsenMomentumForm}
\begin{equation}	\label{eq:dpsiDefinition}
 {\rm d}\psi=\frac{\partial\psi}{\partial x_1}{\rm d}x_1+\frac{\partial\psi}{\partial x_2}{\rm d}x_2+\frac{\partial\psi}{\partial x_3}{\rm d}x_3=\frac{\partial\psi}{\partial x_j}{\rm d}x_j,
\end{equation}
where in the last expression we introduce Einstein's summation convention to make a sum over repeated indices understood. To read off the momentum component $p_j$ from the state $\psi$ at point ${\bf x}$ we let ${\rm d}\psi$ act in an orthonormal base $({\bf e}_1,{\bf e}_2,{\bf e}_3)\sim\left(\frac{\partial}{\partial x_1},\frac{\partial}{\partial x_2},\frac{\partial}{\partial x_3}\right)$ at ${\bf x}$
\begin{equation}	\label{eq:dpsiMomentumRead}
 {\rm d}\psi({\bf e}_j)=\frac{\partial\psi}{\partial x_i}{\rm d}x_i\left(\frac{\partial}{\partial x_j}\right)=\frac{\partial\psi}{\partial x_j}\equiv\frac{i}{\hbar}p_j|_{{\bf x}}.
\end{equation}
Here we used
\begin{equation}	\label{eq:diffFormCoordFieldOrthogonality}
 {\rm d}x_i(\frac{\partial}{\partial x_j})=\delta_{ij}
\end{equation}
with Kronecker delta, $\delta_{ij}=1$ for $i=j$ and zero otherwise.

As an example of using (\ref{eq:dpsiDefinition}) let us consider a plane wave
\begin{equation}	\label{eq:standingWave}
 \psi({\bf x})=\frac{1}{(2\pi)^{3/2}} e^{\frac{i}{\hbar}{\bf p}\cdot{\bf x}},
\end{equation}
normalized over one de Broglie wavelength $\lambda=h/p$ in all three dimensions. Inserting (\ref{eq:standingWave}) in (\ref{eq:dpsiMomentumRead}) we get
\begin{equation}	\label{eq:momentumComponentDerivativeAnalogy}
 \frac{\hbar}{i}{\rm d}\psi({\bf e}_j)=\frac{\hbar}{i}\frac{\partial\psi}{\partial x_j}=\frac{\hbar}{i}\frac{i}{\hbar}p_je^{\frac{i}{\hbar}{\bf p}\cdot{\bf x}}=p_j\psi({\bf x}),
\end{equation}
which corresponds to the usual operator identification in quantum mechanics 
\begin{equation}
 \hat{p}_j=-i\hbar\frac{\partial}{\partial x_j}
\end{equation}
of the momentum operator $\hat{p}_j$ operating on the state $\psi$ with momentum expectation value \cite{SchiffExpectationValue}
\begin{equation}
 <p_j>=-i\hbar\int\psi^\dagger\frac{\partial\psi}{\partial x_j}{\rm d}^3{\bf x}=\int\psi^\dagger p_j\psi{\rm d}^3{\bf x}=p_j
\end{equation}

The introduction of the momentum form (\ref{eq:dpsiMomentumRead}) in the case of a euclidean space with orthonormal base may seem as a mathematical abstraction adding no new information. However, the definition (\ref{eq:dpsiDefinition}) is essential when we enter the realm of a Lie group configuration space. There one has to introduce a base varying from point to point - following the curvature in the intrinsic space. It is therefore of interest to see that the formalism using differential forms (\ref{eq:dpsiDefinition}) gives well-known results in the simple case (\ref{eq:momentumComponentDerivativeAnalogy}).

\begin{figure} 
\begin{center}
\includegraphics[width=0.45\textwidth]{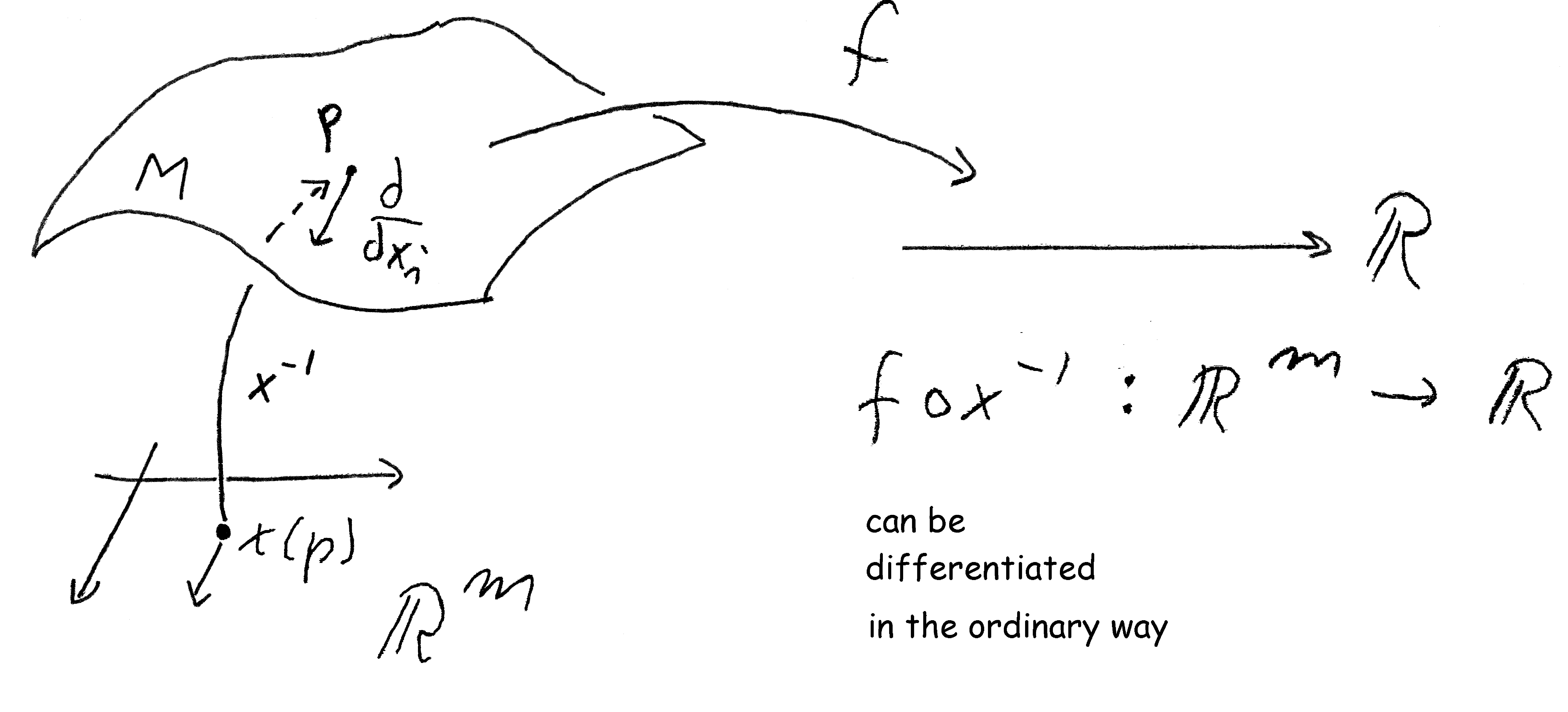}
\caption{Derivation of a real-valued function $f$ at point $u$ in the manifold $M$ is defined by using a local smooth map $x:M\rightarrow \mathbb{R}^m$ to pull back the problem to an ordinary derivation on $\mathbb{R}^m$ by using the pullback function  $f\circ x^{-1}:\mathbb{R}^m\rightarrow \mathbb{R}$. One can then differentiate $f\circ x^{-1}$  in the ordinary way. Figure from ref. \cite{TrinhammerBohrStibiusHiggsPreprint}.}
\label{fig:projectionParameterSpace}
\end{center}
\end{figure}

The strength of the formalism shows if one considers quantum mechanics on curved spaces or more specifically, smooth manifolds. Then one has to rely on local coordinate systems, defined from maps of the manifold to a euclidean space but with globally defined concepts like coordinate fields and coordinate forms. Let us consider a real smooth manifold $M$ of dimension $m$. Let $M$ be embedded in $\mathbb{R}^k, k\geq m$, see fig. \ref{fig:projectionParameterSpace}. In the neighbourhood of each point $u\in M\subset\mathbb{R}^k$ there exists a smooth map $x$ from $M$ to $\mathbb{R}^m$
\begin{equation}
 x: M\rightarrow\mathbb{R}^m.
\end{equation}
This map can be used to induce a local base {$\frac{\partial}{\partial x_j}$} in the tangent space $TM_u$ to $M$ at $u$ by the definition
\begin{equation}	\label{eq:inducedBasis}
 \frac{\partial}{\partial x_j}\equiv {\rm d}(x^{-1})({\bf e}_j),\ \ \ j=1,2,\cdots, m,
\end{equation}
where $\{{\bf e}_j\}$ constitutes an orthonormal base for $\mathbb{R}^m$.

\section{First quantization}

The conjugacy of the coordinate forms ${\rm d}x_j$ to the coordinate fields $\frac{\partial}{\partial x_j}$ expressed in (\ref{eq:diffFormCoordFieldOrthogonality}) carries the characteristic commutation relations between conjugate variables naturally into the formalism. Thus
\begin{equation}	\label{eq:quantizationConditionGeneralized}
 {\rm d}x_i(\frac{\partial}{\partial x_j})=\delta_{ij}\ \ \ \sim\ \ \ [\hat{x}_i,\hat{p}_j]=i\hbar\delta_{ij}
\end{equation}
both express the basic relation of first quantization.

\begin{figure}
\begin{center}
\includegraphics[width=0.4\textwidth]{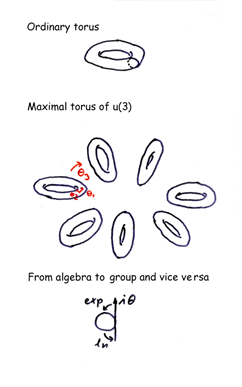}
\caption{An intrinsic space from which quantum fields project out in laboratory space. When the intrinsic space is described with parameters from laboratory space, we see that the intrinsic potentials become periodic functions in parameter space. The complex exponential function maps into the intrinsic space.}
\label{fig:algebraToGroupViceVersa}
\end{center}
\end{figure}

As another example of using (\ref{eq:dpsiDefinition}), let us consider a state $\psi$ on an intrinsic three dimensional torus which we shall denote $U_0(3)$. Here the configuration variable $u$ can be parametrized by three angles $\theta_j\in\mathbb{R}$ as, see fig. \ref{fig:algebraToGroupViceVersa}
\begin{equation}	\label{eq:u0torus}
 u=\left(\begin{matrix} e^{i\theta_1}&0&0\\
 						0&e^{i\theta_2}&0\\
 						0&0&e^{i\theta_3} \end{matrix}\right).
\end{equation}
The map $x$ from $U_0(3)$ to $(i\mathbb{R})^3$ in this case is the inverse exponential, i.e.
\begin{equation}
 x^{-1}=\exp: (i\mathbb{R})^3\rightarrow U_0(3).
\end{equation}
To find the induced base, we need the differential of the exponential mapping. This can be found by using the matrix expansion for the image of a matrix $A\in Gl(3,\mathbb{C})$
\begin{equation}
 \exp A=1+A+\frac{1}{2}A^2+\frac{1}{3!}A^3+\cdots\equiv u
\end{equation}
and rewriting temporarily the result as a vector function $\vec{P}$ with nine coordinates
\begin{equation}
 \vec{P}(A)=\left(P_{11}(A),P_{12}(A),P_{13}(A),P_{21}(A),\cdots,P_{33}(A)\right).
\end{equation}
Here $P_{ij}$ is the $ij$'th element of $\exp A$. Each coordinate is a function of the nine elements $A_{lm}$ of $A$
\begin{equation}	\label{eq:PasFunc9variables}
 P_{ij}(A)=P_{ij}(A_{11},A_{12},\cdots,A_{33}).
\end{equation}
The differential ${\rm d}\exp$ will then be a nine by nine matrix
\begin{equation}	\label{eq:dExpAs9by9matrix}
 {\rm d}\exp=\left(\frac{\partial P_{ij}}{\partial A_{lm}}\right).
\end{equation}
Taking the differential at the origo $A=0$, we get the identity ${\rm d}\exp=1$.

For points in our torus $U_0(3)$ in (\ref{eq:u0torus}), we have in particular
\begin{equation}
 A=\left(\begin{matrix} {i\theta_1}&0&0\\
 						0&{i\theta_2}&0\\
 						0&0&{i\theta_3} \end{matrix}\right)
\end{equation}
and the matrix in (\ref{eq:dExpAs9by9matrix}) taken at origo $\sim A=0, u=1$ will be singular with only three non-zero elements, namely
\begin{equation}
 \frac{\partial P_{11}}{\partial A_{11}}=\frac{\partial P_{22}}{\partial A_{22}}=\frac{\partial P_{33}}{\partial A_{33}}=1
\end{equation}
The singularity is expected since we embedded the original $(i\mathbb{R})^3$ in $\mathbb{C}^9$. We can restrict ourselves to three essential \footnote{We have this term from Morton Hamermesh \cite{HamermeshEssentialVariable}.} variables in (\ref{eq:PasFunc9variables}), i.e.
\begin{equation}	\label{eq:PasFunc3variables}
 u_{ij}(A)=P_{ij}(A_{11},A_{22},A_{33}).
\end{equation}
In other words, we can reduce the expression for the differential to a three by three matrix
\begin{equation}	\label{eq:dExpAs3by3matrix}
 {\rm d}\exp=\left(\begin{matrix} 1&0&0\\
 						0&1&0\\
 						0&0&1 \end{matrix}\right)
\end{equation}
from which we induce the basis {$\partial_j|_e$}, $j=1,2,3$ for the tangent space $TM_e$ at the origo $e=\exp i0$
\begin{equation}
 \partial_j|_e={\rm d}\exp(i{\bf e}_j).
\end{equation}
Traditionally this basis is also signified as
\begin{equation}
 \partial_j|_e=\frac{\partial}{\partial \theta_j}.
\end{equation} 
We can even represent $\partial_j|_e$ by a matrix
\begin{equation}	\label{eq:partialTj}
 \partial_j|_e=iT_j
\end{equation}
where the $iT_j$s are generators of the torus $U_0(3)$ (which is an abelian Lie group)
\begin{equation}	\label{eq:toroidalGeneratorsU3}
 T_1=\left(\begin{matrix} 1&0&0\\
 						0&0&0\\
 						0&0&0 \end{matrix}\right),
 T_2=\left(\begin{matrix} 0&0&0\\
 						0&1&0\\
 						0&0&0 \end{matrix}\right),
 T_3=\left(\begin{matrix} 0&0&0\\
 						0&0&0\\
 						0&0&1 \end{matrix}\right).
\end{equation}

For $u\in U_0(3)$, we have
\begin{equation}
 P_{jj}=e^{i\theta_j}, A_{jj}=i\theta_j
\end{equation}
from which follows
\begin{equation}
 \frac{\partial P_{jj}}{\partial A_{jj}}=\frac{\partial e^{i\theta_j}}{\partial(i\theta_j)}=e^{i\theta_j}
\end{equation}

In the general case, when $u\neq e$ and the generators $iT_k$ may not even be abelian, one introduces left-invariant coordinate fields
\begin{equation}	\label{eq:coordinateFieldGeneral}
 \partial_k|_u=\frac{\rm d}{{\rm d}\alpha}ue^{i\alpha T_k}|_{\alpha=0} =uiT_k.
\end{equation}
This expression can be used for any Lie group, be it abelian or non-abelian with commuting or non-commuting generators $iT_k$. Using the generators one can introduce coordinates $\alpha_k\in\mathbb{R}$ to parametrize any Lie group by writing its elements $u$ as
\begin{equation}	
 u=\exp i\alpha_k T_k.
\end{equation}
The coordinate form ${\rm d}\alpha_k$ is conjugate to the coordinate field $\partial_k$, i.e.
\begin{equation}	\label{eq:coordinateFormFieldConjugacy}
 {\rm d}\alpha_l(\partial_m)=\delta_{lm}.
\end{equation}
The possibility of an unambiguous global conjugacy in (\ref{eq:coordinateFormFieldConjugacy}) is the basis for a consistent quantization on intrinsic Lie group configuration spaces, cf. (\ref{eq:quantizationConditionGeneralized}). It remains to figure out which Lie groups could be of interest to Nature. An obvious association goes to the gauge groups $U(1), SU(2), SU(3)$ of the fundamental quantum interactions known from the standard model of particle physics. But as we shall see in sec. \ref{sec:caseForU3}, $U(3)$ offers itself as an appropriate "mother space" from which the others project under specific conditions.

\section{Momentum transformation}
\label{sec:momentumTransformation}

Before we present gauge transformations in section \ref{sec:localGaugeInvariance}, we want to finish our description of the choice of coordinate system at the point where the state $\psi$ is read off in laboratory space $R^3$. As an example we consider an intrinsic state $\psi$ with configuration variable $u\in U_0(3)$. We want to apply the momentum form ${\rm d}\psi_u$ at a point ${\bf x}$ in laboratory space \footnote{We distinguish dimensionless mathematical spaces $\mathbb{R}^m$ from the dimensionfull coordinate, laboratory space $R^3$.} $R^3$.

We define intrinsic momenta $\pi_j$ in the local, intrinsic base by (c.f. eq. (\ref{eq:dpsiMomentumRead}))
\begin{equation}	\label{eq:intrinsicMomentaDefinition}
 {\rm d}\psi_u(\partial_j)=\partial_j|_u[\psi]=\frac{\rm d}{{\rm d}\alpha}\psi(ue^{\alpha\partial_j})|_{\alpha=0}\equiv \frac{ia}{\hbar}\pi_j(u)
\end{equation}
with momentum dimension corresponding to a length scale $a$ in laboratory space $R^3$. In a fixed base at ${\bf x}\in R^3$ we read off momenta $\pi_j(e)$ by
\begin{equation}	\label{eq:intrinsicMomentaFixedBase}
 {\rm d}\psi_e\left(\frac{\partial}{\partial x_j}\right)= \frac{i}{\hbar}\pi_j(e),\ \ \ j=1,2,3.
\end{equation}
To fix the scale $a$ in (\ref{eq:intrinsicMomentaDefinition}) one needs to know the energy scale $\Lambda=\hbar c/a$ of the phenomenae that one wants to describe. For instance relating $a$ to the classical electron radius $r_{\rm e}=e^2/(4\pi\epsilon_0m_{\rm e}c^2)$ by
\begin{equation}
 \pi a=r_{\rm e}
\end{equation}
and using the projection
\begin{equation}	\label{eq:spaceProjection}
 x_j=a\theta_j
\end{equation}
for a full $U(3)$ intrinsic configuration space, gives satisfactory descriptions of the electron to proton mass ratio \cite{TrinhammerEPL102} and of the baryon spectrum, see figure \ref{fig:NandDeltaSpectrum} \cite{TrinhammerBohrStibiusHiggsPreprint}.

To see how these momenta transform with the choice of intrinsic configuration variable $u$, we exploit the fact that the coordinate fields $\partial_j$ on the manifold are left-invariant as expressed in (\ref{eq:coordinateFieldGeneral}). Rewriting (\ref{eq:intrinsicMomentaDefinition}) we get
\begin{equation}
 \frac{ia}{\hbar}\pi_j(u)=\partial_j|_u[\psi]=u\partial_j|_e[\psi]=u{\rm d}\psi_e(\partial_j)=\frac{ia}{\hbar} u\pi_j(e).
\end{equation}
Summing up we have the transformation property
\begin{equation}	\label{eq:intrinsicMomentumTransformation}
 \pi_j(u)=u\pi_j(e).
\end{equation}
The result (\ref{eq:intrinsicMomentumTransformation}) is not dependent on the intrinsic space being abelian, only on using left-translated coordinate fields (\ref{eq:coordinateFieldGeneral}) and on the differential being linear. We shall see in (\ref{eq:localGaugeInvarianceGeneralDerivative}) that using left-invariant coordinate fields on the intrinsic configurations corresponds to requiring local gauge invariance in laboratory space.

\section{Local gauge transformations}	\label{sec:localGaugeInvariance}

\begin{figure}
\begin{center}
\includegraphics[width=0.45\textwidth]{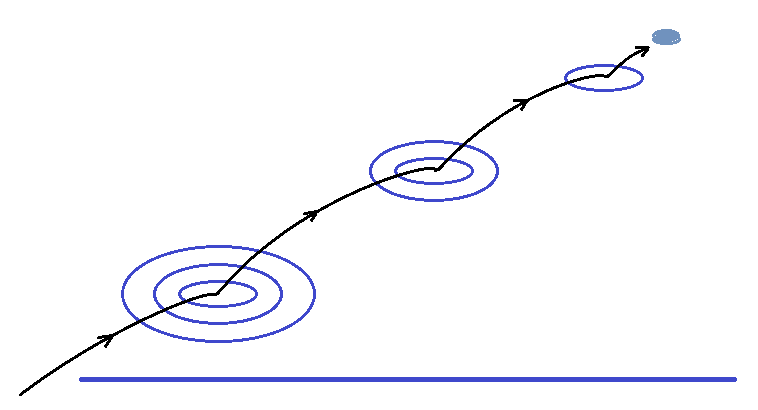}
\caption{Playing ducks and drakes. A stone is thrown at a small angle to a water surface. The horizontal line is the water surface to represent the intrinsic space. The stone (grey-blue) excites waves that are interpreted as field excitations in the point hit by the stone along its direction of flight. The stone may hit the surface several times in one process even in the case of no gravity. This would correspond to multiple scattering on quarks or gluons in the proton with a curved intrinsic configuration space.}
\label{fig:DucksAndDrakesDrawing}
\end{center}
\end{figure}

When the intrinsic momenta
\begin{equation}
 \pi_j(u)=u\pi_j(e|_{{\bf x}})
\end{equation}
and
\begin{equation}
 \pi_j(u')=u'\pi_j(e|_{{\bf x}'})
\end{equation}
are read off from a state $\varphi$ at laboratory space points ${\bf x}$ and ${\bf x}'$ respectively by
\begin{equation}
 \pi_j(u')=\frac{\hbar}{i}{\rm d}\phi_{u'}\left(\frac{\partial}{\partial x'_j}\right)
\end{equation}
and
\begin{equation}
 \pi_j(u)=\frac{\hbar}{i}{\rm d}\varphi_u\left(\frac{\partial}{\partial x_j}\right),
\end{equation}
the origo $e|_{{\bf x}'}$ of the intrinsic configuration space may be induced from a coordinate system at ${\bf x}'$ rotated, translated and/or boosted with respect to that chosen at ${\bf x}$. If the system represented by the state $\varphi$ is not dependent on such changing coordinate choices, we are led to local gauge invariance for the formulation of its dynamics. To see this we consider the kinetic term $\pi_j^\dagger\pi_j$ of some Hamiltonian. We want
\begin{equation}	\label{eq:kineticMomentumTermInvariance}
 \pi_j^\dagger (u')\pi_j(u')=\pi_j^\dagger (u)\pi_j(u)
\end{equation}
independently on the choice of a local phase factor $e^{i\theta({\bf x})}$ on $\varphi$. For $\theta({\bf x})=\theta=\rm constant$, this implies
\begin{equation}
 {\rm d}\varphi^\dagger_{u'}\left(\frac{\partial}{\partial x'_j}\right){\rm d}\varphi_{u'}\left(\frac{\partial}{\partial x'_j}\right)={\rm d}\varphi^\dagger_{u}\left(\frac{\partial}{\partial x_j}\right){\rm d}\varphi_{u}\left(\frac{\partial}{\partial x_j}\right).
\end{equation}
From the identity
\begin{equation}
 {\rm d}\varphi_u\left(\frac{\partial}{\partial x_j}\right)=\partial_j|_u[\varphi]
\end{equation}
we can express the requirement (\ref{eq:kineticMomentumTermInvariance}) as
\begin{equation}
 (\partial_j|_{u'}[\varphi])^\dagger\partial_j|_{u'}[\varphi]=(\partial_j|_{u}[\varphi])^\dagger\partial_j|_{u}[\varphi]
\end{equation}
and use the left-invariance of the coordinate fields
\begin{equation}	\label{eq:leftInvariance}
 \partial_j|_u=u\partial_j|_e
\end{equation}
to get
\begin{equation}
 (u'\partial_j|_e[\varphi])^\dagger u'\partial_j|_e[\varphi]=(u\partial_j|_e[\varphi])^\dagger u\partial_j|_e[\varphi]
\end{equation}
from which follows
\begin{equation}	\label{eq:unitaryCondition}
 (u')^\dagger u'=u^\dagger u.
\end{equation}
In particular we may choose $u=e$ in (\ref{eq:unitaryCondition}) which shows that the configuration variable $u'$, and thus $u$, must be unitary.

For independence on local phase choices
\begin{equation}
 \varphi\rightarrow\varphi'=e^{i\theta({\bf x})}\varphi
\end{equation}
we again consider the kinetic term
\begin{equation}
 (\partial_j|_e[\varphi])^\dagger\partial_j|_e[\varphi]\rightarrow(\partial_j|_e[e^{i\theta({\bf x})}\varphi])^\dagger\partial_j|_e[e^{i\theta({\bf x})}\varphi].
\end{equation}
The term on the right hand side will pick up derivatives of the phase $\theta({\bf x})$. In order to get an invariant formulation one therefore generalizes the derivative $\partial_j$ to
\begin{equation}	\label{eq:generalizedDerivative}
 D_j=\partial_j-iA_j.
\end{equation}
with the gauge fields $A_j$ transforming according to
\begin{equation}
 A'_j({\bf x})=A_j({\bf x})+\partial_j\theta({\bf x}).
\end{equation}
The generalized kinetic term then reads
\begin{equation}
 \left(D_j[\varphi]\right)^\dagger D_j[\varphi]
\end{equation}
and the invariance is expressed as
\begin{equation}	\label{eq:localGaugeInvarianceGeneralDerivative}
 \left(D_j'[\varphi']\right)^\dagger D_j'[\varphi']=\left(D_j[\varphi]\right)^\dagger D_j[\varphi]
\end{equation}
with
\begin{equation}
 D_j'=\partial_j-iA_j'=\partial_j-i(A_j+\partial_j\theta)=D_j-i\partial_j\theta.
\end{equation}
for the transformation of the generalized derivative in (\ref{eq:generalizedDerivative}).

\section{Second quantization}
\label{ch:secondQuantization}

Imagine a scalar state $\varphi$ with intrinsic configuration variable $u$. In section 
\ref{sec:momentumTransformation} we saw how to read off intrinsic momenta $\pi_j(u)$ by applying ${\rm d}\varphi$ to a basis $\frac{\partial}{\partial x_j}$ induced from laboratory space, cf. (\ref{eq:intrinsicMomentaDefinition})
\begin{equation}	\label{eq:momentumComponentsFromDphi}
 \pi_j(u)=-i\hbar\ {\rm d}\varphi_u\left(\frac{\partial}{\partial x_j}\right)=\frac{-i\hbar}{a}\partial_j|_u[\varphi].
\end{equation}
Reading off intrinsic momenta at different laboratory space points ${\bf x}$ and ${\bf x}'$
corresponds to generating conjugate fields $\pi({\bf x})$ and $\pi({\bf x}')$. We take this as the origin of second quantization: Read-offs of intrinsic variables are independent when done at different laboratory space points ${\bf x}$. Below we unfold the details of this conception.

From the commutators
\begin{equation}
 [\hat{x}_i,\hat{p}_j]=i\hbar\delta_{ij}
\end{equation}
in (\ref{eq:quantizationConditionGeneralized}), we introduce raising and lowering operators  in a coordinate representation (see pp. 182 in \cite{SchiffExpectationValue}, see also appendix B in \cite{TrinhammerNeutronProtonMMarXivWithAppendices25Jun2012})
\begin{equation}	\label{eq:creationAnnihilationCoordinateRep}
 \hat{a}_j^\dagger=\frac{1}{\sqrt{2}}\left(\theta_j-i\frac{a}{\hbar}\hat{p}_j\right), \ \ \ \hat{a}_j=\frac{1}{\sqrt{2}}\left(\theta_j+i\frac{a}{\hbar}\hat{p}_j\right)
\end{equation}
to rewrite the momentum component operators as
\begin{equation}	\label{eq:momentumCompAsCreationAnnihilation}
 \hat{p}_j=\frac{-i\hbar}{a}\frac{1}{\sqrt{2}}\left(\hat{a}_j-\hat{a}_j^\dagger\right).
\end{equation}
Note that $a$ without a hat, is the length scale introduced in (\ref{eq:intrinsicMomentaDefinition}) and (\ref{eq:spaceProjection}) for the projection from the intrinsic, toroidal coordinates to laboratory space.

To cast the idea of field generation by momentum read-off into a covariant framework, we 
consider the time-dependent edition of the Schr\"odinger equation
\begin{equation}	\label{eq:timedependentSchroedinger}
 H\Psi(u,t)=i\hbar\frac{\partial}{\partial t}\Psi(u,t),
\end{equation}
where the Hamiltonian is
\begin{equation}	\label{eq:HamiltonianUN}
 H=\frac{\hbar c}{a}\left[-\frac{1}{2}\Delta+V\right]
\end{equation}
with the Laplacian in a polar decomposition \cite{TrinhammerOlafsson}
\begin{equation}  \label{eq:laplacian}  
 \Delta= \sum^N_{j=1} \frac{1}{J^2} \frac{\partial}{\partial \theta_j} J^2 \frac{\partial}{\partial \theta_j} -\frac{1}{\hbar^2}\sum^N_{\substack{ i <  j}} \frac{S^2_{ij} + M^2_{ij}}{8 \sin^2 \frac{1}{2}(\theta_i -\theta_j)}
\end{equation}
for unitary configuration spaces $U(N)$ with $N$ toroidal degrees of freedom. Here the "Jacobian" of the parametrization, the van de Monde determinant, is given by \footnote{Actually $J\equiv \sqrt{J^2}=\sqrt{D^*D}$, $D=\prod_{i<j}^N(e^{i\theta_i}-e^{i\theta_j)}$ is Weyl's expression p. 197 in \cite{Weyl}.}, see p. 197 in \cite{Weyl}
\begin{equation}	\label{eq:JacobianUN}
 J=\prod_{i<j}^N 2\sin\left(\frac{1}{2}(\theta_i-\theta_j)\right)
\end{equation}
and the off-diagonal generators $S_{ij}$ and $M_{ij}$ are given by \cite{TrinhammerOlafsson}
\begin{equation}
 iS_{ij}=\hbar(E_{ij}-E_{ji}),\ \ \ \ iM_{ij}=i\hbar(E_{ij}+E_{ji})
\end{equation}
where $E_{ij}$ in an $N\times N$ matrix representation is the matrix with element $ij$ equal to $1$ and all other elements are zero. For the operators $E_{ij}$ we have the commutation relations
\begin{equation}	\label{eq:uNalgebraEij}
 \left[E_{ij},E_{kl}\right]=\delta_{jk}E_{il}-\delta_{li}E_{kj}.
\end{equation}
For our most interesting case, $N=3$, the Laplacian reads in a more convenient notation
\begin{equation}  \label{eq:laplacianU3}  
 \Delta= \sum^3_{j=1} \frac{1}{J^2} \frac{\partial}{\partial \theta_j} J^2 \frac{\partial}{\partial \theta_j} -\frac{1}{\hbar^2}\sum^3_{\substack{ i <  j \\ k\neq i,j}} \frac{S^2_k + M^2_k}{8 \sin^2 \frac{1}{2}(\theta_i -\theta_j)}
\end{equation}
where the $S_k$s and the $M_k$s, $k=1,2,3$ commute as
\begin{equation}	\label{eq:Kcommutator}
[M_k,M_l]=[S_k,S_l]=-i\hbar\epsilon_{klm}S_m
\end{equation}
and $S_k^2\equiv S_k^\dagger S_k, M_k^2\equiv M_k^\dagger M_k$ like for $S_{ij}^2$ and $M_{ij}^2$ in (\ref{eq:laplacian}).
We note that the $S_k$s commute as intrinsic coordinate angular momenta as known from intrinsic coordinate systems in nuclear physics, see e.g. p. 87 in ref. \cite{BohrMottelsonEDM}.
The polar decomposition in (\ref{eq:laplacianU3}) is analogous to the euclidean Laplacian in polar coordinates
\begin{equation} 	\label{eq:radialLaplacian}
  \Delta_{e,polar}=\frac{1}{r^2}\frac{\partial}{\partial r}r^2\frac{\partial}{\partial r}-\frac{1}{r^2}{\bf L}^2,
\end{equation}
for instance used in solving the hydrogen atom.

The stationary Schr\"odinger equation on $U(3)$
\begin{equation}	\label{eq:schroedingerU3}
 \frac{\hbar c}{a}\left[-\frac{1}{2}\Delta+V\right]\Psi(u)={\cal E}\Psi(u)
\end{equation}
with intrinsic potential inspired by Manton's action from lattice gauge theory \cite{Manton}
\begin{equation}	\label{eq:intrinsicGeodeticPotential}
 V=\frac{1}{2}{\rm Tr}\chi^2=\sum_{j=1}^3w(\theta_j), \ \ \ u=e^{i\chi},
\end{equation}
where \cite{Milnor}
\begin{equation}	\label{eq:periodicPotentialOneDim}
 w(\theta)=\frac{1}{2}(\theta-n2\pi)^2,\ \ \theta\in[(2n-1)\pi,(2n+1)\pi],\ \ n\in\mathbb{Z},
\end{equation}
can be solved from a factorization of the wavefunction into a torodial part $\tau$ and an off-toroidal part $\Upsilon$
\begin{equation}	\label{eq:factorizationHydrogenAnalogueWithVariables}
 \Psi(u)=\tau(\theta_1,\theta_2,\theta_3)\Upsilon(\alpha_4,\alpha_5,\alpha_6,\alpha_7,\alpha_8,\alpha_9).
\end{equation}
The off-toroidal degrees of freedom can be integrated out by a factorization of the measure  \cite{Hurwitz, Haar}. This yields a total potential
\begin{equation}	\label{eq:potentialWtotal}
 W=-1+\frac{1}{2}\cdot\frac{1}{3}\sum^3_{\substack{ i <  j \\ k\neq i,j}} \frac{4}{8 \sin^2 \frac{1}{2}(\theta_i -\theta_j)}+\sum_{j=1}^3w(\theta_j)
\end{equation}
for the minimal value $4$ of $({\bf S}^2+{\bf M}^2)/\hbar^2$ and the periodic potential $w(\theta)$ given in (\ref{eq:periodicPotentialOneDim}). The fractional factor $\frac{1}{3}$ in the centrifugal term comes from exploiting the symmetry under interchange of the toroidal angles $\theta_1,\theta_2,\theta_3$ in the evaluation of the integral of the centrifugal term
\begin{equation}	\label{eq:centrifugalTermC}
 C=\frac{1}{2}\frac{1}{\hbar^2}\sum^3_{\substack{ i <  j \\ k\neq i,j}} \frac{S^2_k + M^2_k}{8 \sin^2 \frac{1}{2}(\theta_i -\theta_j)}
\end{equation}
over the six off-toroidal variables $\alpha_4,\alpha_5,\alpha_6,\alpha_7,\alpha_8,\alpha_9$ in the off-toroidal part $\Upsilon$ of the wavefunction in (\ref{eq:factorizationHydrogenAnalogueWithVariables}). In the toroidal Schr\"odinger equation (\ref{eq:toroidalSchroedinger}), the non-derivative terms from the $U(3)$ Laplacian $\Delta$ in (\ref{eq:schroedingerU3}) have been included in the total potential $W$ leaving us with the equivalent dimensionless Schr\"odinger eguation
\begin{equation}	\label{eq:toroidalSchroedinger}
 \left[-\frac{1}{2}\Delta_{\rm e}+W\right]R(\theta_1,\theta_2,\theta_3)={\rm E}R(\theta_1,\theta_2,\theta_3)
\end{equation}
with dimensionless eigenvalues ${\rm E}\equiv{\cal E}/\Lambda,\ \Lambda\equiv\frac{\hbar c}{a}$ and with a euclidean Laplacian $\Delta_{\rm e}$ containing only the second order derivatives in the toroidal angles $\theta_j$
\begin{equation}	
 \Delta_{\rm e}=\frac{\partial^2}{\partial\theta_1^2}+\frac{\partial^2}{\partial\theta_2^2}+\frac{\partial^2}{\partial\theta_3^2}.
\end{equation}
Accurate eigenvalues of (\ref{eq:toroidalSchroedinger}) can be found by a Rayleigh-Ritz method from expansions of the measure scaled toroidal wave function $R\equiv J\tau$ on Slater determinants like
\begin{equation}	\label{eq:fpqr}
 f_{pqr}=\begin{vmatrix}
  \cos p\theta_1 & \cos p\theta_2 & \cos p\theta_3\\
  \sin q\theta_1 & \sin q\theta_2 & \sin q\theta_3\\
  \cos r\theta_1 & \cos r\theta_2 & \cos r\theta_3
 \end{vmatrix},
\end{equation}
for electrically neutral states and
\begin{equation}	\label{eq:gpqr}
 g_{pqr}=e^{i\frac{\theta_1}{2}}e^{i\frac{\theta_2}{2}}e^{i\frac{\theta_3}{2}}
 \begin{vmatrix}
  \cos p\theta_1 & \cos p\theta_2 & \cos p\theta_3\\
  \sin q\theta_1 & \sin q\theta_2 & \sin q\theta_3\\
  \cos r\theta_1 & \cos r\theta_2 & \cos r\theta_3
 \end{vmatrix}
\end{equation}
for electrically charged states. These basis can be integrated analytically wherefore accurate eigenvalues of the Hamiltonian can be found, see section \ref{sec:RayleighRitz}.

\section{The case for $U(3)$}
\label{sec:caseForU3}

The Lie group $U(3)$ has nine generators which can be related to nine kinematic generators from laboratory space $R^3$, namely the three momentum operators which correspond to the three toroidal generators $T_j,j=1,2,3$, the three rotation operators which correspond to the three intrinsic angular momentum operators $S_j$ and finally the three Laplace-Runge-Lenz operators which correspond to the three "mixing" operators $M_j$.

Let us therefore consider the particular case for $N=3$ \cite{TrinhammerIQM1}, where the potential in (\ref{eq:HamiltonianUN}) is time-independent and depends only on the toroidal angles of $U(3)$. We factorize the time-independent wavefunction $\Psi$ into a toroidal part $\tau$ and an off-toroidal part $\Upsilon$ analogous of the radial part and the spherical harmonics introduced in solving the hydrogen atom, i.e. we write
\begin{equation}	\label{eq:factorizationHydrogenAnalogue}
 \Psi(u)=\tau\Upsilon.
\end{equation}
In that way the measure-scaled, time-dependent wave function $\Phi(u,t)=J\Psi(u,t)$ for a time-independent, toroidally symmetric potential $V(u,t)=V(\theta_1,\theta_2,\theta_3)$ becomes
\begin{equation}	\label{eq:timedependentWavefunction}
 \Phi(u,t)=e^{-i{\cal E}t/\hbar}R(\theta_1,\theta_2,\theta_3)\Upsilon\equiv {\cal R}(\theta_0,\theta_1,\theta_2,\theta_3)\Upsilon
\end{equation}
with {\it measure-scaled} toroidal wavefunction
\begin{equation}	\label{eq:RmeasureScaledWavefunction}
 R(\theta_1,\theta_2,\theta_3)=J(\theta_1,\theta_2,\theta_3)\tau(\theta_1,\theta_2,\theta_3)
\end{equation}

In (\ref{eq:timedependentWavefunction}) we scaled the time projection by the same length scale $a$ as we used for the space projections (\ref{eq:spaceProjection}) and thus define for the toroidal "time angle" $\theta_0$ to be determined by
\begin{equation}	\label{eq:timeProjection}
 a\theta_0=ict
\end{equation}
where $t$ is the time parameter in spacetime and $c$ is the speed of light in empty space. Further, the time derivative corresponds to the time coordinate field generator
\begin{equation}	\label{eq:timeCoordinateField}
 \frac{\partial}{\partial\theta_0}=\frac{a}{ic}\frac{\partial}{\partial t}=-H/\Lambda\equiv iT_0
\end{equation}
where $\Lambda=\hbar c/a$ is the energy scale related to the length scale $a$ involved in the projection (\ref{eq:spaceProjection}) to laboratory space.
We can generalize this to suit the left-invariance in (\ref{eq:leftInvariance}) such that for $\tilde{u}=e^{i\theta_0T_0}u$ we have
\begin{equation}	\label{eq:leftInvarianceTilde}
 \partial_0 |_{\tilde{u}}=\frac{d}{d\alpha}(\tilde{u}\exp\alpha iT_0)|_{\alpha=0}=\tilde{u}iT_0
\end{equation}
with the generator $iT_0$ and corresponding time form $d\theta_0$ fulfilling $d\theta_0(\partial_0)=1$. In that way the dynamics inherent in the time-dependent Schr\"odinger equation (\ref{eq:timedependentSchroedinger}) can be embedded in $U(1)\otimes U(3)$ based on four-dimensional representations like
\begin{equation}	\label{eq:Trepresentation}
 T_0=\begin{Bmatrix} i&0&0&0\\ 0&0&0&0\\ 0&0&0&0\\ 0&0&0&0 \end{Bmatrix}, \ \ 
 T_3=\begin{Bmatrix} 0&0&0&0\\ 0&0&0&0\\ 0&0&0&0\\ 0&0&0&1 \end{Bmatrix}.
\end{equation}
and
\begin{equation}
 S_3=\begin{Bmatrix} 0&0&0&0\\ 0&0&-i&0\\ 0&i&0&0\\ 0&0&0&0 \end{Bmatrix}, \ \ 
 M_3=\begin{Bmatrix} 0&0&0&0\\ 0&0&1&0\\ 0&1&0&0\\ 0&0&0&0 \end{Bmatrix}.
\end{equation}

Embedding in a product $U(1)\otimes U(3)$ with the time dimension separated from the intrinsic configuration space $U(3)$ allows for time not to be a dynamical quantum variable \cite{HelgeKragh} and at the same time to have a four-dimensional formulation of the fields in spacetime projection.

We now consider projections along the torus $U(1)\otimes U_0(3)$ given by
\begin{equation}
 \tilde{u}=e^{i\theta_0T_0+i\boldsymbol\theta\cdot\bf T}
\end{equation}
where we define
\begin{equation}	\label{eq:toroidalMomentumExpandedOnBaseGenerators}
 {\bf T}=q_1T_1+q_2T_2+q_3T_3=\frac{a}{\hbar}\bf p
\end{equation}
with the three $iT_j$ as the toroidal generators of $U(3)$, seen in a $3\times3$ matrix representation in (\ref{eq:toroidalGeneratorsU3}) and $\boldsymbol{\theta}=(\theta_1,\theta_2,\theta_3)$. The space projection $dR$ for $R$ - which would result from using the definition in (\ref{eq:dpsiDefinition}) and lead to the momentum components in (\ref{eq:momentumComponentsFromDphi}) - is then replaced by the spacetime projection
\begin{equation}	\label{eq:dSspacetimeProjection}
 d{\cal R}\equiv \frac{a}{-i\hbar}\pi_\mu d\theta_\mu.
\end{equation}

If we want to project the structure inherent in the solution $\cal R$ on a given base at a particular event in the Minkowski spacetime we must consider the directional derivative at a fixed base $\{ iT_\mu\}$, i.e.
\begin{equation}
 \left(iT_\mu\right)_{\tilde{u}}\left[{\cal R}\right]= d{\cal R}_{\tilde{u}}(iT_\mu).
\end{equation}
From the left-invariance (\ref{eq:leftInvariance}) and (\ref{eq:leftInvarianceTilde}) of the coordinate fields $\partial_\mu$, $\mu=0,1,2,3$ we have
\begin{gather}	\label{eq:directionalDerivativeDS}
 d{\cal R}_{\tilde{u}}(iT_\mu)=d{\cal R}_{\tilde{u}}(\tilde{u}^{-1}\partial_\mu)=\tilde{u}^{-1}d{\cal R}_{\tilde{u}}(\partial_\mu)\\ \nonumber
 =\tilde{u}^{-1}\frac{a}{-i\hbar}\pi_\mu(\tilde{u})
\end{gather}
where the latter expression uses (\ref{eq:dSspacetimeProjection}) and
\begin{equation}
 {\rm d}\theta_\mu(\partial_\nu)=\delta_{\mu\nu}.
\end{equation}
From the pull-back ${\cal R}^*$ of $\cal R$ from $U(1)\otimes U_0(3)$ to $\mathbb{R}\otimes\mathbb{R}^3$ given by
\begin{equation}	\label{eq:pullBack}
 {\cal R}^*={\cal R}\circ \exp :\ \ \mathbb{R}\oplus\mathbb{R}^3\rightarrow\ \ U(1)\otimes U_0(3)\rightarrow \mathbb{C}
\end{equation}
we also have the directional derivative using (\ref{eq:directionalDerivativeDS})
\begin{equation}	\label{eq:directionalDerivativesFixedBase}
 d{\cal R}_{\tilde{u}}(iT_\mu)=\tilde{u}^{-1}\frac{\partial{\cal R}^*}{\partial\theta_\mu}\left(\frac{{\cal E}t}{\hbar},\frac{p_1x^1}{\hbar},\frac{p_2x^2}{\hbar},\frac{p_3x^3}{\hbar}\right).
\end{equation}
We use $\theta_0=ict/a$ from (\ref{eq:timeProjection}), introduced $q_0={\cal E}/\Lambda=\frac{a}{\hbar}\frac{\cal E}{c}$ and get for the phase factor
\begin{align}
 \tilde{u}^{-1}=\exp(q_0\theta_0-i{\bf q}\cdot{\boldsymbol\theta})=\exp\left(i\frac{{\cal E}t}{\hbar}-i\frac{{\bf p\cdot x}}{\hbar}\right)\\ \nonumber
 \equiv\exp(i\omega t-i{\bf k\cdot x})\equiv e^{ik\cdot x}
\end{align}
with $\hbar\omega=\cal E$ and $\hbar\bf k=p$.

In the pull-back (\ref{eq:pullBack}) we have used the coordinate fields as induced base
\begin{equation}
 \partial_\mu |_{\tilde{u}}=\frac{\partial}{\partial\theta_\mu} |_{\tilde{u}}=d(\exp)_{\exp^{-1}(\tilde{u})}(e_\mu)
\end{equation}
where $\{ e_\mu \}$ is a base in the parameter space, i. e. a base at the event in Minkowski space, see eq. (\ref{eq:inducedBasis}) and fig. \ref{fig:projectionParameterSpace} where the manifold $M$ in the present case could be $U(1)\otimes U(3)$, the inverse map $x^{-1}=\exp$ and the complex-valued function $f$ would be the measure-scaled wavefunction  $\cal R$ introduced in (\ref{eq:timedependentWavefunction}). 

For the mapping between spacetime and the torus, we have in particular the following corresponding bases
\begin{equation}
 \partial_\mu |_e=iT_\mu=e_\mu
\end{equation}
and can write at each event $x$
\begin{equation}
 x=x^\mu e_\mu
\end{equation}
with contravariant spacetime coordinates $x^\mu$ and covariant base and with Einsteins summation convention as throughout. For the induced base $\{ iT_\mu\}$ at the origo $e=I$ in the $4D$ torus we may choose a representation as in (\ref{eq:Trepresentation}) with
\begin{equation}
 iT_0=\begin{Bmatrix}-1&0&0&0\\ 0&0&0&0\\ 0&0&0&0\\ 0&0&0&0 \end{Bmatrix}, \ \ 
 iT_1=\begin{Bmatrix} 0&0&0&0\\ 0&i&0&0\\ 0&0&0&0\\ 0&0&0&0 \end{Bmatrix},
\end{equation}
and 
\begin{equation}
 iT_2=\begin{Bmatrix} 0&0&0&0\\ 0&0&0&0\\ 0&0&i&0\\ 0&0&0&0 \end{Bmatrix}, \ \ 
 iT_3=\begin{Bmatrix} 0&0&0&0\\ 0&0&0&0\\ 0&0&0&0\\ 0&0&0&i \end{Bmatrix}.
\end{equation}
With matrix multiplication and trace-taking as metric among the $iT_\mu $s this corresponds to that of the $e_\mu$s with scalar products
\begin{equation}
 x^\mu x_\nu=x^\mu g_{\mu\nu}x^\nu=(x^0)^2-(x^1)^2-(x^2)^2-(x^3)^2
\end{equation}
from a metric tensor with non-zero components $g_{00}=1, g_{11}=g_{22}=g_{33}=-1$.
We get accordingly
\begin{equation}
 \left(\theta_\mu iT_\mu\right)\left(\theta_\nu iT_\nu\right)=(\theta_0)^2-(\theta_1)^2-(\theta_2)^2-(\theta_3)^2.
\end{equation}
We see that the generators $\{ iT_\mu\}$ carry the Minkowski metric intrinsically. Thus as
\begin{equation}	\label{eq:cbaseMinkowski}
 e_\mu \cdot e_\nu = g_{\mu\nu}
\end{equation}
we likewise have
\begin{equation}	\label{eq:TbaseMinkowski}
 {\rm Tr}\left(iT_\mu iT_\nu\right)=g_{\mu\nu}.
\end{equation}
One may want to check joggling indices for the Minkowski base (\ref{eq:cbaseMinkowski}) with the base vectors visible like in the following example for the scalar product $x^\mu y_\mu$ written as
\begin{gather}
 x\cdot y=x^\mu e_\mu\cdot y_\nu e^\nu=x^\mu e_\mu\cdot g_{\nu\sigma}y^\sigma e^\nu\\ \nonumber
 =x^\mu y^\sigma e_\mu\cdot g_{\nu\sigma}e^\nu=x^\mu y^\sigma e_\mu\cdot g_{\sigma\nu}e^\nu\\ \nonumber
 =x^\mu y^\sigma e_\mu\cdot e_\sigma=(x^0y^0)-(x^1y^1)-(x^2y^2)-(x^3y^3).
\end{gather}
Here we used that the metric tensor $g_{\mu\nu}$ is symmetric in $\mu\nu$.

\section{Generation of a quantum field}

We now return to the interpretation of momentum components as directional derivatives in (\ref{eq:directionalDerivativesFixedBase}). By comparison with (\ref{eq:directionalDerivativeDS}), we infer intrinsic momenta
\begin{equation}	\label{eq:intrinsicMomentumComponents}
 p_\mu=\frac{-i\hbar}{a}\frac{\partial{\cal R}^*}{\partial\theta_\mu}|_e.
\end{equation}
Using (\ref{eq:momentumCompAsCreationAnnihilation}) and (\ref{eq:directionalDerivativesFixedBase}) we have for a fixed basis projection to space
\begin{gather}	\label{eq:dRasAnnihilationCreationExpression}
 {\rm d}{\cal R}_{\tilde{u}}(iT_j)\\ \nonumber
 =e^{ik\cdot x}\frac{1}{\sqrt{2}}\left(\hat{a}_j({\bf k})-\hat{a}^\dagger_j({\bf k}))\right){\cal R}^*|_{(\omega t,k_1x^1,k_2x^2,k_3x^3)},\\ \nonumber j=1,2,3.
\end{gather}
We interpret (\ref{eq:dRasAnnihilationCreationExpression}) as Fourier components of a  conjugate momentum field 
\begin{equation}
 \pi(x)=\int\frac{{\rm d}^3{\bf k}}{\sqrt{(2\pi)^3}}\sqrt{\frac{{\cal E}({\bf k})}{2}}\left(\hat{a}({\bf k})-\hat{a}^\dagger({\bf k}))\right)e^{ik\cdot x}
\end{equation}
to be excited at the spacetime coordinate $x$ where the intrinsic momenta are read off according to (\ref{eq:intrinsicMomentumComponents}). We incorporated the $1/\sqrt{2}$ prefactor on the annihilation and creation operators into the standard normalization of the momentum field and omitted a factor $-i\hbar/a$. The above expansion compares closely with standard expressions in the construction of quantum fields \cite{LancasterBlundellAnnihilationCreationInQuantumField, WeinbergAnnihilationCreationInQuantumField}. Note only, that $\hat{a}$ and $\hat{a}^\dagger$ have same sign phase factors $e^{ikx}$. Still, as we shall see, we get a standard propagator.

If we uphold the canonical relation
\begin{equation}
 \dot{\phi}(x)=\pi^\dagger(x),
\end{equation}
where 'dot' represents derivation with respect to time $t$, we get for the field $\phi$ conjugate to the momentum field $\pi$, that
\begin{equation}	\label{eq:phiFieldOfX}
 \phi(x)=-i\int\frac{{\rm d}^3{\bf k}}{\sqrt{(2\pi)^3}}\frac{1}{\sqrt{2{\cal E}({\bf k})}}\left(\hat{a}({\bf k})-\hat{a}^\dagger({\bf k})\right)e^{-ik\cdot x}.
\end{equation}
According to (\ref{eq:dRasAnnihilationCreationExpression}), this field is meant to act on the Fock space spanned by the pulled back solutions ${\cal R}^*$ for the toroidal part of the wavefunction in (\ref{eq:timedependentWavefunction})

For (\ref{eq:phiFieldOfX}) to represent a useful expression on which to build a quantum field theory, we must check that we can get a standard expression for the free propagator as on pp. 156 in \cite{LancasterBlundellAnnihilationCreationInQuantumField}
\begin{equation}	\label{eq:propagatorDefinition}
 \Delta(x,y)=<0|T\phi(x)\phi^\dagger(y)|0>,
\end{equation}
where $|0>$ is the vacuum state and $T$ is Wick's time ordering prescription
\begin{equation}	\label{eq:propagatorHeavisideSum}
 T\phi(x)\phi^\dagger(y)=\theta(x^0-y^0)\phi(x)\phi^\dagger(y)+\theta(y^0-x^0)\phi^\dagger(y)\phi(x)
\end{equation}
expressed by the help of the Heaviside step function
\begin{equation}
 \theta(x^0-y^0)=1\ \ \ {\rm for}\ \ x^0>y^0,\ \ \ \theta(x^0-y^0)=0\ \ \ {\rm for}\ \ \ x^0<y^0.
\end{equation}
We follow Lancaster and Blundell. For this, we first rewrite (\ref{eq:phiFieldOfX}) to get
\begin{gather}	\label{eq:phiFieldDaggerOfY}
 \phi^\dagger (y)|0>\\ \nonumber
 =i\int\frac{{\rm d}^3{\bf k}}{\sqrt{(2\pi)^3}}\frac{1}{\sqrt{2{\cal E}({\bf k})}}\left(\hat{a}^\dagger({\bf k})|0>-\hat{a}({\bf k})|0>\right)e^{ik\cdot y}.
\end{gather}
The annihilation operator $\hat{a}$ gives $0$ on the vacuum state $|0>$ and thus
\begin{equation}	\label{eq:phiFieldDaggerWithKstate}
 \phi^\dagger (y)|0>=i\int\frac{{\rm d}^3{\bf k}}{\sqrt{(2\pi)^3}}\frac{1}{\sqrt{2{\cal E}({\bf k})}}|{\bf k}>e^{ik\cdot y}.
\end{equation}
To get $<0|\phi(x)$ we exchange $y$ with $x$ and $k$ with $k'$ in (\ref{eq:phiFieldDaggerWithKstate}) and take the conjugate to get
\begin{equation}	\label{eq:phiFieldWithKprimeState}
 <0|\phi(x)=-i\int\frac{{\rm d}^3{\bf k'}}{\sqrt{(2\pi)^3}}\frac{1}{\sqrt{2{\cal E}({\bf k}')}}<{\bf k}'|e^{-ik'\cdot x}.
\end{equation}
With
\begin{equation}
 <{\bf k}'|{\bf k}>=\delta^3({\bf k}'-{\bf k})
\end{equation}
we have for the first term of (\ref{eq:propagatorHeavisideSum})
\begin{gather}
 <0|\phi(x)\phi^\dagger(y)|0>\\ \nonumber
 =\int\frac{{\rm d}^3{\bf k'}{\rm d}^3{\bf k}}{(2\pi)^3\sqrt{2{\cal E}({\bf k}')2{\cal E}({\bf k})}}\delta^3({\bf k}'-{\bf k})e^{-ik'\cdot x+k\cdot y}
\end{gather}
in the time-ordered expression. Exploiting the delta function to do the ${\rm d}^3{\bf k}'$ integral gives
\begin{equation}
 <0|\phi(x)\phi^\dagger(y)|0>=\int\frac{{\rm d}^3{\bf k}}{(2\pi)^3(2{\cal E}({\bf k}))}e^{-ik\cdot(x-y)}.
\end{equation}
Calculation of the second term $<0|\phi^\dagger(y)\phi(x)|0>$ in the time-ordered expression is done similarly. Still following Lancaster and Blundell we consider
\begin{gather}	\label{eq:phiFieldOfXonVacuum}
 \phi(x)|0>\\ \nonumber
 =-i\int\frac{{\rm d}^3{\bf k}}{\sqrt{(2\pi)^3}}\frac{1}{\sqrt{2{\cal E}({\bf k})}}\left(\hat{a}({\bf k})|0>-\hat{a}^\dagger({\bf k})|0>\right)e^{-ik\cdot x}.
\end{gather}
From this we get
\begin{equation}	\label{eq:phiFieldWithKstate}
 \phi(x)|0>
 =i\int\frac{{\rm d}^3{\bf k}}{\sqrt{(2\pi)^3}}\frac{1}{\sqrt{2{\cal E}({\bf k})}}|{\bf k}>e^{ik\cdot x}
\end{equation}
and use the trick of variable change $x\rightarrow y$, $k\rightarrow k'$ and conjugation to get
\begin{equation}	\label{eq:phiDaggerFieldWithKprimestate}
 <0|\phi^\dagger(y)=-i\int\frac{{\rm d}^3{\bf k}'}{\sqrt{(2\pi)^3}}\frac{1}{\sqrt{2{\cal E}({\bf k}')}}<{\bf k}'|e^{-ik'\cdot y}.
\end{equation}
Putting together (\ref{eq:phiDaggerFieldWithKprimestate}) and (\ref{eq:phiFieldWithKstate}) and integrating over ${\bf k}'$, we get
\begin{equation}
 <0|\phi^\dagger(y)\phi(x)|0>=\int\frac{{\rm d}^3{\bf k}}{(2\pi)^3(2{\cal E}({\bf k}))}e^{ik\cdot(x-y)}.
\end{equation}
\begin{widetext}
For the propagator (\ref{eq:propagatorDefinition}) we then get the expression
\begin{equation}	\label{eq:propagatorHeaviside}
 \Delta(x,y)=\int\frac{{\rm d}^3{\bf k}}{(2\pi)^3(2{\cal E}({\bf k}))}\left[\theta(x^0-y^0)e^{-ik\cdot(x-y)}+\theta(y^0-x^0)e^{ik\cdot(x-y)}\right].
\end{equation}
\end{widetext}
This is identical to Lancaster and Blundell, see p. 158 in \cite{LancasterBlundellAnnihilationCreationInQuantumField}. The order among $x$ and $y$ in the propagator phases originates in the transformation by $\tilde{u}^{-1}$ in (\ref{eq:directionalDerivativesFixedBase}). In other words, it depends on the orientation chosen, when parametrizing the toroidal angles in $U(1)\otimes U_0(3)$. The choice is a convention, and we suggest that once a choice has been made for particle creation the opposite choice should be identified with antiparticle creation. The consistency of this interpretation is supported by the standard expression for the propagator in (\ref{eq:propagatorHeaviside}).

From here on the machinery of standard quantum field theory takes over - in so far as projections from intrinsic dynamics are needed to derive observable phenomenology as is the case e.g. in scattering processes among particles in laboratory space. On the other hand, from the intrinsic viewpoint, we should exploit the possibility to derive relations from the intrinsic conception where spacetime projections are {\it not} needed, e.g. in baryon mass spectroscopy from (\ref{eq:U3hamiltonian}) - or in other relations concerning the concept of mass where it can be related to intrinsic structure like in section \ref{sec:electroweakMixingAngle}.

\section{Spinor coefficients}
\label{sec:spinorCoefficients}

Next we discuss the spin part. Anthony Duncan \cite{DuncanUspinor} derives - from general considerations of covariance under transformations of the homogeneous Lorentz group - the following expression for a local covariant field of any spin $j$ and Lorentz representation
\begin{widetext}
\begin{equation}
 \phi_{ab}^{AB}(x)=\int\frac{{\rm d}^3k}{(2\pi)^{3/2}\sqrt{2E(k)}}\sum_\sigma\left(u_{ab}^{AB}(\vec{k},\sigma)e^{-ik\cdot x}\hat{a}(\vec{k},\sigma)+(-)^{2B}(-)^{j-\sigma}u_{ab}^{AB}(\vec{k},-\sigma)e^{ik\cdot x}\hat{a}^{c\dagger}(\vec{k},\sigma)\right).
\end{equation}
\end{widetext}
Here the $u$ spinor coefficient function for spin $\sigma$ is
\begin{equation}	\label{eq:uSpinorDuncan}
  u^{AB}_{ab}({\bf k},\sigma)=\sum_{a',b'}(e^{-\xi \bf k\cdot A})_{aa'}(e^{+\xi \bf k\cdot B})_{bb'}<ABa'b'|j\sigma>,
\end{equation}
where $<ABa'b'|j\sigma>$ are Clebsch-Gordan coefficients and 
\begin{equation}	\label{eq:AandBdefinition}
  A_j\equiv\frac{1}{2}(J_j-iK_j)\ \ {\rm and}\ \ B_j\equiv\frac{1}{2}(J_j+iK_j)
\end{equation}
with rotation generators $J_j$ and boost generators $K_j$ for the homogeneous Lorentz group and with boost parameter $\xi$ determined by $\cosh\xi=\frac{E({\bf k})}{m}$. See also Steven Weinberg pp. 230 in \cite{WeinbergAnnihilationCreationInQuantumField}. We should be able to use similar definitions as in (\ref{eq:AandBdefinition}) to decouple our $S$ and $M$ algebras in two mutually commuting $su(2)$ algebras
\begin{equation}
  [A'_i,A'_j]=i\hbar\epsilon_{ijk}A'_k,\ \ [B'_i,B'_j]=i\hbar\epsilon_{ijk}B'_k,\ \ [A'_i,B'_j]=0 
\end{equation}
since our $S_k$s and $M_k$s share commutation algebra
\begin{equation}
  [M_k,M_l]=[S_k,S_l]=-i\hbar\epsilon_{klm}S_m,\ \ [S_k,M_l]=\pm i\hbar\epsilon_{klm} M_m
\end{equation}
with the generators $J_j$ and $K_j$ from the homogeneous Lorentz group
\begin{equation}
  -[K_i,K_j]=[J_i,J_j]=i\hbar\epsilon_{ijk}J_k\ \ [J_i,K_j]=i\hbar\epsilon_{ijk}K_k.
\end{equation}
Our $S_k$s correspond to the $J_k$s apart from a sign change in our $S_k$s reflecting their role as intrinsic, body fixed, angular momentum operators (see e.g. p. 87 in \cite{BohrMottelsonEDM}). The procedure of decoupling the algebras corresponds to the off-toroidal part $\Upsilon$ in (\ref{eq:factorizationHydrogenAnalogueWithVariables}) being factorizable into
\begin{equation}
  \Upsilon(\alpha_4,\alpha_5,\alpha_6,\alpha_7,\alpha_8,\alpha_9)={\cal A}(\alpha_1,\alpha_2,\alpha_3){\cal B}(\beta_1,\beta_2,\beta_3),
\end{equation}
where $\alpha_j$ and $\beta_j, j=1,2,3$ (being complicated functions of $\alpha_4,\cdots,\alpha_9$) parametrize Lie groups $SU_A'(2)$ and $SU_B'(2)$ generated by $\{ A'_j\}$ and $\{ B'_j\}$ respectively.

We have called the set $\{ M_j\}$ an intrinsic edition of the boosts from the Lorentz algebra. We have also likened them to components of a Laplace-Runge-Lenz vector \cite{TrinhammerBohrStibiusHiggsPreprint} - Whichever association one prefers, it is essential to note that the structural information in $\Upsilon$ is blurred when $\Phi$ is represented by $d\cal R$ in spacetime in stead of by the full $d\Phi$. When one inserts the spinors (\ref{eq:uSpinorDuncan}) from the rotation and boost algebra of spacetime, the structural details from the intrinsic spin and Laplace-Runge-Lenz algebra will be lost - mainly because the intrinsic configuration space is not a linear vector space but rather a curved manifold with a Lie algebra. Thus a full-fledged spacetime quantum field can only be an approximate representation of the intrinsic dynamics carried by $\Phi={\cal R}\Upsilon$. The remaining details will have to be described by possibly adding higher order terms to a spacetime Lagrangian in such a way that the higher order terms can emulate the curved structure of the configuration manifold. However, it is our conjecture that the most important terms are already secured by the equivalence of the {\it exponential mapping} between algebra and intrinsic group configuration manifold and the {\it exponential expansion} inherent in the Feynman rules of quantum field theory, most clearly expressed in the path integral formulation of the transition kernel \cite{PathIntegralsPedestrians}
\begin{equation}
 K({\bf x}_1,t_1;{\bf x}_0,t_0)=\int_{{\bf x}_0,t_0}^{{\bf x}_1,t_1} {\cal D}[{\bf x}(t)]e^{\frac{i}{\hbar}S[{\bf x}(t)]}
\end{equation}
from spacetime point $({\bf x}_0,t_0)$ to $({\bf x}_1,t_1)$ under the influence of the action $S$ of the Lagrangian density along possible trajectories ${\bf x}(t)$.

Lagrangians of free fields correspond to the linear approximations based on the algebra as opposed to the group structure which in spacetime projections manifests itself as higher order interaction terms. We think this to be in line with Steven Weinberg's considerations in the following citation: "...there began to be doubts whether the quantum field theory of the Standard Model was truly a fundamental theory or just the first term in an effective field theory in which there appears every possible interaction allowed by symmetries, the nonrenormalizable as well as the renormalizable ones, perhaps an effective field theory that arises from a deeper underlying theory that might not even be a quantum field theory at all. Of course I'm thinking here about string theory, but that's not the only possibility...". After these considerations, however, Weinberg ends by a "renewed optimism for quantum field theory as part of a description of nature at the most fundamental level". \cite{WeinbergPraha}. Our point is exactly this: The intrinsic configuration space limits the structure as to which quantum fields will actually come to life in laboratory space. But once they live in laboratory space, they can be handled by quantum field theory.

\section{Intrinsic variable}
\label{ch:intrinsicVariable}

We distinguish between intrinsic and interior. Interior refers to something inside a certain spacetime region; the region may be of finite or infinite size but {\it not} pointlike (a point has no interior). Interior variables might be positions of the electrons and the nucleus "inside" an atom. Here the positions of the electrons relative to the nucleus (and to each other) are interior variables of the system and serve as configuration variables based on which one may formulate a dynamical theory like the Schr\"odinger equation in non-relativistic quantum mechanics. However, being variables in spacetime, such position variables are subject to the rules of the general theory of relativity and one has therefore been forced to seek a coexistence between quantum mechanics and relativity, i.e. quantum field theory to get a proper description (so far restricted to quantum mechanics and the special theory of relativity). Intrinsic variables on the other hand are variables in an intrinsic space {\it not} affected by gravity. The intrinsic space should therefore {\it not} be likened to extra spacetime dimensions as in string theory. In stead, the intrinsic space should be thought of as "orthogonal" to spacetime. In other words, an intrinsic configuration space like $U(3)$ as introduced in sec. \ref{sec:caseForU3} is thought to be excitable at every spacetime point $P$ as illustrated in fig. \ref{fig:MaldacenaWithTori}. One might liken this excitability in every point to an effective mean field theory, but as dynamics can be defined and treated in the intrinsic space to yield particle resonance spectra without reference to field theory, we prefer to consider the idea of intrinsic configuration space as a fundamental outset. We saw in section \ref{ch:secondQuantization} how to generate quantum fields from this outset, if need be (and need {\it definitely} exists, e.g. for all kinds of scattering phenomenae, which per construction take place in spacetime). In sections \ref{sec:partonDistributionFunctions}, \ref{sec:protonSpinStructureFunction} and \ref{sec:protonMagneticMoment} we shall derive parton distribution functions, spin structure function and magnetic dipole moment for the proton as an application of fields generated from intrinsic configurations. But first we focus on the intrinsic variables and what can be derived from them.

\section{Intrinsic potential - Bloch phases}
\label{sec:intrinsicPotentialBlochPhase}

Let us consider a one-dimensional example of intrinsic dynamics. We let the configuration variable $u=e^{i\theta}$ belong to the Lie group $U(1)$, i.e.
\begin{equation}
 u=e^{i\theta},\ \ \ \theta\in\mathbb{R}
\end{equation}
where $\theta$ is a dynamical variable with the canonical commutation relation
\begin{equation}	\label{eq:commutationCanonical}
 \left[\theta,\frac{\partial}{\partial\theta}\right]=-1.
\end{equation}

\begin{figure}
\begin{center}
\includegraphics[width=0.45\textwidth]{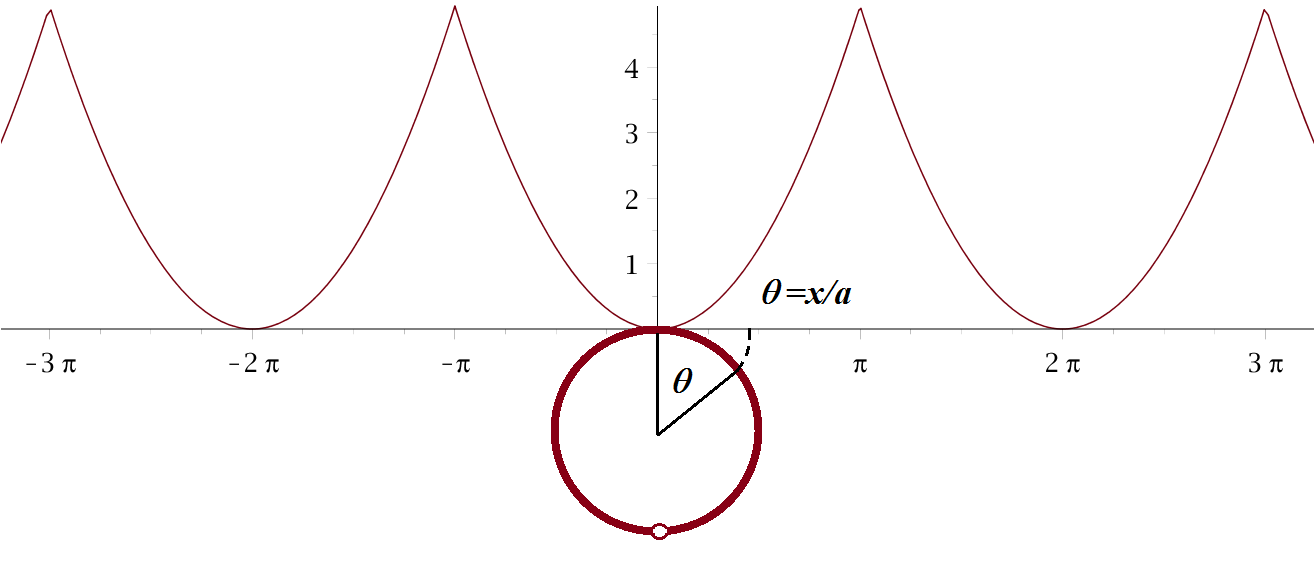}
\caption{The intrinsic potential chosen in (\ref{eq:intrinsicPotential}) is a chopped harmonic oscillator, i.e. half the square of the euclidean measure folded onto the intrinsic configuration space, the $U(1)$ circle. The dynamical variable $\theta$ projects into laboratory space by introduction of a length scale $a$ in (\ref{eq:spaceProjectionOneDimXandP}) \cite{TrinhammerEPL102}. The compact nature of the configuration space manifests itself in a periodic parametric potential. This opens for Bloch degrees of freedom in the wavefunction (\ref{eq:BlochWaveOneDim}).}
\label{fig:intrinsicPotential}
\end{center}
\end{figure}

The compact nature of the configuration variable space forces the potential $V$ for a hamiltonian description of such a system to be periodic in the parametrizing variable $\theta$. We take as an example the potential to be half the square of the euclidean measure folded onto the configuration space \footnote{I thank Hans Plesner Jacobsen for mentioning this interpretation to me \cite{PlesnerJacobsen}.}, i.e.
\begin{equation}	\label{eq:intrinsicPotential}
 V(u)\equiv w(\theta)=\frac{1}{2}(\theta-n2\pi)^2, \ \ \ \theta\in[(2n-1)\pi,(2n+1)\pi], \ \ \ n\in\mathbb{Z}.
\end{equation}
This is the well-known harmonic oscillator in the intrinsic case, see fig. \ref{fig:intrinsicPotential}.

In a dimensionful projection with length scale $a$, the commutation relation (\ref{eq:commutationCanonical}) corresponds to
\begin{equation}
 \left[a\theta,\frac{-i\hbar}{a}\frac{\partial}{\partial\theta}\right]=i\hbar
\end{equation}
with spatial position and momentum operators respectively
\begin{equation}	\label{eq:spaceProjectionOneDimXandP}
 x=a\theta,\ \ \ \ p=\frac{-i\hbar}{a}\frac{\partial}{\partial\theta}.
\end{equation}

It should be stressed, however, that the momentum operator does {\it not} describe momentum in laboratory space. It describes intrinsic momentum. We assume this intrinsic momentum to represent an intrinsic kinetic term for the energy of the system. This term is to be added to the potential energy term to get a Hamiltonian $H$ for the total energy $\cal E$ of the system (in the laboratory rest frame)
\begin{equation}
 H\Psi(u)=\Lambda\left[-\frac{1}{2}\frac{\partial^2}{\partial\theta^2}+w(\theta)\right]\Psi(u)={\cal E}\Psi(u)
\end{equation}
described by the complex wavefunction
\begin{equation}
 \Psi:U(1)\rightarrow {\mathbb{C}}
\end{equation}
and an energy scale $\Lambda$, which we may relate to the length scale $a$ in the projection (\ref{eq:spaceProjectionOneDimXandP}) by taking
\begin{equation}
 \Lambda=\frac{\hbar c}{a}.
\end{equation}
In other words, we look for solutions of the Schr\"odinger equation
\begin{equation}	\label{eq:oneDimSchroedinger}
 \frac{\hbar c}{a}\left[-\frac{1}{2}\frac{\partial^2}{\partial\theta^2}+w(\theta)\right]\Psi(u)={\cal E}\Psi(u).
\end{equation}

The Hamiltonian in the one-dimensional equation (\ref{eq:oneDimSchroedinger}) is a particular case of (\ref{eq:HamiltonianUN}) and solutions can be found with methods from solid state physics. For eq. (\ref{eq:oneDimSchroedinger}) it all boils down to solving the dimensionless equation
\begin{equation}	\label{eq:oneDimSchroedingerDimLess}
 \left[-\frac{1}{2}\frac{\partial^2}{\partial\theta^2}+w(\theta)\right]b_i(\theta)=e_ib_i(\theta)
\end{equation}
with periodic potential $w$. The dimensionless eigenvalue $e_i\equiv{\cal E}/\Lambda$, and we call $b_i$ a one-dimensional {\it parametric} wavefunction
\begin{equation}
 b_i:{\mathbb{R}}\rightarrow{\mathbb{C}}.
\end{equation}
In mathematical terms, $b_i$ is the pull-back of $\Psi$ to parameter space, i.e.
\begin{equation}
 b_i=\Psi^*=\Psi\circ\exp:{\mathbb{R}}\rightarrow U(1)\rightarrow {\mathbb{C}}.
\end{equation}

An arbitrary solution to (\ref{eq:oneDimSchroedingerDimLess}) can be written as a Bloch wavefunction \cite{AshcroftMerminBlochTheorem}
\begin{equation}	\label{eq:BlochWaveOneDim}
 b_i(\theta)=e^{i\kappa\theta}f_\kappa(\theta)
\end{equation}
where $\kappa$ is real and $f$ has the $2\pi$-periodicity of the potential $w$
\begin{equation}
 \kappa\in{\mathbb{R}}\ \ \ \ {\rm and}\ \ \ f(\theta+n2\pi)=f(\theta), \ \ \ n\in{\mathbb{Z}}.
\end{equation}
The solution (\ref{eq:BlochWaveOneDim}) is the result of a special one-dimensional case of  Bloch's theorem \cite{AshcroftMerminBlochTheorem}

{\it Bloch's theorem}

The eigenstate $\psi$ of the one-electron Hamiltonian $H=-\hbar^2\nabla^2/2m+V(\bf r)$, where $V({\bf r + R})=V({\bf r})$ for all $\bf R$ in a Bravais lattice \footnote{The Bravais lattice is the lattice of atoms in an infinite three-dimensional crystaline structure.}, can be chosen to have the form of a plane wave times a function with the periodicity of the Bravais lattice
\begin{equation}	\label{eq:BlochWaveFunction3D}
 \psi_{n\boldsymbol\kappa}({\bf r})=e^{i{\boldsymbol\kappa}\cdot{\bf r}}f_{n\boldsymbol\kappa}({\bf r}),\ \ \ {\bf r}=(x,y,z)
\end{equation}
where
\begin{equation}
 f_{n\boldsymbol\kappa}({\bf r}+{\bf R})=f_{n\boldsymbol\kappa}({\bf r})
\end{equation}
for all $\bf R$ in the Bravais lattice
\begin{equation}
 {\bf R}=n_1{\bf a_1}+n_2{\bf a_2}+n_3{\bf a_3},\ \ \ n_1,n_2,n_3\in\mathbb{Z}
\end{equation}
spanned by the unit cell vectors ${\bf a_1},{\bf a_2},{\bf a_3}$.\footnote{The index $n$ in (\ref{eq:BlochWaveFunction3D}) is just an arbitrary numbering of the state like $i$ in the one-dimensional analogue (\ref{eq:oneDimSchroedingerDimLess}). There is no summation over the repeated index $i$ on the right hand side of (\ref{eq:oneDimSchroedingerDimLess}).}\\

\begin{figure}
\begin{center}
\includegraphics[width=0.4\textwidth]{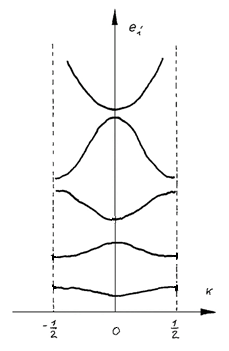}
\caption{Reduced zone scheme, see p. 160 in \cite{AshcroftMerminBlochTheorem}, for the eigenvalues $e_i$ of a one-dimensional harmonic oscillator (\ref{eq:oneDimSchroedingerDimLess}) in intrinsic space. The variation of $e_i$ with Bloch wave vector $\kappa$ for the lowest lying levels is exaggerated for clarity.}
\label{fig:reducedZoneScheme}
\end{center}
\end{figure}

In solid state physics, the {\it Bloch phases} $e^{i{\boldsymbol\kappa}\cdot{\bf r}}$ lead to a band structure of alternating allowed {\it energy bands} and forbidden {\it energy gaps}, see fig. \ref{fig:reducedZoneScheme} for a one-dimensional case. Each energy band encompasses a number of states for varying values of ${\boldsymbol\kappa}=(\kappa_1,\kappa_2,\kappa_3)$ equal to the number of atoms in the actual crystal. The number of states in each band is often huge, of the order of Avogadro's number for a $\rm cm^3$ volume crystal, and thus in solid state physics $\kappa_j$ is often treated as a continuous variable although in principle it is a discrete variable. For intrinsic quantum mechanics, the "crystal" is truly infinite - the angular variable $\theta$ in (\ref{eq:spaceProjectionOneDimXandP}) is not limited to an interval on the real axis, opposite to the position variable $x$ on a finite crystal lattice. Thus, {\it a priori} one would expect $\kappa$ in (\ref{eq:BlochWaveOneDim}) to be truly continuous. But that is not at all so. For an intrinsic wavefunction $\psi$ we have to require $|\psi(u)|^2$ to be single valued in parameter space in order to maintain its probability density interpretation on $U(N)$, i.e. we can only allow for
\begin{equation}	\label{eq:kappaSpectrum}
 \kappa_j=0,\pm\dfrac{1}{2}, j=1,2,\cdots N
\end{equation}
where $N$ is the number of toroidal degrees of freedom of the intrinsic space $U(N)$.

\begin{figure}
\begin{center}
\includegraphics[width=0.45\textwidth]{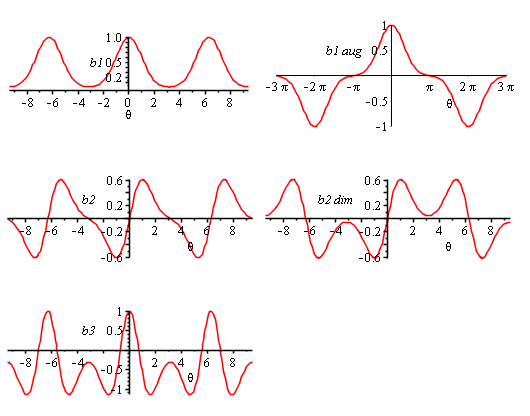}
\caption{One-dimensional wavefunctions as solutions of (\ref{eq:oneDimSchroedingerDimLess}) with $2\pi$-periodicity (left) and $4\pi$-periodicity (right). The wavefunctions shown are real valued functions found by iterative integration using Sturm-Louiville theory, and therefore not constructed as Bloch wavefunctions. A neutronic state can be constructed as a Slater determinant of the three functions to the left. By a combination of diminishing the eigenvalue in level two (middle right) with an augmented eigenvalue in level one (upper right) one gets a protonic state with period doubled parametrization. See also eq. (\ref{eq:mNmPapproxTheory}) for the neutron to proton mass shift.}
\label{fig:oneDimWavefunctions}
\end{center}
\end{figure}

In fig. \ref{fig:reducedZoneScheme} we show the eigenvalues $e_i$ of (\ref{eq:oneDimSchroedingerDimLess}) as a function of a continuous Bloch wave vector $\kappa$ in a reduced zone scheme. In fig. \ref{fig:oneDimWavefunctions} we show eigenfunctions found by iterative integration based on Sturm-Louiville theory.

\section{Quantum numbers from the Laplacian}
\label{sec:quantumNumbers}

The off-diagonal degrees of freedom carry quantum numbers via the off-toroidal generators in the Laplacian (\ref{eq:laplacian}). For our "generic" case (\ref{eq:laplacianU3}), where the configuration variable $u\in U(3)$, we have six off-toroidal generators $S_k, M_k, k=1,2,3$ with commutation relations
\begin{equation}
[M_k,M_l]=[S_k,S_l]=-i\hbar\epsilon_{klm}S_m
\end{equation}
as stated in (\ref{eq:Kcommutator}). The operators $S_k$ commute as body-fixed angular momentum. This gives the well-known eigenvalues \cite{DiracSpinSpectrum}
\begin{equation}	\label{eq:SandS2spectrum}
 S_k/\hbar=-s, -s+1,\cdots, s-1, s,\ \ \ \ {\bf S}^2=S_1^2+S_2^2+S_3^2=s(s+1)\hbar^2.
\end{equation}
With the related degrees of freedom being intrinsic, we allow for half odd-integer eigenvalues
\begin{equation}
 s=\frac{1}{2},\frac{3}{2},\frac{5}{2},\cdots
\end{equation}
With this choice, the Hamiltonian (\ref{eq:HamiltonianUN}) describes fermionic entities. For the case $u\in U(3)$ we interpret these to be baryons whereas the case $u\in U(2)$ seems to be relevant for leptons.\footnote{For $u\in U(1)$ there is no spontaneous decay from $2\pi$ to $4\pi$ periodicity in the ground state wherefore such a structure cannot be caught topologically as an intrinsic configuration. The variable $u\in U(1)$ remains a "free" phase factor.} For $u\in U(3)$, the spectrum for ${\bf M}^2=M_1^2+M_2^2+M_3^2$ needs some algebra to derive \cite{TrinhammerArXiv2011v3}. We now give the main steps.

In the intrinsic interpretation the presence of the components of ${\bf S}=(S_1,S_2,S_3)$ and ${\bf M}=(M_1,M_2,M_3)$ in the Laplacian opens for the inclusion of spin and non-neutral flavour. It can be shown \cite{TrinhammerOlafsson} that the components commute with the Laplacian as they should since the Laplacian is a Casimir operator. They also commute with the geodetic potential $\dfrac{1}{2}{\rm Tr}\ \chi^2$, so
\begin{equation}
 [S_k,H]=[M_k,H]=0
\end{equation}
where
\begin{equation}	\label{eq:U3hamiltonian}
 H=\Lambda\left[-\frac{1}{2}\Delta+\frac{1}{2}{\rm Tr}\ \chi^2\right]
\end{equation}
for $u=e^{i\chi}\in U(3)$. Further
\begin{equation}
 [S_k,{\bf S}^2]=[S_k,{\bf M}^2]=[M_k,{\bf S}^2]=[M_k,{\bf M}^2]=0.
\end{equation}
Thus we may choose ${\bf S}^2, S_3, {\bf M}^2$ as a set of mutually commuting generators which commute with the Hamiltonian $H$. As just mentioned, the well-known eigenvalues of ${\bf S}^2$ and $S_3$ in (\ref{eq:SandS2spectrum}) derives \cite{DiracSpinSpectrum} from the commutation relations (\ref{eq:Kcommutator}). Here we choose to interpret $\bf S$ as an interior angular momentum operator and allow for half-integer eigenvalues of $S_k$. The Hamiltonian is independent of the eigenvalue $s_3\hbar$ of $S_3$ as it should be because there is no preferred direction in the intrinsic space. Instead of choosing eigenvalues of $S_3$ we may choose $I_3$, the isospin 3-component.
To determine the spectrum for ${\bf M}^2$, we introduce a canonical body fixed "coordinate" representation, (see pp. 210 in \cite{SchiffExpectationValue})
\begin{gather}	\label{eq:CoordinateRepSk}
 S_1=a\theta_2 p_3-a\theta_3 p_2=\hbar\lambda_7\\ \nonumber
 S_2=a\theta_1 p_3-a\theta_3 p_1=\hbar\lambda_5\\ \nonumber
 S_3=a\theta_1 p_2-a\theta_2 p_1=\hbar\lambda_2
\end{gather}
The remaining Gell-Mann generators $\lambda_1,\lambda_3,\lambda_4,\lambda_6,\lambda_8$ are traditionally collected into a quadrupole moment tensor $\bf Q$, but we need to distinguish between the two diagonal components
\begin{gather}
 Q_0/\hbar=\frac{1}{2\sqrt{3}}(\theta_1^2+\theta_2^2-2\theta_3^2)+\frac{1}{2\sqrt{3}}\frac{a^2}{\hbar^2}(p_1^2+p_2^2-2p_3^2)=\lambda_8\\ \nonumber
 Q_3/\hbar=\frac{1}{2}(\theta_1^2-\theta_2^2)+\frac{1}{2}\frac{a^2}{\hbar^2}(p_1^2-p_2^2)=\lambda_3
\end{gather}
and the three off-diagonal components which we have collected into $\bf M$
\begin{gather}	\label{eq:CoordinateRepMk}
 M_3/\hbar=\theta_1\theta_2+\frac{a^2}{\hbar^2}p_1p_2=\lambda_1\\ \nonumber
 M_2/\hbar=\theta_3\theta_1+\frac{a^2}{\hbar^2}p_3p_1=\lambda_4\\ \nonumber
 M_1/\hbar=\theta_2\theta_3+\frac{a^2}{\hbar^2}p_2p_3=\lambda_6 
\end{gather}
The "mixing" operator $\bf M$ is a kind of Laplace-Runge-Lenz "vector" of our problem, (compare with pp. 236 in \cite{SchiffExpectationValue}). This is felt already in its commutation relations (\ref{eq:Kcommutator}). We shall see in the end (\ref{eq:Mspectrum}) that conservation of ${\bf M}^2$ corresponds to conservation of particular combinations of hypercharge and isospin. For the spectrum in projection space we calculate the $SU(3)$ Casimir operator, (compare with pp. 210 in \cite{SchiffExpectationValue})
\begin{equation}	\label{eq:Casimir}
 C_1=\frac{1}{\hbar^2}\left({\bf S}^2+{\bf M}^2+Q_0^2+Q_3^2\right)=-3+\frac{1}{3}\left(\frac{2H_{\rm e}}{\Lambda}\right)^2
\end{equation} 
where the Hamiltonian $H_{\rm e}$ of the euclidean harmonic oscillator is given by
\begin{equation}	\label{eq:euclideanHamilton}
 2H_{\rm e}=\frac{ca}{\hbar}{\bf p}^2+\frac{\hbar c}{a}\boldsymbol{\theta}^2
\end{equation}
and the energy scale $\Lambda=\frac{\hbar c}{a}$. To derive (\ref{eq:Casimir}) we used repeatedly the commutation relations
\begin{equation}
 [a\theta_i,p_j]=i\hbar\delta_{ij}.
\end{equation}

We now use the creation and annihilation operators (\ref{eq:creationAnnihilationCoordinateRep}) from section \ref{ch:secondQuantization}
\begin{equation}
 \hat{a}_j^\dagger=\frac{1}{\sqrt{2}}\left(\theta_j-i\frac{a}{\hbar}\hat{p}_j\right), \ \ \ \hat{a}_j=\frac{1}{\sqrt{2}}\left(\theta_j+i\frac{a}{\hbar}\hat{p}_j\right)
\end{equation}
with commutation relations as before
\begin{equation}
 [\hat{a}_i,\hat{a}_j^\dagger]=\delta_{ij},\ \ \ \ [\hat{a}_i^\dagger,\hat{a}_j^\dagger]=[\hat{a}_i,\hat{a}_j]=0
\end{equation}
and we want to settle the interpretation of the two diagonal operators $Q_0$ and $Q_3$. We find
\begin{gather}	\label{eq:hyperchargeIsospinChargeFromQoperators}
 Y/\hbar\equiv\frac{Q_0/\hbar}{\sqrt{3}}=\frac{N}{3}-\hat{a}_3^\dagger \hat{a}_3\\ \nonumber
 2I_3/\hbar\equiv Q_3/\hbar=\hat{a}_1^\dagger \hat{a}_1-\hat{a}_2^\dagger \hat{a}_2\\ \nonumber
 Q_2/\hbar=\hat{a}_1^\dagger \hat{a}_1-\hat{a}_3^\dagger \hat{a}_3
\end{gather}
where the number operator
\begin{equation}
 N=\sum_{j=1}^3\hat{a}_j^\dagger \hat{a}_j
\end{equation}
and
\begin{equation}
 Q_2/\hbar\equiv\frac{\sqrt{3}Q_0+Q_3}{2\hbar}=\frac{1}{2}\left(\theta_1^2-\theta_3^2\right)+\frac{1}{2}\frac{a^2}{\hbar^2}\left(p_1^2-p_3^2\right).
\end{equation}
From (\ref{eq:hyperchargeIsospinChargeFromQoperators}) we get
\begin{equation}\label{eq:GellMannNeeman}
 3Y=2Q_2-2I_3.
\end{equation}
Provided we can interpret $Q_2$ as a charge operator, this is the well-known Gell-Mann, Ne'eman, Nakano, Nishijima relation between charge, hypercharge and isospin \cite{GellMannGellMannNakanoNishijimaRelation,
NeemanGellMannNakanoNishijimaRelation,
DasOkuboGellMannNakanoNishijimaRelation,
GasiorowiczGellMannOkuboMassRelation}. Inserting (\ref{eq:hyperchargeIsospinChargeFromQoperators})
 in (\ref{eq:Casimir}) and rearranging, we get
\begin{equation}
 {\bf M}^2=\frac{4}{3}\hbar^2\left(\frac{H_{\rm e}}{\Lambda}\right)^2-{\bf S}^2-3\hbar^2-3Y^2-4I_3^2.
\end{equation}

The spectrum of the three-dimensional euclidean isotropic harmonic oscillator Hamiltonian in (\ref{eq:euclideanHamilton}) is well-known and follows from separation of the variables into three independent one-dimensional oscillators with the spectrum $(n_i+\frac{1}{2})\Lambda$ \cite{MessiahIsotropicHarmoicOscillatorPD,
GriffithsIsotropicHarmonicOscillator3D}, see also p. 241 in \cite{SchiffExpectationValue}. If we  assume the standard interpretations in (\ref{eq:GellMannNeeman}) with $Q_2$ as a charge operator, we have a relation among  baryonic quantum numbers ($y\sim 3Y/\hbar$) from which to determine the spectrum of ${\bf M}^2/\hbar^2$, namely
\begin{gather}	\label{eq:Mspectrum}
 M^2=\frac{4}{3}\left(n+\frac{3}{2}\right)^2-s(s+1)-3-\frac{1}{3}y^2-4i_3^2,\\ \nonumber n=n_1+n_2+n_3,\ \ \ n_j=0,1,2,3,\cdots
\end{gather}
Since $\bf M$ is hermittean, $M^2$ must be non-negative. With $s=\frac{1}{2}, y=1, i_3=\pm\frac{1}{2}$ as for the nucleon, the lowest possible value for $n$ is 1 (where $M^2=\frac{13}{4}$). Instead of (\ref{eq:Mspectrum}) we may write
\begin{gather}	\label{eq:SplusMsquaredSpectrum}
 \left({\bf S}^2+{\bf M}^2\right)/\hbar^2=\frac{4}{3}\left(n+\frac{3}{2}\right)^2-3-\frac{1}{3}y^2-4i_3^2,\\ \nonumber n=0,1,2,3\cdots
\end{gather}
This form is useful for generating baryon spectra as seen in (\ref{eq:potentialWtotal}) and (\ref{eq:centrifugalTermC}). This latter edition can be cast into an Okubo-form by choosing a different set of mutually commuting operators. We want to replace the three-component of isospin by isospin itself. This is possible because
\begin{equation}	\label{eq:isospinFromIsospin3}
 I^2=I_1^2+I_2^2+I_3^2=\frac{1}{4}(S_3^2+M_3^2)+I_3^2
\end{equation}
and $[S_3^2+M_3^2,I^2]=0$. We write
\begin{equation}
 G_3^2=S_3^2+M_3^2
\end{equation}
and rearrange (\ref{eq:isospinFromIsospin3}) and (\ref{eq:Mspectrum}) to get
\begin{equation}	\label{eq:S2M2spectrumIntrinsic}
 s(s+1)+M^2=\frac{4}{3}(n+\frac{3}{2})^2-3+g_3^2-\frac{1}{3}y^2-4i(i+1).
\end{equation}
Here $g_3^2$ is an eigenvalue of $G_3^2/\hbar^2$, i.e. a single quantum number. For a given value of $g_3^2$ we may group the spectrum in (\ref{eq:S2M2spectrumIntrinsic}) according to $n+y=\rm constant$ and get the Okubo structure
\begin{equation}	\label{eq:S2M2OkuboEdition}
 s(s+1)+M^2=a'+b'y+c'\left[\frac{1}{4}y^2-i(i+1)\right]
\end{equation}
for the nominator in the centrifugal potential in (\ref{eq:laplacianU3}). Equation (\ref{eq:S2M2OkuboEdition}) is the famous Okubo mass formula that reproduces the Gell-Mann, Okubo, Ne'eman mass relations within the baryon $\rm N$-octet and $\Delta$-decuplet independently of the values of $a',b',c'$ \cite{GellMannOkuboMassRelation,
OkuboMassRelation,NeemanOkuboMassRelation, FondaGhirardiOkuboMassRelation,
GasiorowiczGellMannOkuboMassRelation}\footnote{Due to the $\boldsymbol{\theta}$-dependence in the centrifugal term in (\ref{eq:laplacianU3}) the spacing within higher multiplets will not be the same as for the lowest multiplet. So far only the lowest multiplets have been experimentally confirmed with candidates in all positions. One might undertake the task of calculating higher multiplets within the intrinsic viewpoint and compare with quark model calculations. The most prominent difference, though, has already been demonstrated as a solution to the missing resonance problem in fig. \ref{fig:NandDeltaSpectrum} when compared with quark model calculations fig. 15.5 p. 285 in \cite{RPP2016}. Note that the parametric eigenvalues in fig. \ref{fig:reducedZoneScheme} for higher levels go with the square of the level number whereas ordinary harmonic oscillator levels go linearly.}.

\section{Flavour in colour. $SU(N)$ in $U(N)$}
\label{sec:flavourIncolour}

In the present section we investigate the relationship between flavour and colour as seen from the intrinsic viewpoint. We are aware that in the Standard Model these concepts are treated with independent symmetry groups $SU_f(3)$ and $SU_c(3)$ respectively. The latter is taken as the gauge group of strong interactions whereas the former is an approximate symmetry group inferred from spectroscopic phenomenology. The three colour charges $r,g,b$ (red, green, blue) and six flavours $u,d,s,c,b,t$ (up, down, strange, charm, beauty, top) are ascribed to quark fields which carry both colour and flavour. It is the colour group that is at the basis of baryon interactions, represented by quantum chromo dynamics, QCD in the standard model \cite{RPP2016}. The spectroscopic flavour group $SU_f(3)$ was instrumental in coming to terms with the concept of quarks. Its most successful prediction was that of the $\Omega^-$ resonance with triple strangeness \cite{GellMannOmegaMinusPrediction, BarnesEtAlOmegaMinusObservation}.

From section \ref{sec:quantumNumbers} we see that a Hamiltonian on $U(3)$ has enough structure to carry both colour, spin and flavour degrees of freedom. We interpret the three toroidal degrees of freedom as colour with generators $T_j, j=1,2,3$. Spin and flavour are carried by the off-diagonal generators of the Laplacian, $S_k, k=1,2,3$ and $M_k, k=1,2,3$ respectively with the flavours intermingled with colour and spin as expressed in (\ref{eq:Mspectrum}). In chapter \ref{ch:coordinateSystem} and \ref{ch:secondQuantization} we saw that taking $u\in U(3)$ as intrinsic configuration variable implies local $U(3)$ gauge invariance in laboratory space of the fields projected from the wavefunction $\Psi(u)$.

The flavour group $SU_f(3)$ taken at face value predicts many more baryons than are observed. Thus the flavour group has lost some of its spectroscopic relevance\footnote{It is still used for {\it a postiori} naming of discovered resonances, but not so much for predictions of such resonances.}. Not so, however, for scattering experiments. In particular the analysis of scattering data from proton-proton collisions as in the large hadron collider, LHC, at CERN, needs detailed information on the up and down quark momentum distributions in the proton - given that these scattering data are interpreted within a Standard Model setting.

To see how flavours come about in connection with scattering, we need to express flavour generators in the colour basis and later to apply these expressions in our scheme for reading off intrinsic momenta, see chapter \ref{ch:coordinateSystem}. We shall find in section \ref{sec:partonDistributionFunctions} that
\begin{equation}	\label{eq:flavourGenerators}
 T_u=\frac{2}{3}T_1-T_3,\ \ \ T_d=-\frac{1}{3}T_1-T_3
\end{equation}
generates respectively u and d quark parton distribution functions from a protonic state. To support more formally these relations we need to consider the various group algebras.

Let us consider first the general case $SU(N)$ in $U(N)$ \footnote{It is deliberate that we do not write $SU(N)\subset U(N)$ since we want to ascribe different interpretations to some of the generators of the two groups. Note e.g that above we used the $SU((3)$ Casimir operator to find the spectrum of some of the common generators in the $U(3)$ Laplacian. In that connection we used an $SU(3)$ edition for hypercharge but for the flavour representation in $U(3)$ we shall need a different edition. Secondly $U(N)$ in itself is related to the configuration variable and thus contains all the nine degrees of freedom in the dynamical model whereas $SU(3)$ is used for spectroscopic multiplet organization. Although the $SU(3)$ multiplets follow naturally from the $U(3)$ Laplacian they are {\it not} exact algebraic reproductions of the $U(3)$ spectra as mentioned in the note on (\ref{eq:S2M2OkuboEdition}).}. We follow Das and Okubo, see pp. 71 in \cite{DasOkuboGellMannNakanoNishijimaRelation}. The generators of $U(N)$ may be defined from annihilation and creation operators $a^\dagger_\mu$ and $a_\nu$ for the $N$-dimensional harmonic oscillator by 
\begin{equation}
 X^\mu_\nu=a^\dagger_\mu a_\nu
\end{equation}
with commutation algebra
\begin{equation}
 \left[X^\mu_\nu,X^\alpha_\beta\right]=\delta^\alpha_\nu X^\mu_\beta-\delta^\mu_\beta X^\alpha_\nu
\end{equation}
with all greek indices running from $1$ to $N$. This algebra is identical to (\ref{eq:uNalgebraEij}). For $SU(N)$ one introduces the trace operator
\begin{equation}
 X=X^\alpha_\alpha,
\end{equation}
summing over $\alpha=1,\cdots N$ to get the $SU(N)$ generators
\begin{equation}	\label{eq:suNgenerators}
 A^\mu_\nu=X^\mu_\nu-\frac{1}{N}\delta^\mu_\nu X.
\end{equation}
With this definition it becomes clear that the two groups share algebraic structure
\begin{equation}
 \left[A^\mu_\nu,A^\alpha_\beta\right]=\delta^\alpha_\nu A^\mu_\beta-\delta^\mu_\beta A^\alpha_\nu.
\end{equation}
because the trace operator $X$ commutes with all the individual $X_\nu^\mu$ in $A_\nu^\mu$. For our particular case, $SU(3)$, the diagonal isospin three component operator
\begin{equation}
 I_3=\dfrac{1}{2}(A_1^1-A_2^2)
\end{equation}
will remain common with $U(3)$ whereas other diagonal generators like the charge $Q$ and the hypercharge $Y$ operators
\begin{equation}	\label{eq:chargeHyperchargeSUN}
 Q=A_1^1,\ \ \ Y=-A^3_3
\end{equation}
as defined by Das and Okubo, (p. 221 in \cite{DasOkuboGellMannNakanoNishijimaRelation}) will contain the characteristic fraction $1/3$ for $SU(3)$ in (\ref{eq:suNgenerators}). These fractions of $1/3$ are carried into the quark fractional charge units according to the charge operator mentioned by Das and Okubo as the Gell-Mann-Nakano-Nishijima formula
\begin{equation}	\label{eq:GellMannNakanoNishijima}
 Q=I_3+\frac{Y}{2}=A^1_1.
\end{equation}

By comparing (\ref{eq:chargeHyperchargeSUN}) with (\ref{eq:flavourGenerators}) we infer the $U(3)$ identifications for "flavour in colour"
\begin{equation}
 Q'=X_1^1\sim T_1,\ \ \ Y'=-X_3^3\sim -T_3.
\end{equation}
This identification works well for the proton as it yields seemingly correct parton distribution functions, proton spin function and proton magnetic moment, see sections \ref{sec:partonDistributionFunctions}, \ref{sec:protonSpinStructureFunction} and \ref{sec:protonMagneticMoment}. To what extent it generalizes to strange baryons remains to be investigated.

\section{Neutral pentaquark predictions}
\label{sec:neutralPentaquarks}

\begin{figure}
\begin{center}
\includegraphics[width=0.45\textwidth]{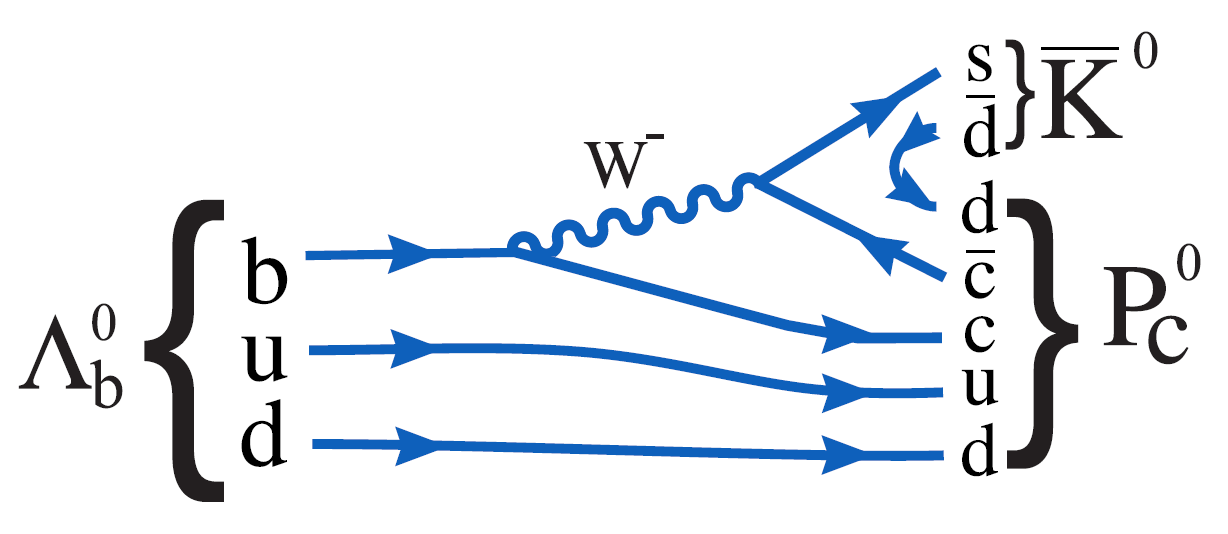}
\caption{Feynman diagram for neutral pentaquark observation. The neutral pentaquark $P_c^0$ may later break up into a $J/\Psi$ and a neutron $n$ or a into a $J/\Psi$ and $\Delta^0$, see eqs. (\ref{eq:pentaquarkNeutron}) and  (\ref{eq:pentaquarkDelta0}). Figure adapted from \cite{LHCbPentaquarkObservation}.}
\label{fig:neutralPentaquark}
\end{center}
\end{figure}

In the summer of 2015, LHCb announced unexpected narrow baryon resonances interpreted as hidden charm charged pentaquarks, $P_c^+(4380)$ and $P_c^+(4450)$ \cite{LHCbPentaquarkObservation}. The observed resonances fell in the neighbourhood of our predictions for singlet neutral flavour resonances \cite{TrinhammerNeutronProtonMMarXivWithAppendices25Jun2012,TrinhammerBohrStibiusHiggsPreprint}  near the open charm threshold, see table \ref{tab:singlets} - only our prediction concerns electrically neutral states which can be calculated accurately from expansions of the measure-scaled wavefunction $R$ on a base set
\begin{equation}	\label{eq:fpqr0}
 f_{pqr}^0=\begin{vmatrix}
  \cos p\theta_1 & \cos p\theta_2 & \cos p\theta_3\\
  \cos q\theta_1 & \cos q\theta_2 & \cos q\theta_3\\
  \cos r\theta_1 & \cos r\theta_2 & \cos r\theta_3
 \end{vmatrix},
\end{equation}
where $p,q,r$ are integer $p=0,1,2,\cdots,q=p+1,p+2,,\cdots,r=q+1,q+2,\cdots$. The set (\ref{eq:fpqr0}) is equivalent to the set (\ref{eq:fpqr}) except it does not invite period-doubling to decrease the individual level energies whereby it would inflict charge creating topological changes. We call such states {\it neutral flavour neutral charge singlets}. Even though one would not expect them to have charged partners, anyhow they seem to couple to neighbouring "ordinary" neutral flavour resonances of isospin $\frac{1}{2}$. For instance we consider $N(1440)$, $N(1535)$ and $N(1650)$ to be the result of a mixing of a singlet $\Phi_{135}$ and two doublets $\Phi_{125}$ and $\Phi_{134}$. We suppose the observed charged pentaquarks are such mixing partners of neutral pentaquarks which should show up around the energies listed in the four bottom lines of table \ref{tab:singlets}. Another interesting singlet state is $\Phi_{137}$ which corresponds to the state at $2051\ \rm MeV$  in table \ref{tab:singlets}. Such a state lies in a "desert" area which implies weaker coupling to neighbouring resonances. It is therefore interesting to note that no clear, electrically charged $N$ resonance is observed in this area whereas a {\it neutral charge} resonance $N(2040)$ {\it has} been observed \cite{AblikimEtAlNeutralNobservation}. Since the partial wave analysis establishing the baryon resonances naturally must rely on charged particles - because these are the easiest to observe - a lone neutral charge resonance cannot be expected to be granted a four star status in the Particle Data Group listings. We therefore encourage the search for neutral pentaquarks $P_c^0$ around the energies $4228, 4499, 4652, 4723\ \rm MeV$ listed at the bottom of table \ref{tab:singlets}. We have previously suggested to look for such resonances in \cite{TrinhammerNeutronProtonMMarXivWithAppendices25Jun2012} and had the opportunity to discuss the possibilities at LHCb with Sheldon S. Stone at the EPS-HEP 2015 conference in Vienna. Our immediate suggestion of looking at invariant mass in $p\pi^-$ spectra would drown in the background at LHCb \footnote{"You won't see it!", Sheldon said. "Because of background?" I asked. "Yes" he replied. Later during the conference I mentioned to him the possibility of $\Sigma_c^+(2455)D^-$ which he considered doable once a factor five higher statistics has been reached.}. Later I asked about another possibility
\begin{equation}	\label{eq:pentaquarkNeutron}
 \Lambda_b^0\rightarrow\overline{K}^0+P_c^0\rightarrow\overline{K}^0+J/\Psi+n,
\end{equation}
see fig. \ref{fig:neutralPentaquark}. But neutrons are elusive in accelerator experiment detectors, so instead Sheldon Stone suggested the following channel
\begin{equation}	\label{eq:pentaquarkDelta0}
 \Lambda_b^0\rightarrow\overline{K}^0+P_c^0\rightarrow\overline{K}^0+J/\Psi+\Delta^0\rightarrow\overline{K}^0+J/\psi+p+\pi^-,
\end{equation}
because the $\Delta^0$ as well as the other intermediates break up into charged particles which are easily detectable. However, also this channel would need a factor five increase  of the statistics as of summer 2016 [private email of 10 July 2016]. Figure \ref{fig:neutralPentaquark} shows a quark structure interpretation for $P_c^0$ production in $\Lambda_b^0$ decay which can be reached at LHCb.

Other ways to look for neutral charge, neutral flavour baryon singlets could be as narrow resonances in photoproduction on neutrons and in $\pi^-p$ scattering.

\begin{table}
\begin{center}
\caption{Scarce singlet states. Eigenvalues based on Slater determinants (\ref{eq:fpqr0}) of three cosines up to order 20 analogous to (\ref{eq:fpqr}). The first column shows eigenvalues of from an approximate Hamiltonian \cite{TrinhammerNeutronProtonMMarXivWithAppendices25Jun2012, TrinhammerBohrStibiusHiggsPreprint} and the third column shows eigenvalues of the exact equation (\ref{eq:toroidalSchroedinger}). The toroidal labels refer to the band labels used to construct Slater determinants of three one-dimensional states picked from the reduced zone "tower" shown in fig. \ref{fig:reducedZoneScheme}. A singlet 579-like resonance is predicted at 4499 MeV in the free charm system $\Sigma_c^+(2455)D^-$  slightly above its threshold at 4324 MeV. The rest masses are predicted from a common fit of the nucleon ground state 939.6 MeV to the ground state 4.38 of (\ref{eq:schroedingerU3}) resp. (\ref{eq:toroidalSchroedinger}) with no period doublings \cite{TrinhammerNeutronProtonMMarXivWithAppendices25Jun2012}. \vspace{3mm}}
\label{tab:singlets}
\begin{tabular}{c c c c c} \hline\hline \\
Singlet & Toroidal & Singlet & Rest mass\\
approximate \cite{TrinhammerNeutronProtonMMarXivWithAppendices25Jun2012, TrinhammerBohrStibiusHiggsPreprint} & label & exact (\ref{eq:toroidalSchroedinger}) & MeV/c$^2$\\ \hline \\
7.1895&1 3 5&7.1217&1526 \\ 
9.3568&1 3 7&9.5710&2051 \\ 
11.1192&1 5 7&11.2940&2420 \\ 
12.7175&1 3 9&13.2505&2839 \\ 
13.0927&3 5 7&13.2811&2846 \\ 
14.4494&1 5 9&14.9641&3206 \\ 
16.4086&3 5 9&16.9213&3626 \\ 
16.6605&1 7 9&17.3006&3707 \\
17.1769&1 3 11&18.0090&3859 \\
18.6320&3 7 9&19.2577&4126 \\
18.9214&1 5 11&19.7327&4228 \\
20.3774&5 7 9&20.9940&4499 \\
20.8910&3 5 11&21.7110&4652 \\
21.0766&1 7 11&22.0409&4723 \\ \hline\hline
\end{tabular}
\end{center}
\end{table}

\section{Parton distribution functions}	\label{sec:partonDistributionFunctions}

Parton distributions derive from a probability amplitude interpretation of the external derivative of the wavefunction taken along specific generators $T=a_1T_1+a_2T_2+a_3T_3$ (\ref{eq:toroidalMomentumExpandedOnBaseGenerators}) with the three colour generators $T_j$ given as in (\ref{eq:partialTj}) and (\ref{eq:intrinsicMomentaFixedBase}) by
\begin{equation}	\label{eq:toroidalGeneratorsAtOrigo}
  iT_j=\frac{\partial}{\partial\theta_j}=\partial_j|_e \ \ \ {\rm and\ where}\ \ \ p_j=-i\hbar\frac{1}{a}\frac{\partial}{\partial\theta_j}
\end{equation}
are parametric momentum operators. To unfold this we factorized the wavefunction into a toroidal part $\tau$ and an off-torus part $\Upsilon$ to get $\Psi(u)=\tau\Upsilon$. With the measure-scaled toroidal wavefunction $R=J\tau$ in (\ref{eq:RmeasureScaledWavefunction}) the exterior derivative expanded on torus forms $d\theta_j$ with colour components $c_j$ reads
\begin{equation}
 dR=c_jd\theta_j
\end{equation}
as in (\ref{eq:dpsiDefinition}) \cite{GuilleminPollack}. The colour components transform according to the fundamental representation of $SU(3)$ \cite{TrinhammerEPL102} as follows from (\ref{eq:intrinsicMomentumTransformation}). At a given point $u$ they are extracted by the colour generators $\partial_j$ which act as left-invariant vector fields, thus
\begin{equation}
 c_j(u)=\partial_j|_u[R]=uiT_j[R]=(iT_j)_u[R]=dR_u(iT_j).
\end{equation}
In particular along a track $\theta iT$ we have
\begin{equation}
 c_j(u)=dR_{u=\exp(\theta iT)}(iT_j).
\end{equation}
We get the total quark probability amplitude as a sum over these components, i.e. for a derivative along the track generated by $\theta iT$  \cite{TrinhammerEPL102}
\begin{gather}	
     \sum^3_{j=1}dR_{u = \exp(\theta i T)}(iT_j)
      =\sum_{j=1}^3\frac{d}{dt}R(ue^{tiT_j}) {\text{\Large{$\mid$}}}_{t=0} \nonumber \\
      =\sum_{j=1}^3\frac{\partial R}{\partial\theta_j} {\text{\Large{$\mid$}}}_{(\theta_1,\theta_2,\theta_3)=(\theta a_1,\theta a_2, \theta a_3)} \cdot \frac{\partial(a_j\theta+t)}{\partial t} {\text{\Large{$\mid$}}}_{t=0} \nonumber \\
       =\left(  \frac{\partial R^*}{\partial \theta_1}+ \frac{\partial R^*}{\partial \theta_2}+ \frac{\partial R^*}{\partial                    \theta_3} \right){\text{\Large{$\mid$}}}_{(\theta_1, \theta_2, \theta_3)
  =(\theta \cdot a_1, \theta \cdot a_2, \theta \cdot a_3)}     \nonumber \\
 \equiv D(\theta \cdot a_1, \theta \cdot a_2, \theta \cdot a_3).	\label{eq:dR}
\end{gather}
Here we used the chain rule and $R^*$ is the pull-back (\ref{eq:pullBack}) of $R$ to parameter space.

When a momentum fraction $xP$ is read off from the system, it means that the device reading off this momentum leaves the interaction zone with momentum change $(1-x)P$ such that momentum conservation holds. The device doing the read-off is typically an impacting particle like the electron in deep inelastic scattering and the electron momentum in the end is registered in the detector. To determine the relation between the toroidal angle $\theta$ and the momentum fraction $x$ in the intrinsic dynamics, we used a derivation inspired by Alessandro Bettini \cite{TrinhammerEPL102, Bettini}. Citing ourselves: "Imagine a proton at rest with four-momentum $P = ({\text{\bf{0}}}, E_0)$. We boost it virtually to energy $E$ by impacting upon it a massless four-momentum $q = ({\text{\bf{q}}}, E - E_0) $ which we assume to hit a parton $xP$. After impact the parton represents a virtual mass $xE$. Thus
\begin{equation}	\label{eq:RelativisticBoost}
    (xP_{\mu} + q_{\mu})\cdot(xP^{\mu} + q^{\mu}) = x^2E^2,
\end{equation}
from which we get the parton momentum fraction $x = \frac{2E_0}{E + E_0}$, or the boost parameter \cite{TrinhammerEPL102}
\begin{equation}	\label{eq:boostParameter}
  \xi(x)\equiv \frac{E - E_0}{E} = \frac{2 - 2x}{2-x}." 
\end{equation}

\begin{figure} 
\begin{center}
\includegraphics[width=0.45\textwidth]{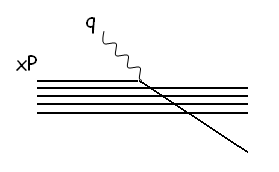}
\caption{Boosting a proton with massless momentum $q$ by scattering on a parton of momentum $xP$ where $P$ is the proton momentum and $x$ the parton momentum fraction.}
\label{fig:partonBoost}
\end{center}
\end{figure}

\begin{figure}
\begin{center}
\includegraphics[width=0.3\textwidth]{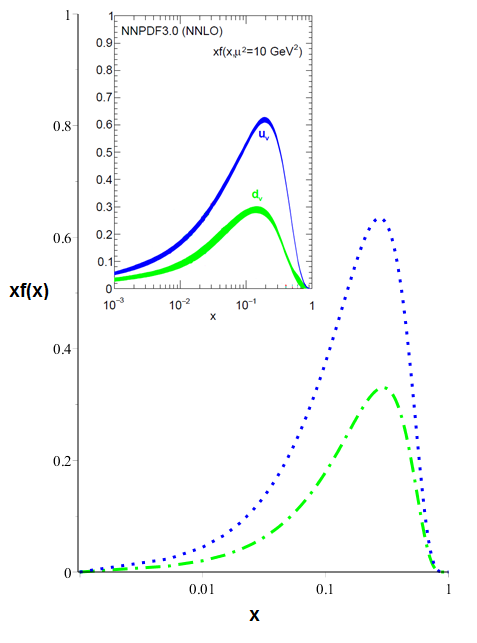}
\caption{Parton distribution functions (\ref{eq:TqDistribution}) for u (dotted blue) and d (dashdotted green) quarks of the proton from an approximate period doubled state (\ref{eq:protonicApproximateFirstOrder}) \cite{TrinhammerEPL102} compared to distribution functions extracted from experiments (insert from \cite{RPP2016} with other distributions erased). Note that the quark content evolves towards smaller $x$-values for higher energy scales $Q^2$. The distributions from the intrinsic dynamics would correspond to $Q^2=m_{\rm p}^2c^4\approx 1\ \rm GeV^2$ whereas the insert is for $Q^2=10\ \rm GeV^2$ \cite{RPP2016}. Figure updated from ref. \cite{TrinhammerEPL102}.} 
\label{fig:PDFwithInsert}
\end{center}
\end{figure}

\begin{figure}
\begin{center}
\includegraphics[width=0.45\textwidth]{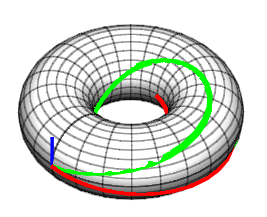}
\caption{Helical traces from flavour generators $T_u$ (upper green) and $T_d$ (lower red) on the colour torus showing the intermingling of flavour and colour degrees of freedom in the present intrinsic description on a common $U(3)$ manifold where colours are toroidal degrees of freedom and flavours enter via the Laplacian on the manifold.}
\label{fig:helicalsOnTorus}
\end{center}
\end{figure}

The boost $q_{\mu}\sim E-E_0$, see fig. \ref{fig:partonBoost}, corresponds to $(1-x)E\sim (1-x)T$ in the directional derivative. In other words, boosting with $\sim (1-x)T$ we probe on $xP$. With $\xi$ \footnote{Note that $\xi$ is not the same boost as in (\ref{eq:uSpinorDuncan}).} and $x$ inversely related we identify $\theta T=\pi\xi T$ and get the corresponding distribution function $f_T(x)$ determined by squaring the sum of probability amplitudes over colour 
\begin{equation}	\label{eq:fT}
   f_T(x)dx=\left(\sum^3_{j=1} dR_{u=\exp(\theta iT)}(iT_j)\right)^2 d\theta.
\end{equation}
We did this \cite{TrinhammerEPL102} for the toroidal part of the wavefunction for a first order approximation
\begin{equation}	\label{eq:protonicApproximateFirstOrder}
  b(\theta_1, \theta_2, \theta_3)  = \frac{1}{N} \begin{vmatrix}  1&1 & 1\\
\sin \frac{1}{2} \theta_1 & \sin \frac{1}{2} \theta_2 & \sin \frac{1}{2} \theta_3 \\
 \cos \theta_1 & \cos \theta_2 & \cos \theta_3\end{vmatrix}\,
\end{equation}
to a protonic state with normalization constant $N$. Note that $b$ is antisymmetric under interchange of the three colour degrees of freedom $\theta_j$. For charge fraction $e_u=+\frac{2}{3}$, respectively $e_d=-\frac{1}{3}$, eq. (\ref{eq:fT}) leads to the distribution functions seen in fig. \ref{fig:PDFwithInsert}
\begin{equation}	\label{eq:TqDistribution}
 f_{Tq}(x)= 
\left[D\left(e_q \cdot\pi \xi(x),  0, 
    (-1) \cdot\pi \xi(x)\right) \right]^2 \cdot \pi\frac{d\xi}{dx}
\end{equation}
where the directional derivative from (\ref{eq:dR}) is given by
\begin{equation}	
D(\theta_1, \theta_2, \theta_3) = \frac{\partial b}{\partial\theta_1}+\frac{\partial b}{\partial\theta_2}+\frac{\partial b}{\partial\theta_3} \label{eq:Dkernel}  
\end{equation}
and the generators $T_u$ and $T_d$ for the flavour directional derivatives are intermingled with the three colour generators as seen in (\ref{eq:flavourGenerators})
\begin{equation}
  T_u=\frac{2}{3}T_1-T_3,\ \ \ T_d=-\frac{1}{3}T_1-T_3.\nonumber
\end{equation}
Equation (\ref{eq:flavourGenerators}) states a linear relationship among the generators. This linearity just means that the flavour tracks run steadily, but helically on the torus through the exponential mapping onto the curved Lie group manifold, see fig. \ref{fig:helicalsOnTorus}. It should not be misunderstood as a linear relation between flavour and colour variables. The colour generators are conjugate to continuous dynamical toroidal angular variables (\ref{eq:toroidalGeneratorsAtOrigo}). The well-known discrete quantum numbers of spectroscopic flavour multiplet grouping of baryons, on the other hand, is mapped via the Lie algebra structure from the Laplacian (\ref{eq:laplacian}) as seen from the Okubo-like mass relation (\ref{eq:S2M2OkuboEdition}).

\section{Proton spin structure function} \label{sec:protonSpinStructureFunction}

\begin{figure} 
\begin{center}
\includegraphics[width=0.45\textwidth]{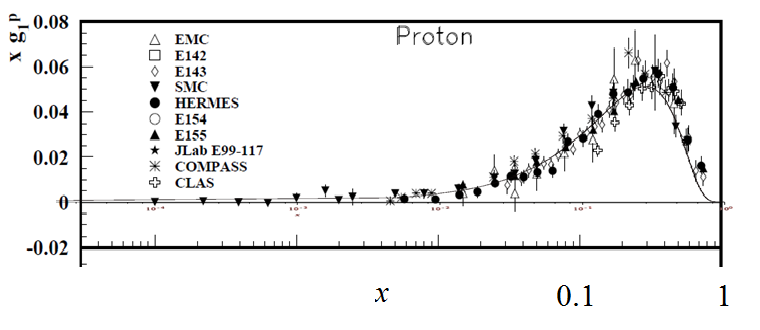}
\caption{Intrinsic proton spin structure function (solid) of the present work overlaid on the $g_1^{\rm P}$ spin structure data from various experimental groups \cite{RPP2014} as a function of the momentum fraction $x$ of the scattering centre in the proton.}
\label{fig:spinFuncWorldData}
\end{center}
\end{figure}

From the intrinsic point of view the proton is an entire, indivisible object. We do not consider the spin to be a combination of three independent constituent quark spins. This means that the usual parton model expressions \cite{EllisStirlingWebber} for quark distributions $q=q(x)$
\begin{equation}
 q=q\uparrow+q\downarrow\ \ \ {\rm and}\ \ \ \Delta q=q\uparrow-q\downarrow
\end{equation}
have $q\downarrow=0$ such that
\begin{equation}
 g_1(x)=\frac{1}{2}\sum_qe_q^2\left[\Delta q(x)+\Delta \bar{q}(x)\right]
\end{equation}
with $\bar{q}(x)=0$ would simplify to
\begin{equation}
 g_1(x)=\frac{1}{2}\sum_qe_q^2q(x).
\end{equation}
Now, we consider colour and flavour to be intermingled as seen in (\ref{eq:flavourGenerators}). For the unpolarized flavour distribution functions we summed over colour components in the derivative $D$ in (\ref{eq:RelativisticBoost}). For the polarized case, however, the spin $\frac{1}{2}$ parton scattering is along a specific flavour and colour degree of freedom. We therefore average the flavour distributions over the three colours to get spin distributions. Thus the spin structure function is a sum of the two distributions (\ref{eq:TqDistribution}) averaged in colour and weighted by their corresponding interaction strengths, $e_q^2$. The resulting spin structure function reads
\begin{equation}	\label{eq:spinStructureFunc}
  g_{\rm int}^{\rm P}(x)=\frac{1}{2}\left[e_u^2\cdot\frac{1}{3}f_{Tu}(x)+e_d^2\cdot\frac{1}{3}f_{Td}(x)\right].
\end{equation}
This expression provides the curve in figs. \ref{fig:spinFuncWorldData} and \ref{fig:protonSpinStructureFunctionCOMPASS}. It should be stressed that (\ref{eq:spinStructureFunc}) contains no fitting parameters - not even a normalization to fit the data. The normalization constant $N$ is set by restricting the measure-scaled wavefunction (\ref{eq:protonicApproximateFirstOrder}) to $[0,\pi]^3$. Thus
\begin{equation}
  1=\frac{1}{N^2}\int_0^\pi\int_0^\pi\int_0^\pi b^2(\theta_1,\theta_2,\theta_3)d\theta_1 d\theta_2 d\theta_3,
\end{equation}
which settles $N^2=\frac{3}{2}\pi^3-\frac{44}{3}\pi$. The range of integration corresponds to the parametrization $\theta=\pi\xi$ above.

The example (\ref{eq:protonicApproximateFirstOrder}) for a first approximation is not integrable over the second term in the Laplacian (\ref{eq:laplacian}). But we can generalize the structure of the $b$ state (\ref{eq:protonicApproximateFirstOrder}) to a set of expansion states
\begin{equation}	\label{eq:BlochSlaterGeneral}
  b_{pqr} = \begin{vmatrix}  e^{-i\frac{\theta_1}{2}}\cos p\theta_1&e^{-i\frac{\theta_2}{2}}\cos p\theta_2 & e^{-i\frac{\theta_3}{2}}\cos p\theta_3\\
\sin(q-\frac{1}{2})\theta_1 & \sin(q-\frac{1}{2})\theta_2 & \sin(q-\frac{1}{2})\theta_3 \\
 e^{-i\frac{\theta_1}{2}}\cos r\theta_1&e^{-i\frac{\theta_2}{2}}\cos r\theta_2 & e^{-i\frac{\theta_3}{2}}\cos r\theta_3\end{vmatrix}\,
\end{equation}
individually integrable in the full parameter space $[-2\pi,2\pi]^3$ suitable for the period doubling in the wavefunction. With the proton state expanded on this (incomplete) set for $p=0,1,2,\cdots P-1, q=1,2,3,\cdots P, r=p+1,\cdots P$ we get for $P=15$ with $1800$ base functions the dimensionless ground state eigenvalue ${\cal E}_{\rm p}/\Lambda=4.41$ close to the value $4.46$ found by expansion on Slater determinants from solutions to the one-dimensional Scchr\"odinger equation (\ref{eq:oneDimSchroedingerDimLess}). Another integrable protonic base can be constructed from the $g_{pqr}$s in (\ref{eq:gpqr}) and their complex conjugate.

\section{Proton magnetic moment}	\label{sec:protonMagneticMoment}

We distribute quark masses $m_{\rm u}$ and $m_{\rm d}$ to constitute the proton mass $m_{\rm p}$ as
\begin{equation}
 m_{\rm p}=2m_{\rm u}+m_{\rm d}
\end{equation}
where the quark masses are found by integrating the flavour distributions (\ref{eq:TqDistribution})
\begin{equation}
 m_{\rm u}=\int_0^1 xm_{\rm p}f_{T_{\rm u}}(x)dx\ \ \ {\rm and}\ \ \ m_{\rm d}=\int_0^1 xm_{\rm p}f_{T_{\rm d}}(x)dx.
\end{equation}
With quark magnetic moments $\mu_{\rm q}$ and the nuclear magneton $\mu_{\rm N}$ given by
\begin{equation}
 \mu_{\rm q}=\frac{1}{2}\frac{e_{\rm q}\hbar}{m_{\rm q}}\ \ \ {\rm and}\ \ \ \mu_{\rm N}=\frac{1}{2}\frac{e\hbar}{m_{\rm p}}
\end{equation}
we find from the constituent quark model expression \cite{DonoghueGolowichHolsteinMDM}
\begin{equation}
 \mu_{\rm p}=\frac{4}{3}\mu_{\rm u}-\frac{1}{3}\mu_{\rm d}
\end{equation}
the following result for the proton magnetic dipole moment
\begin{equation}	\label{eq:protonMagneticMoment}
 \mu_{\rm p}=2.779\cdots\mu_{\rm N}.
\end{equation}
The result agrees with the experimental result \cite{RPP2014} 
\begin{equation}
 \mu_{\rm p,exp}=2.792847356(23)\mu_{\rm N}
\end{equation}
within half a percentage. Note that the state (\ref{eq:protonicApproximateFirstOrder}) used for generating the flavour distributions is only a first approximation. For the neutron we would distribute the mass as $m_{\rm n}=m_{\rm u}+2m_{\rm d}$ and use \cite{DonoghueGolowichHolsteinMDM} $\mu_{\rm n}=\frac{4}{3}\mu_{\rm d}-\frac{1}{3}\mu_{\rm u}$ to get the neutron magnetic dipole moment $\mu_{\rm n}=-2.187\cdots\mu_{\rm N}$. This compares less well with the experimental result \cite{RPP2014} $\mu_{\rm n,exp}=-1.9140427(5)\mu_{\rm N}$ because the neutron magnetic dipole moment is more sensitive to the d-quark distribution and the d-quark distribution is less accurately derived from the approximate state (\ref{eq:protonicApproximateFirstOrder}), see fig. \ref{fig:PDFwithInsert}. We urge for a more accurate calculation for a protonic state expanded on 
\begin{equation}
 \Phi_{pqr}^{{\rm p}-}=\frac{1}{2}\left(g_{pqr}+g_{pqr}^*\right)
\end{equation}
with $g_{pqr}$s from (\ref{eq:gpqr}) in stead of the approximate state (\ref{eq:protonicApproximateFirstOrder}) from which the parton distributions in (\ref{fig:PDFwithInsert}) are derived.

\section{Electroweak mixing}
\label{sec:electroweakMixingAngle}

The aim of unifying the electromagnetic interactions with weak interactions into one, common electroweak $U(2)$ gauge field theory, was carried through via the introduction of the Higgs mechanism \cite{EnglertBrout, HiggsSep1964, HiggsOct1964, GuralnikHagenKibble} with its Higgs field $\phi$ to mediate the necessary spontaneous symmetry break among the four gauge fields, the massless photon $\gamma$ known from electrodynamics and three massive, intermediate vector bosons $W^\pm$ and $Z$ supposed to undertake the weak interactions. We have exemplified the symmetry break by the neutron to proton decay and used that decay to set the energy scale (\ref{eq:phi0PhysicsOverview}) of the unified electroweak interactions \cite{TrinhammerBohrStibiusHiggsPreprint}. We shall here discuss the different masses of the four particles related to the four gauge fields. The field particles start out {\it a priori} on an equal footing being all massless. In the Standard Model, the symmetry break is handled by accepting two different coupling constants $g'$ for a $U(1)$ sector and $g$ for an $SU(2)$ sector. We first present the Standard Model derivation of the respective masses and then present a derivation in subsection \ref{subsec:MixingAngleQuarkGenerators} based on one common coupling constant.

Following pp. 437 in \cite{LancasterBlundellAnnihilationCreationInQuantumField}, we start out by half odd-integer phase factor transformations under respectively $U(1)$ and $SU(2)$ transformations in accordance with paired half odd-integer Bloch phase factors in (\ref{eq:gpqr})
\begin{gather}
 \left(\begin{matrix}\phi^+\\ \phi^0\end{matrix}\right)\rightarrow
 \left(\begin{matrix}e^{i\frac{\beta}{2}}&0\\ 0&e^{i\frac{\beta}{2}}\end{matrix}\right)
 \left(\begin{matrix}\phi^+\\ \phi^0\end{matrix}\right),\\ \nonumber
 \left(\begin{matrix}\phi^+\\ \phi^0\end{matrix}\right)\rightarrow
 \left(\begin{matrix}e^{i\frac{\boldsymbol\alpha\cdot\boldsymbol\tau}{2}}&0\\ 0&e^{i\frac{\boldsymbol\alpha\cdot\boldsymbol\tau}{2}}\end{matrix}\right)
 \left(\begin{matrix}\phi^+\\ \phi^0\end{matrix}\right).
\end{gather}
This corresponds to the Higgs field having weak hypercharge $Y=+1$ and weak isospin $I=\frac{1}{2}$. Pairing of the Bloch phase factors in the spontaneous symmetry break in the baryonic sector, e.g. in the neutron decay
\begin{equation}
 n\rightarrow p+e+\overline{\nu}_e
\end{equation}
is necessary to keep the centrifugal term (\ref{eq:centrifugalTermC}) integrable.
With generators of weak hypercharge and isospin transformations chosen respectively as
\begin{gather}	\label{eq:hyperchargeAndIsospinGenerators}
 Y=-2I_0=\left(\begin{matrix}-1&0\\ 0&-1\end{matrix}\right),\ \ \ 2I_1=\left(\begin{matrix}0&1\\ 1&0\end{matrix}\right),\\ \nonumber 2I_2=\left(\begin{matrix}0&-i\\ i&0\end{matrix}\right),\ \ \ 2I_3=\left(\begin{matrix}1&0\\ 0&-1\end{matrix}\right)
\end{gather}
we maintain the Gell-Mann-Nakano-Nishijima relation among charge, isospin and hypercharge quantum numbers (\ref{eq:GellMannNakanoNishijima}) also in the electroweak case
\begin{equation}
 q=i_3+\frac{y}{2}.
\end{equation}

Breaking of invariance under the local {\it a priori} $U(2)$ gauge transformation
\begin{equation}
 \left(\begin{matrix}\phi^+\\ \phi^0\end{matrix}\right)\rightarrow 
 e^{i\Lambda_k(x)\frac{\tau_k}{2}}\left(\begin{matrix}\phi^+\\ \phi^0\end{matrix}\right)
\end{equation}
with generators $i\tau_k/2\equiv iI_k$ corresponds to an introduction of two separate coupling constants $g',g$ related to the respective generators $I_0=\frac{1}{2}\tau_0$ and ${\bf I}=(I_1,I_2,I_3)=\frac{1}{2}\boldsymbol\tau$ in the generalized derivative (\ref{eq:generalizedDerivative})
\begin{equation}	\label{eq:generalizedDerivativeBrokenSymmetry}
 D_\mu=\partial_\mu{\bf 1}-ig'B_\mu I_0-ig{\bf W}_\mu\cdot{\bf I}.
\end{equation}
The presence of a Higgs potential to shift the vacuum expectation value of $\phi$ away from zero generates mass terms for three of four gauge bosons depending on the relative values of $g'$ and $g$. This is seen by applying (\ref{eq:generalizedDerivativeBrokenSymmetry}) to the Higgs field expanded around its value
\begin{equation}
 \phi_0=\left(\begin{matrix} 0\\ \varphi_0 \end{matrix}\right)
\end{equation}
at a minimum of the Higgs potential $V_{\rm H}(\phi^\dagger\phi)$ which in our edition has a constant term in order to fit the intrinsic potential, see fig. \ref{fig:higgsPotentialFit}
\begin{gather}	\label{eq:HiggsPotentialWithConstant}
 V_{\rm H}(\phi)=\frac{1}{2}\delta^2\varphi_0^2-\frac{1}{2}\mu^2\phi^2+\frac{1}{4}\lambda^2\phi^4,\\ \nonumber
  \delta^2=\frac{1}{4}\varphi_0^2,\ \ \mu^2=\frac{1}{2}\varphi_0^2,\ \ \lambda^2=\frac{1}{2}.
\end{gather}
The constant term is omitted in the Standard Model edition of the Higgs potential. Then one squares to get the generalized kinetic term contribution $(D^\mu\phi)^\dagger D_\mu\phi$ to the Lagrangian of the Higgs field
\begin{equation}
 L_\phi=(D^\mu\phi)^\dagger D_\mu\phi-V_{\rm H}(\phi^\dagger\phi).
\end{equation}

In the present section we focus on the masses of the particle excitations of the gauge boson fields. We therefore restrict ourselves to mass terms of these
\begin{gather}	\label{eq:massProductForGaugeBosons}
 (D^\mu\phi)^\dagger D_\mu\phi-V_{\rm H}(\phi^\dagger\phi)=\\ \nonumber
 (\phi_0\left[-ig'B^\mu I_0-ig{\bf W}^\mu\cdot {\bf I}\right])^\dagger\left[-ig'B_\mu I_0-ig{\bf W}_\mu\cdot{\bf I}\right]\phi_0\\ \nonumber
 +\ {\rm kinetic\ terms}+{\rm cross\ terms}+{\rm Higgs\ field\ terms}.
\end{gather}
Suppressing the Lorentz indices $\mu$ on the gauge fields, we write the operation of the fractional $U(2)$ generators $I_k,k=0,1,2,3$ on the Higgs field as
\begin{gather}
 \left[-ig'B I_0-ig{\bf W}\cdot{\bf I}\right]\phi_0\\ \nonumber
 =\left[-ig'BI_0-ig\left(W^{(1)}I_1+W^{(2)}I_2+W^{(3)}I_3\right)\right]\phi_0.
\end{gather}
We want the set of four gauge fields to contain the massless $U(1)$ gauge field $A_\gamma$ of quantum electrodynamics but we have no guarantee that $A_\gamma$ equals the {\it a priori} $U(1)$ gauge field $B$ because both generators $I_0$ and $I_3$ are diagonal. We thus anticipate a transformation from the {\it a priori} fields $B,W^{(3)}$ into spacetime fields $Z,A_\gamma$ given by
\begin{equation}
 \left(\begin{matrix} Z\\ A_\gamma \end{matrix}\right)=
 \left(\begin{matrix}
 \cos\theta_{\rm W}&-\sin\theta_{\rm W}\\
 \sin\theta_{\rm W}&\cos\theta_{\rm W} \end{matrix}\right)
 \left(\begin{matrix}  W^{(3)}\\ B \end{matrix}\right)
 \equiv \Theta\left(\begin{matrix}  W^{(3)}\\ B \end{matrix}\right).
\end{equation}
The condition on the {\it electroweak mixing angle} $\theta_{\rm W}$ is that $A_\gamma$ remains massless after the spontaneous symmetry break in accordance with the infinite range of electromagnetic interactions. This requirement puts a constraint on the ratio between the two coupling constants $g,g'$. To find the constraint, we write the mass term coefficient on $\phi_0$ from the diagonal generators in (\ref{eq:hyperchargeAndIsospinGenerators}) as
\begin{gather}
 (gW^{(3)},g'B)\left(\begin{matrix}I_3\\ I_0  \end{matrix}\right)=(W^{(3)},B)\left(\begin{matrix}gI_3\\ g'I_0\end{matrix}\right)\\ \nonumber
 =(W^{(3)},B)\Theta\Theta^{-1}\left(\begin{matrix}gI_3\\ g'I_0\end{matrix}\right).
\end{gather}
Expressed in the "rotated" fields $Z, A_\gamma$ this means
\begin{equation}
 (gW^{(3)},g'B)\left(\begin{matrix}I_3\\ I_0  \end{matrix}\right)=\left(\begin{matrix}Z\\ A_\gamma \end{matrix}\right)^T \Theta^{-1}\left(\begin{matrix} gI_3\\ g'I_0 \end{matrix}\right).
\end{equation}
From this we read off the $Z$ and $A_\gamma$ field generators
\begin{gather}	\label{eq:IzIgammaI3I0}
 I_Z=g\cos\theta_{\rm W}I_3+g'\sin\theta_{\rm W}I_0\\ \nonumber
 I_\gamma=-g\sin\theta_{\rm W}I_3+g'\cos\theta_{\rm W}I_0
\end{gather}
and require
\begin{equation}	\label{eq:mGammaConstraint}
 \frac{1}{2}m_\gamma^2c^4=(\phi_0I_\gamma)^\dagger I_\gamma\phi_0=0
\end{equation}
which is fulfilled provided
\begin{equation}
 g\sin\theta_{\rm W}+g'\cos\theta_{\rm W}=0\ \ \ {\rm or}\ \ \ \tan\theta_{\rm W}=-\frac{g'}{g}.
\end{equation}
The electroweak mixing angle $\theta_{\rm W}$ remains an ad hoc parameter in the Standard Model which is why the $Z$ (and $W$) masses could not be predicted accurately. Given $\theta_{\rm W}$, the $Z$ mass follows from an analogous calculation to that of fixing $m_\gamma=0$, i.e.
\begin{equation}	\label{eq:mZSM}
 \frac{1}{2}m_{\rm Z}^2c^4=(\phi_0I_Z)^\dagger I_Z\phi_0=\frac{g^2+g'^2}{4}\varphi_0^2.
\end{equation}
We namely have in the "lower" component of $I_Z$
\begin{gather}
 -\frac{1}{2} g\cos\theta_{\rm W}+\frac{1}{2} g'\sin\theta_{\rm W}\\ \nonumber
 =-\frac{1}{2} g\frac{g}{\sqrt{g^2+g'^2}}+\frac{1}{2} g'\frac{-g'}{\sqrt{g^2+g'^2}}\\ \nonumber
 =-\frac{1}{2}\sqrt{g^2+g'^2}.
\end{gather}
To determine the absolute coupling strengths $g,g'$ we look again at the photon field generator $I_\gamma$. It couples to $I_3$ with the strength $-g\sin\theta_{\rm W}$ and to $I_0$ with the strength $g'\cos\theta_{\rm W}$. If we assume both these strengths to equal the elementary unit of charge $e$ characteristic of quantum electro dynamics, we get
\begin{equation}
 g=-|e|/\sin\theta_{\rm W},\ \ \ g'=|e|/\cos\theta_{\rm W}
\end{equation}
where we have chosen a sign convention such that $g,g'>0$ and $\sin\theta_{\rm W}<0$. With these we get for the $Z$ mass
\begin{equation}
 \frac{1}{2}m_{\rm Z}^2c^4=\frac{e^2}{4\sin^2\theta_{\rm W}\cos^2\theta_{\rm W}}\varphi_0^2
\end{equation}

As for the remaining gauge field components $W^{(1)},W^{(2)}$ on the off-diagonal generators $I_1$ and $I_2$, these are  collected into charged boson fields in
\begin{equation}	\label{eq:WplusWminusBrokenCase}
 W^{\pm}_\mu\equiv\frac{1}{\sqrt{2}}\left(W^{(1)}_\mu\mp iW^{(2)}_\mu\right)
\end{equation}
which expand on
\begin{equation} \label{eq:IplusIminus}
 I_\pm=(I_1\pm iI_2).
\end{equation}
With these rephrasings similar to p. 248 in \cite{FlorianScheckElectroweakAndStrongInteractions}, we have
\begin{equation}	\label{eq:WbaseShift}
 g\left(W^{(1)}I_1+W^{(2)}I_2\right)=\frac{g}{\sqrt{2}}\left(W^+I_++W^-I_-\right).
\end{equation}
To get the masses of $W^\pm$ we exploit the isospin algebra
\begin{equation}
 \frac{1}{2}\left(I_+I_-+I_-I_+\right)=I_1^2+I_2^2={\bf I}^2-I_3^2.
\end{equation}
Applied to the Higgs field this yields
\begin{equation}
 \left({\bf I}^2-I_3^2\right)\phi_0=\left(\frac{1}{2}\left(\frac{1}{2}+1\right)-\left(\frac{1}{2}\right)^2\right)\phi_0=\frac{1}{2}\phi_0
\end{equation}
and - after some lines of algebra squaring (\ref{eq:WbaseShift}) - we get
\begin{equation}	\label{eq:mWSM}
 m_{\rm W}^2c^4=g^2\frac{1}{2}\varphi_0^2=\frac{g^2v^2}{4}\rightarrow m_{\rm W}c^2=80.36(2)\ \rm GeV .
\end{equation}
For the numerical value, see note after eq. (\ref{eq:mWmZbruteResult}).

\begin{widetext}

\subsection{Mixing angle from quark generators}
\label{subsec:MixingAngleQuarkGenerators}

We hint at the origin of the electroweak mixing angle. First we note that calculation of the $Z$ and $W$ masses in (\ref{eq:mZSM}) and (\ref{eq:mWSM}) rely on the eigenvalues of their respective generators on the Higgs field in its $U(2)$ representation. In particular we note that the generators (\ref{eq:IzIgammaI3I0}) together with (\ref{eq:IplusIminus}) from (\ref{eq:hyperchargeAndIsospinGenerators}) do {\it not} have a common normalization. To the contrary, they are scaled by different combinations of the coupling constants $g,g'$. In the language of intrinsic quantum mechanics this is a sign of rescaled intrinsic momentum which in the end manifests itself in different masses of the related particles.

We express the $I_0$ and $I_3$ of  (\ref{eq:IzIgammaI3I0}) in the equivalent base of $T_u, T_d$ expressed in the 2-dimensional representation space of the Higgs field, where
\begin{equation}
 \phi_0=\left(\begin{matrix}
  0\\ \varphi_0  \end{matrix} \right)
\end{equation}
and where we suppress the "inactive"  second component of the quark generators $T_1$ and $T_3$ from (\ref{eq:toroidalGeneratorsU3}) to have two-dimensional editions of $T_u$ and $T_d$ from (\ref{eq:flavourGenerators})
\begin{equation}
 T_u=\left(\begin{matrix} \frac{2}{3}&0\\ 0&-1 \end{matrix}\right), \ \ T_d=\left(\begin{matrix} -\frac{1}{3}&0\\ 0&-1 \end{matrix}\right).
\end{equation}
We then have
\begin{equation}
 2I_0=\frac{2}{3}T_u-\frac{5}{3}T_d,\ \ \ 2I_3=\frac{4}{3}T_u-\frac{1}{3}T_d.
\end{equation}
We now substitute $\theta_{\rm W}$ in (\ref{eq:IzIgammaI3I0}) by $\theta_{ud}$ defined by
\begin{equation}	\label{eq:cosThetaUD}
 \cos^2\theta_{ud}={\rm Tr}\ T_uT_d=\frac{7}{9}.
\end{equation}
This yields the $Z$ boson mass from rewriting $I_{\rm Z}$ in (\ref{eq:IzIgammaI3I0}) and letting it operate on $\phi_0$. Thus
\begin{equation}
 2I_{\rm Z}=g\cos\theta_{ud}\left(\frac{4}{3}T_u-\frac{1}{3}T_d\right)+g'\sin\theta_{ud}\left(\frac{2}{3}T_u-\frac{5}{3}T_d\right)
\end{equation} 
operating on $\phi_0$ means
\begin{equation}
 \left(\frac{2I_{\rm Z}}{g\cos\theta_{ud}}\right)\left(\begin{matrix}  0\\\varphi_0 \end{matrix}\right)=\left[\left(\frac{4}{3}T_u-\frac{1}{3}T_d\right)+\frac{g'}{g}\tan\theta_{ud}\left(\frac{2}{3}T_u-\frac{5}{3}T_d\right)\right]\left(\begin{matrix}  0\\\varphi_0 \end{matrix}\right).
\end{equation}
Exploiting $\tan\theta_{ud}=-\frac{g'}{g}$ from the zero mass constraint on the photon field in (\ref{eq:mGammaConstraint}) we rewrite to get
\begin{equation}
 \left(\frac{2I_{\rm Z}}{g\cos\theta_{ud}}\right)=\left(\frac{4}{3}-\frac{2}{3}\tan^2\theta_{ud}\right)T_u-\left(\frac{1}{3}-\frac{5}{3}\tan^2\theta_{ud}\right)T_d
\end{equation}
and thus
\begin{equation}
  \left(\frac{2I_{\rm Z}}{g\cos\theta_{ud}}\right)\left(\begin{matrix}  0\\\varphi_0 \end{matrix}\right)=\left[-\left(\frac{4}{3}-\frac{2}{3}\tan^2\theta_{ud}\right)+\left(\frac{1}{3}-\frac{5}{3}\tan^2\theta_{ud}\right)\right]\phi_0
\end{equation}
which reduces to
\begin{equation}
 \left(\frac{2I_{\rm Z}}{g\cos\theta_{ud}}\right)\left(\begin{matrix}  0\\\varphi_0 \end{matrix}\right)=\left[-1-\tan^2\theta_{ud}\right]\phi_0=\frac{-1}{\cos^2\theta_{ud}}\phi_0.
\end{equation}
Multiplying by $\cos\theta_{ud}$ and squaring we get
\begin{equation}
 \left[\left(\frac{2I_{\rm Z}}{g}\right)\left(\begin{matrix}  0\\\varphi_0 \end{matrix}\right)\right]^2=\frac{1}{\cos^2\theta_{ud}}\varphi_0^2.
\end{equation}
With
\begin{equation}
 \frac{1}{2}m_{\rm Z}^2c^4=\left(I_{\rm Z}\phi_0\right)^2=\left(\frac{g}{2}\right)^2\left[\left(\frac{2I_{\rm Z}}{g}\right)\left(\begin{matrix}  0\\ \varphi_0 \end{matrix}\right)\right]^2
\end{equation}
we get
\begin{equation}	\label{eq:mZintrinsic}
 m_{\rm Z}^2c^4=2\left(\frac{g}{2}\right)^2\frac{1}{\cos^2\theta_{ud}}\varphi_0^2=\frac{g^2v^2}{4\cos^2\theta_{ud}}
\end{equation}
in accordance with standard expressions (\ref{eq:mZSM}). For $\cos^2\theta_{ud}=\frac{7}{9}$ eq. (\ref{eq:mZintrinsic}) yields
\begin{equation}	\label{eq:mZbruteResult}
 m_{\rm Z}c^2=91.11(2)\ \rm GeV.
\end{equation}
Combining with (\ref{eq:mWSM}) we compare with measured masses \cite{RPP2018}
\begin{equation}	\label{eq:mWmZbruteResult}
 \frac{m_{\rm W}^2}{m_{\rm Z}^2}=\frac{\alpha_{\rm W}}{\alpha_{\rm Z}}\frac{1}{1/\cos^2\theta_{ud}}=0.77757(16)\approx\left(\frac{80.379(12)\ \rm GeV}{91.1876(21)\ \rm GeV}\right)^2=0.7771(3).
\end{equation}
Here we used $\alpha_{\rm W}^{-1}=127.984(20)$ obtained by sliding \cite{TrinhammerBohrStibiusEPS2015} from $\alpha_{\rm Z}^{-1}=127.950(17)$ \cite{RPP2016}. Elsewhere \cite{TrinhammerBohrStibiusHiggsPreprint, TrinhammerBohrStibiusEPS2015} we have suggested how to derive $\varphi_0$ from fitting the Higgs potential (\ref{eq:HiggsPotentialWithConstant}) to the intrinsic geodetic potential (\ref{eq:intrinsicGeodeticPotential}). There we found
\begin{equation}
 \frac{v}{\sqrt{2}}=\varphi_0=\frac{2\pi}{\alpha_{\rm W}}\frac{\pi}{\alpha_{\rm e}}m_{\rm e}c^2
\end{equation}
with the fine structure coupling $\alpha_{\rm W}$ taken at $W$ energies and $\alpha_{\rm e}$ at electronic energies respectively. For a fundamental treatment it is not satisfactory to calculate the $Z$ and $W$ masses from fine structure couplings taken at their respective {\it a priori} unknown masses. Thus one might treat the problem iteratively. First we might use $\alpha_{\rm e}=1/137.0359...$ \cite{RPP2018} for all fine structure couplings in our formulae to get a first iteration for the mass values. Then we might use sliding scale techniques \cite{TrinhammerBohrStibiusEPS2015} to evaluate, iteratively, the fine structure coupling at the relevant energy scale. Since the fine structure coupling only changes logarithmically with energy, this iteration quickly converges (in domains where sliding scale makes sense, i. e. at energies high compared to baryonic energy scales). At least the results in (\ref{eq:mZbruteResult}) and (\ref{eq:mWmZbruteResult}) show the consistency of using $T_u$ and $T_d$ as relevant generators.

It is as if the selection of the mixing angle $\theta_{\rm W}$ is guided by the fixation of the quark generators from the strong interaction sector. This may be a coincidence but we rather think that it is a consequence of the interrelation between the electroweak and strong interactions as they meet in the neutron to proton decay and in other weak baryonic decays. In our intrinsic conception, the interrelation between strong and electroweak degrees of freedom is shaped specifically by the requirement of paired Bloch phase factors with half odd-integer Bloch wave vectors which select a $U(2)$ subgroup in the baryonic $U(3)$ configuration space. It is this suspicion that guided the Ansatz on $\cos\theta_{ud}$ in (\ref{eq:cosThetaUD}). We have likened the action of the generators to momentum read offs in eq. (\ref{eq:momentumComponentsFromDphi}) and visualized it in fig. \ref{fig:DucksAndDrakesDrawing}. The larger the momentum of the generator, the larger is the intrinsic mass generated.

\section{Rayleigh-Ritz solution for a trigonometric base}
\label{sec:RayleighRitz}

We want to find the eigenvalues E of the following equation
\begin{equation}   \label{eq:schroedingerAppendixDimLess} 
 [ -\Delta_e+W]R(\theta_1,\theta_2,\theta_3) = 2 \mbox{E}R(\theta_1,\theta_2,\theta_3),
\end{equation}
which is the full eq. (\ref{eq:toroidalSchroedinger}).

In the Rayleigh-Ritz method \cite{Bruun} one expands the eigenfunction on an orthogonal set of base functions with a set of expansion coefficients, multiply the equation by this expansion, integrates over the entire variable volume and end up with a matrix problem in the expansion coefficients from which a set of eigenvalues can be got. Thus with the approximation
\begin{equation}	\label{eq:RNexpansion}   
R_N = \sum^N_{l=1} a_l f_l
\end{equation}
we have the integral equation
\begin{equation}  	\label{eq:RNintegralEquation}  
  \int^{\pi}_{-\pi}\int^{\pi}_{-\pi}\int^{\pi}_{-\pi} R_N \cdot ( -\Delta_e+W) 
   R_N d\theta_1 d\theta_2d\theta_3
  =\int^{\pi}_{-\pi}\int^{\pi}_{-\pi}\int^{\pi}_{-\pi} R_N \cdot
  2 \mbox{E} R_N d\theta_1 d\theta_2 d\theta_3.
\end{equation}
The counting variable $l$ in (\ref{eq:RNexpansion}) is a suitable ordering of the set of tripples $p$, $q$, $r$ in (\ref{eq:fpqr}) such that we expand on an orthogonal set. The eq. (\ref{eq:RNintegralEquation}) can be interpreted as a vector eigenvalue problem, where {\bf{a}} is a vector, whose elements are the expansion coefficients $a_l$. Thus (\ref{eq:RNintegralEquation}) is equivalent to the eigenvalue problem
\begin{equation}  \label{eq:aTGa}  
  \mbox{\bf{a}}^T \mbox{\bf{Ha}} = 2 \mbox{E} \mbox{\bf{a}}^T \mbox{\bf{Fa}},
\end{equation}
where the matrix elements of {\bf{H}} and {\bf{F}} are given by
\begin{equation}   \label{eq:Glm} 
    H_{lm} \equiv \int^{\pi}_{-\pi}\int^{\pi}_{-\pi}\int^{\pi}_{-\pi} f_l \cdot  ( -\Delta_e+W)  f_m 
   d\theta_1 d\theta_2 d\theta_3
\end{equation}
and
 \begin{equation}    \label{eq:Flm}
    F_{lm} \equiv \int^{\pi}_{-\pi}\int^{\pi}_{-\pi}\int^{\pi}_{-\pi} f_l f_m 
   d\theta_1 d\theta_2 d\theta_3. 
\end{equation}				

When the set of expansion functions is orthogonal, (\ref{eq:aTGa}) implies
\begin{equation}    \label{eq:RREigenvalueMatrixEdition}
  {\mbox{\bf{Ha}}}= {\mbox{2E{\bf{Fa}}}},
\end{equation}
from which we get a spectrum of $N$ eigenvalues determined as the set of components of a vector {\bf{E}} generated from the eigenvalues of the matrix ${\bf{F}}^{-1}{\bf{H}}$, i.e.
\begin{equation}    
  {\mbox{\bf{E}}}= \frac{1}{2} {\mbox{eig}} ({\mbox{\bf{F}}^{-1}} {\mbox{\bf{H}}}).
\end{equation}

The lowest lying eigenvalues will be better and better determined for increasing values of $N$ in (\ref{eq:RNexpansion}). For the base (\ref{eq:fpqr}) the integrals (\ref{eq:Glm}) and (\ref{eq:Flm}) can be solved analytically, and as (\ref{eq:fpqr}) is an educated guess based on the solutions of the one-dimensional problem (\ref{eq:oneDimSchroedingerDimLess}) like the ones shown in fig. \ref{fig:oneDimWavefunctions}, it improves the convergence in $N$ for the general problem in (\ref{eq:schroedingerAppendixDimLess}) and (\ref{eq:RREigenvalueMatrixEdition}).

The exact expressions to be used in constructing {\bf{H}} and {\bf{F}} are given below for the base (\ref{eq:fpqr}). For $r > p,\, u > s$ and $q,\,t \ge 1$ we have the following orthogonality relations
\begin{equation}  \label{eq:fpqrfstuOrtogonal}  
   <f_{pqr}\! \mid \! f_{stu}\! > \equiv
\int^{\pi}_{-\pi}\int^{\pi}_{-\pi}\int^{\pi}_{-\pi}
   f_{pqr}(\theta_1,\theta_2,\theta_3) \cdot f_{stu} (\theta_1,\theta_2,\theta_3)
   d\theta_1 d\theta_2 d\theta_3
 =6 \pi^3 \delta_{ps}\delta_{qt}\delta_{ru}.
 \end{equation}
Here for a convenient notation we have generalized the Kronecker delta
\begin{equation}   
     \delta_{ij} = 
 \begin{cases} 
  1  \, \, \, & \text{for} \, \, \,  i=j \, \wedge \, i \ne 0  \\
  2  & \text{for} \, \, \,  i=j \, \wedge \, i = 0 \\
  0   & \text{for}\, \, \,  i \ne j
 \end{cases}.   
\end{equation}
The Laplacian yields
\begin{equation}  \label{eq:fpqrfstuLaplacian}  
  <f_{pqr} \! \mid \! \frac{\partial^2}{\partial \theta^2_1} + \frac{\partial^2}{\partial \theta^2_2} +   
 \frac{\partial^2}{\partial \theta^2_3} \! \mid \! f_{stu}>
  = ( - p^2 - q^2 - r^2) \cdot 
  6 \pi^3 \delta_{ps}\delta_{qt}\delta_{ru}.
\end{equation}
The matrix elements for the geodetic potential couples the individual base functions and follows from a more lengthy calculation below yielding the following expression
\begin{eqnarray} \label{eq:fpqrfstuIntrinsicPotential}   
 &<\! f_{pqr} \! \mid \! \theta_1^2 +\theta_2^2 +\theta_3^2 \! \mid \! f_{stu}\!> =  \nonumber \\ 
 & \nonumber \\
&\text{for }p,q,r,s,t,u >0: \nonumber \\
&\delta_{ps}\delta_{qt}\delta_{ru} \cdot 6 \pi^3 \left(\pi^2 + \frac{1}{2p^2}- 
\frac{1}{2q^2}+\frac{1}{2r^2} \right)\nonumber  \\
&+(1-\delta_{ps})\delta_{qt} \delta_{ru} \cdot 6 \pi^3 \cdot 4 \frac{p^2+s^2}{(p^2-s^2)^2} \cdot(-1)^{p+s} \nonumber\\
&+(1-\delta_{qt})\delta_{ps} \delta_{ru} \cdot 6 \pi^3 \cdot 4 \frac{2qt}{(q^2-t^2)^2} \cdot(-1)^{q+t} \nonumber\\
&+(1-\delta_{ru})\delta_{ps} \delta_{qt} \cdot 6 \pi^3 \cdot 4 \frac{r^2+u^2}{(r^2-u^2)^2} \cdot(-1)^{r+u} \nonumber\\
&-(1-\delta_{pu})\delta_{qt} \delta_{rs} \cdot 6 \pi^3 \cdot 4 \frac{p^2+u^2}{(p^2-u^2)^2} \cdot(-1)^{p+u} \nonumber\\
&-(1-\delta_{rs})\delta_{pu} \delta_{qt} \cdot 6 \pi^3 \cdot 4 \frac{r^2+s^2}{(r^2-s^2)^2} \cdot(-1)^{r+s} \nonumber\\
& \nonumber\\
&\text{for } p=0 \, \wedge \, s \ne 0 \, \wedge \, u> s: \nonumber \\
& 24 \pi^3 \left[ \delta_{qt} \delta_{ru} \frac{1}{s^2} (-1)^s - \delta_{qt} \delta_{rs} \frac{1}{u^2}(-1)^{u} \right] \nonumber\\
& \nonumber\\
&\text{for } p=0 \, \wedge \, s=0: \nonumber \\
& \delta_{qt} \delta_{ru} \cdot 6 \pi^3 ( 2 \pi^2 - \frac{1}{q^2} + \frac{1}{r^2} ) \nonumber \\
& +(1-\delta_{qt} ) \delta_{ru} \cdot 48 \pi^3 \frac{2 q t}{(q^2-t^2)^2}\cdot (-1)^{q+t}  \nonumber \\
& +(1-\delta_{ru} ) \delta_{qt} \cdot 48 \pi^3 \frac{r^2 + u^2}{(r^2-u^2)^2}\cdot (-1)^{r+u}. 
\end{eqnarray}

Finally the integrals needed for the matrix elements of the centrifugal potential can be solved by a change of variables.  Exploiting the periodicity of the trigonometric functions the domain of integration can be selected to suit the new set of variables, see fig. \ref{figC1} and the subsection \ref{sec:elementaryIntegralsRR} on elementary integrals.

The result is
\begin{align}	\label{eq:fpqrfstuCentrifugal}
 &<\! f_{pqr} \! \mid \frac{1}{\sin^2 \frac{1}{2} (\theta_1 - \theta_2)} + \frac{1}{\sin^2 \frac{1}{2} (\theta_2 - \theta_3)} + \frac{1}{\sin^2 \frac{1}{2} (\theta_3 - \theta_1)} \mid f_{stu} > \nonumber \\
&= 3 <\! f_{pqr} \! \mid \frac{1}{\sin^2 \frac{1}{2} (\theta_1 - \theta_2)}  \mid f_{stu} > \nonumber \\
&= 3 \pi^3[ \delta_{ps}(\delta_{r-q,u-t}nn(r+q,u+t) -  \delta_{r-q,u+t}nn(r+q,u-t))\nonumber \\
&+  \delta_{ps}(-\delta_{r+q,u-t}nn(r-q,u+t) +  \delta_{r+q,u+t}nn(r-q,u-t))\nonumber \\
&+  \delta_{pu}(\delta_{r-q,s+t}nn(r+q,s-t) -  \delta_{r-q,s-t}nn(r+q,s+t))\nonumber \\
&+ \delta_{pu}(-\delta_{r+q,s+t}nn(r-q,s-t) +  \delta_{r+q,s-t}nn(r-q,s+t))\nonumber \\
&+   \delta_{qt}(\delta\delta_{p+r,s+u}nn(r-p,u-s) + \delta  \delta_{p+r,u-s}nn(r-p,u+s))\nonumber \\
&+   \delta_{qt}(\delta\delta_{r-p,s+u}nn(p+r,u-s) + \delta  \delta_{r-p,u-s}nn(p+r,u+s))\nonumber \\
&+  \delta_{rs}(\delta_{p+q,u-t}nn(p-q,u+t) -  \delta_{p+q,u+t}nn(p-q,u-t))\nonumber \\
&+   \delta_{rs}(-\delta_{p-q,u-t}nn(p+q,u+t) +  \delta_{p-q,u+t}nn(p+q,u-t))\nonumber \\
&+  \delta_{ru}(\delta_{p+q,s+t}nn(p-q,s-t) -  \delta_{p+q,s-t}nn(p-q,s+t))\nonumber \\
&+  \delta_{ru}(-\delta_{p-q,s+t}nn(p+q,s-t) +  \delta_{p-q,s-t}nn(p+q,s+t))],
\end{align}
where two more shorthand notations have been introduced

\begin{equation}   
     \delta \delta_{ij} = 
 \begin{cases} 
  1  \, \, \, & \text{for} \, \, \,  i=j \, \wedge \, i \ne 0  \\
  -1  & \text{for} \, \, \,  i=-j \, \wedge \, i \ne 0 \\
  0   & \text{otherwise}
 \end{cases}   
\end{equation}
 and
\begin{equation}   
     nn(i,j) = 
 \begin{cases} 
  \mid i+j \mid - \mid i-j \mid   \, \, \, & \text{for} \, \, \,  i+j \equiv 0 \mod 2  \\
  0  & \text{otherwise.} 
 \end{cases}   
\end{equation}
The factor $nn$ originates from the following rule \cite{Amtrup}
\begin{equation}  \label{eq:sinMxSinNxRRedition} 
     \int^{\pi}_{-\pi} \frac{\sin mx\cdot \sin nx}{\sin^2 x}dx =
 \begin{cases} 
  (\mid m+n \mid - \mid m-n \mid ) \pi  \, \, \, & \text{for} \, \, \,  m-n \equiv 0 \mod 2  \\
  0  & \text{for} \, \, \,  m-n \equiv 1 \mod 2.
 \end{cases}   
\end{equation}
The integrals in (\ref{eq:sinMxSinNxRRedition}) pop up after the aforementioned change of variables which exploits the following trigonometric relations
\begin{gather}
\cos px \cos ry - \cos rx \cos py = \nonumber \\  \sin nu \sin mt + \sin mu \sin nt,  \nonumber \\ 
u =\frac{x+y}{2}, \,\, t = \frac{x-y}{2},\,\, n = r+p,\,\, m = r-p,
\end{gather}
and
\begin{equation}	\label{eq:cosSinMinusSinCos}
\cos px \sin qy - \sin qx \cos py = \cos nu \sin mt - \cos mu \sin nt, \quad n=p+q, \,\,m=p-q. 
\end{equation}

\begin{figure}
\begin{center}
\includegraphics[width=0.4\textwidth]{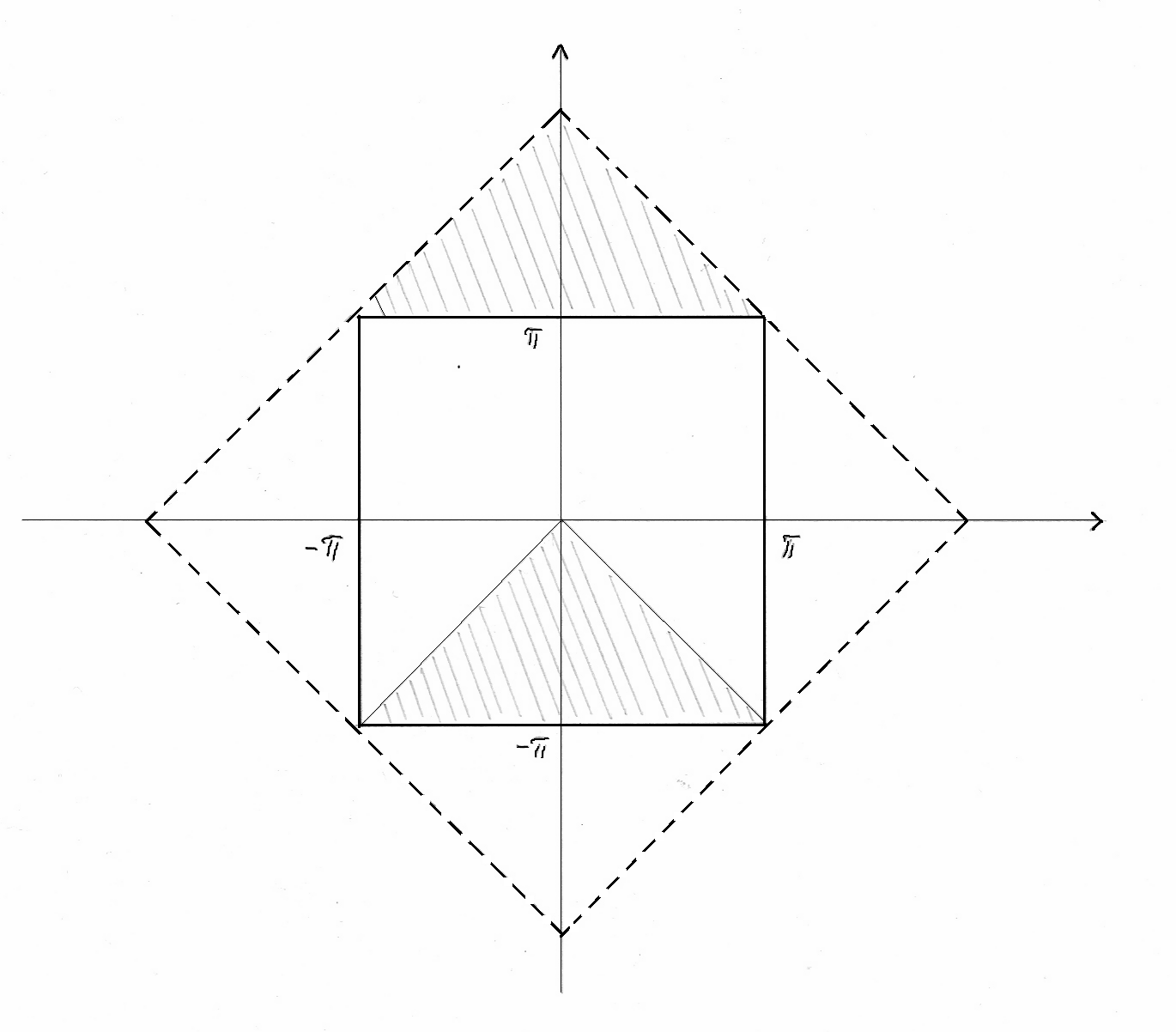}
\caption{A change of variables from the horizontal/vertical $(x,y)$ to a 45 degrees inclined system of coordinates $(u,t)=(\dfrac{x+y}{2},\dfrac{x-y}{2})$ needed in order to find the matrix elements of the centrifugal potential. The seemingly singular denominator in the centrifugal potential is then found to be integrable, see (\ref{eq:sinMxSinNxRRedition}) to (\ref{eq:cosSinMinusSinCos}).  The domain of integration is expanded to suit the new set of variables. This is possible because of the periodicity of the trigonometric functions such that functional values on the hatched area outside the original domain of integration $[-\pi,\pi]\times[-\pi,\pi]$ are identical by parallel transport from the hatched area within that same area. Figure and caption from \cite{TrinhammerBohrStibiusHiggsPreprint}.}
\label{figC1}
\end{center}
\end{figure}

\subsection{Elementary integrals for matrix elements in the Rayleigh-Ritz method}
\label{sec:elementaryIntegralsRR}

We solve here exemplar integrals for the trigonometric basis needed to prove the orthogonality relation (\ref{eq:fpqrfstuOrtogonal}), the expectation value of the geodetic potential (\ref{eq:fpqrfstuIntrinsicPotential}) and the centrifugal potential (\ref{eq:fpqrfstuCentrifugal}). First the orthogonality relation for   $p>0, \, q>0, \, r>p$  ($p=0$ is left for the reader). With a slight change in notation for the angular variables we seek the scalar product 
\begin{equation}  \label{eq:fgScalarProduct} 
   <f,g> \equiv \int^{\pi}_{-\pi}  \int^{\pi}_{-\pi}  \int^{\pi}_{-\pi}  fg dx_i dx_j dx_k
\end{equation}			
between base functions (\ref{eq:fpqr}) like
\begin{gather}
f_{pqr}(x_i, x_j, x_k) = \epsilon_{ijk} \cos p x_i \sin q x_j \cos rx_k \quad {\text{and}}  \nonumber \\
g_{stu}(x_l, x_m, x_n) = \epsilon_{lmn} \cos p x_l \sin q x_m \cos rx_n. 
\end{gather}

The three-dimensional integral in (\ref{eq:fgScalarProduct}) factorizes into three one-dimensional integrals and using the orthogonality of the trigonometric functions on $[-\pi, \pi]$   we readily have
\begin{equation}
     <f_{pqr},g_{stu}> \, = ( \delta_{ps}  \delta_{ru}- \delta_{pu} \delta_{rs}) \delta_{qt} \cdot 6 \pi^3,
\end{equation}
which for $p>0, \, q>0, \,  r>p$ and $s>0, \, t>0, \, u>s$  reduces to (\ref{eq:fpqrfstuOrtogonal}). To obtain the expectation value (\ref{eq:fpqrfstuIntrinsicPotential}) of the geodetic potential we use the same kind of factorization together with the following list of elementary integrals
\begin{gather} 
    \int^{\pi}_{-\pi} x^2 \cos px \cos sx \, ds = (-1)^{p+s} 4 \pi \frac{p^2 + s^2}{(p^2-s^2)^{2}}, \quad p \ne s \\
    \int^{\pi}_{-\pi} x^2 \cos sx \, dx= \frac{4 \pi}{s^2} (-1)^{s}, \quad p=0, \, s \ne 0 \hspace{17mm} \\
    \int^{\pi}_{-\pi} x^2 \cos^2 px \, dx = \frac{\pi^3}{3} + \frac{\pi}{2 p^2}, \quad p=s \hspace{25mm} \\ 
    \int^{\pi}_{-\pi} x^2 \sin^2qx \, dx = \frac{\pi^3}{3} - \frac{\pi}{2 q^2}, \quad q=t \hspace{26mm} \\
    \int^{\pi}_{-\pi} x^2 \sin qx \, \sin tx \, dx = \! (-1)^{q+t}\, 4 \pi \frac{2qt}{(q^2 - t^2)^{2}}, \hspace{2mm}
    | q | \ne | t |\\ 
    \int^{\pi}_{-\pi} x^2 \cos px \, \sin tx \, dx =0. \hspace{42mm}
\end{gather}

For the Laplacian and for the centrifugal potential we make another slight change in notation for our angular variables and rewrite our base functions (\ref{eq:fpqr}) as a sum of subdeterminants
\begin{gather}	\label{eq:subdeterminants}
 f_{pqr}(x,y,z) = 
\begin{vmatrix}
 \cos px & \cos py & \cos pz \\
\sin qx & \sin qy & \sin qz \\
 \cos rx & \cos ry & \cos rz   
\end{vmatrix} \hspace{5.9cm} \nonumber \\
= \cos pz 
\begin{vmatrix}
\sin qx & \sin qy\\
\cos rx & \cos ry
\end{vmatrix} \hspace{13mm} \nonumber \\
- \sin qz 
\begin{vmatrix}
\cos px & \cos py\\
\cos rx & \cos ry
\end{vmatrix} \hspace{12.8mm} \nonumber \\
+ \cos rz 
\begin{vmatrix}
\cos px & \cos py\\
\sin qx & \sin qy
\end{vmatrix}. \hspace{11.8mm}
\end{gather}
For the Laplacian we then get terms like
\begin{gather}
   \frac{\partial^2}{\partial z^2} f_{pqr} (x,y,z) =-  p^2 \cos pz
\begin{vmatrix}
\sin qx  \sin qy\\
\cos rx  \cos ry
\end{vmatrix}
\nonumber \\
 + q^2 \sin qz 
\begin{vmatrix}
\cos px  \cos py\\
\sin qx  \sin qy
\end{vmatrix} \nonumber \\
-r^2 \cos rz
 \begin{vmatrix}
\cos px  \cos py\\
\sin qx  \sin qy
\end{vmatrix}.
\end{gather}
Again the integral for the expectation value factorizes into one-dimensional integrals where the orthogonality of the trigonometric functions can be exploited to obtain (\ref{eq:fpqrfstuLaplacian}).

For the centrifugal term the three-dimensional integral does not readily factorize. We need a change of variables which suits the mixing of the variables in the denominators. Due to the arbitrary labelling of our angles we have
\begin{equation}	\label{eq:centrifugalTrippleToSingle}
 < f_{pqr} \! \mid  \! \frac{1}{\sin^2 \frac{1}{2}(x-y)} +\frac{1}{\sin^2 \frac{1}{2}(y-z)} +
\frac{1}{\sin^2 \frac{1}{2}(z-x)} \! \mid \! f_{stu} > 
 = 3  <f_{pqr} \! \mid  \! \frac{1}{\sin^2 \frac{1}{2}(x-y)} \! \mid \! f_{stu}>.
\end{equation}
With
\begin{equation}
  f_{pqr} f_{stu} = 
\begin{vmatrix}
    \cos px & \cos py & \cos pz \\
\sin qx & \sin qy & \sin qz \\
\cos rx & \cos ry & \cos rz 
\end{vmatrix}
\begin{vmatrix}
    \cos sx & \cos sy & \cos sz \\
\sin tx & \sin ty & \sin tz \\
\cos ux & \cos uy & \cos uz 
\end{vmatrix},
\end{equation}
we can use subdeterminant expressions in to get e.g.\ a factor $3 \delta_{qt} \pi$  from the $z$-integration of the term involving the two sines while the product of the two corresponding subdeterminants is used for a shift of variables, see below. We have
\vspace*{-3mm}
 \begin{equation}
\begin{vmatrix}
   \cos px& \cos py\\
   \cos rx & \cos ry  
\end{vmatrix}
\begin{vmatrix}
   \cos sx& \cos sy\\
   \cos ux &\cos uy  \nonumber 
\end{vmatrix}\\
\end{equation}
\vspace*{-6mm}
 \begin{eqnarray}	\label{eq:subdeterminantProduct}
&=& \!\!\! (\cos px \cos ry - \cos rx \cos py)(\cos sx \cos uy - \cos ux \cos sy) \hspace{3.2cm} \nonumber \\
&=& \frac{1}{2} \frac{1}{2} [ \cos(px-ry) + \cos (px + ry) - \cos (rx - py) - \cos (rx + py)] \nonumber \\
&\cdot & \!\!\!\!\! [( \cos(sx-uy) + \cos (sx + uy) - \cos (ux - sy) - \cos (ux -sy)] \nonumber \\
&=& \!\! \![- \sin \frac{1}{2}(px-ry+ rx-py)\sin \frac{1}{2}(px-ry-(rx-py)) \nonumber \\
&-& \!\!\!\!  \sin \frac{1}{2}(px+ry+ rx+py)\sin \frac{1}{2}(px+ry-(rx+py))] \nonumber \\
&\cdot &\!\!\!\!\! [ - \sin \frac{1}{2}(sx-uy+ux-sy)\sin \frac{1}{2}(sx-uy-(ux-sy))\nonumber \\
&-&\!\!\!\! \sin \frac{1}{2} (sx + uy + ux +sy) \sin \frac{1}{2} (sx +uy - (ux+sy))]\nonumber \\
&=&  \!\!\![\sin nv \sin mw - \sin nw \sin mv][ \sin kv \sin lw  - \sin kw \sin lv],
\end{eqnarray}
where
\begin{equation}
   n= p+r, \,\, m = p-r, \,\, k = s+u, \,\, l=s-u, \,\, w = \frac{x+y}{2} \,\,\,\, \text{and}\,\,\,\,   v = \frac{x-y}{2}.
\end{equation}
Since both nominator and denominator in (\ref{eq:centrifugalTrippleToSingle}) are trigonometric functions we can exploit their periodicity to enlarge the domain of integration and make a shift of variables to $w$ and $v$. An integration over the original domain is namely half the value of an integration over the enlarged domain in fig. \ref{figC1}, thus
\begin{gather}	\label{eq:coordinateTransformForCentrifugal}
\int^\pi_{-\pi} \int^\pi_{-\pi} dx dy = \frac{1}{2} \int^{2\pi}_{- 2\pi} dw' \int^{v_2'(w')}_{v_1'(w')} dv'\nonumber \\ =  \frac{1}{2}  \frac{1}{2}  \int^{2\pi}_{- 2\pi} dw'  \int^{2\pi}_{- 2\pi} dv' = \int^\pi_{-\pi} dw \int^\pi_{-\pi} dv,  \nonumber
\end{gather}
\vspace*{-3mm}
\begin{equation}
   \quad \text{where} \quad w' = x+y \quad \text{and} \quad v' =x+y.
\end{equation}
The factor $ \frac{1}{2}$   in the second expression is just from the change of coordinates  $dw'du' = 2dxdy$ and the domain of integration is still not enlarged, but limited by piecewise linear functions $v'_1$  and  $v'_2$ . In the third expression then we double the area of integration to lift the coupling between $w'$  and $v'$. In the last expression we just rescale our variables to suit our needs in (\ref{eq:subdeterminantProduct}). 

With the coordinate transformations we can use (\ref{eq:sinMxSinNxRRedition}) to get the final result		
\begin{equation}	\label{eq:centrifugalIntegral}
   3 < f_{pqr} \! \mid \! \frac{1}{\sin^2 \frac{1}{2} (x-y)} \! \mid f_{stu} > \nonumber 
\end{equation} 
 \vspace*{-2mm}
\begin{equation}
  = 3 \delta_{qt} \cdot \pi  \int^\pi_{-\pi} dw  \int^\pi_{-\pi} dv \,
     \frac{[\sin nv \sin mw - \sin nw \sin mv] [\sin kv \sin lw - \sin kw \sin lv] }{\sin^2 v} \nonumber
\end{equation} 
\vspace*{-2mm}
\begin{equation}
= 3 \delta_{qt} \cdot \pi [ \delta_{ml} \cdot \pi ( \mid n+k\mid - \mid n-k \mid ) \cdot \pi - 
  \delta_{mk} \cdot \pi ( \mid n+l \mid - \mid n-l \mid ) \cdot \pi \nonumber
\end{equation} 
\vspace*{-3mm}
\begin{equation}
   \hspace*{1cm}  - \delta_{nl} \cdot \pi ( \mid m+k\mid - \mid m-k \mid ) \cdot \pi + 
  \delta_{nk} \cdot \pi ( \mid m+l \mid - \mid m-l \mid ) \cdot \pi ],
\end{equation}
which is a specific example of the general result (\ref{eq:fpqrfstuCentrifugal}).

\end{widetext}

\section{Future experiments and observations}
\label{sec:futureExperiments}

On neutral pentaquarks, proton radius, precise Higgs mass, Higgs self-couplings, Higgs couplings to gauge bosons, beta decay neutrino mass and dark energy to baryon ratio.

We give just a short report on results from the intrinsic view that may show up in future experiments.

In section \ref{sec:neutralPentaquarks} we suggested to look for neutral electric charge neutral flavour baryon singlets, see table \ref{tab:singlets}. From a quark model perspective such states would be interpreted as neutral pentaquarks and as such might be visible in the LHCb set-up at CERN, where charged pentaquarks have been discovered \cite{LHCbPentaquarkObservation}. We also expect neutral flavour singlets to be visible as narrow resonances in direct production experiments like $\pi^-$ scattering on protons.

In \cite{TrinhammerBohrProtonRadiusIQM2} we argue for the following relation between the proton electric charge radius $r_{\rm p}$ and its Compton wavelength $\lambda_{\rm p}=h/(m_{\rm p}c)$
\begin{equation}
 r_{\rm p}\frac{\pi}{2}=\lambda_{\rm p}
\end{equation}
where the factor $\frac{\pi}{2}$ originates from a projection like in (\ref{eq:spaceProjection}), but for a period doubled state to describe the nucleon after neutron decay. This yields $r_{\rm p}=0.841235642(38)\ \rm fm$ in accordance with the value $r_{\rm p}^\mu=84087(39)\ \rm fm$ \cite{RPP2016} from Lamb shift spectroscopy on muonic hydrogen but is at tension with the results $r_{\rm p}^{e}=0.8751(61)\rm fm$ \cite{RPP2016} from proton form factors in electron scattering on protons and from spectroscopy on ordinary hydrogen. We look forward to simultaneous scattering of muons and electrons on protons to clarify the discrepancy, e.g. in experiments like the PSI-MUSE proposal from the Paul Scherrer Institute \cite{PSI-MUSEproposal}.

\begin{widetext}

\begin{figure}
\begin{center}
\includegraphics[width=0.9\textwidth]{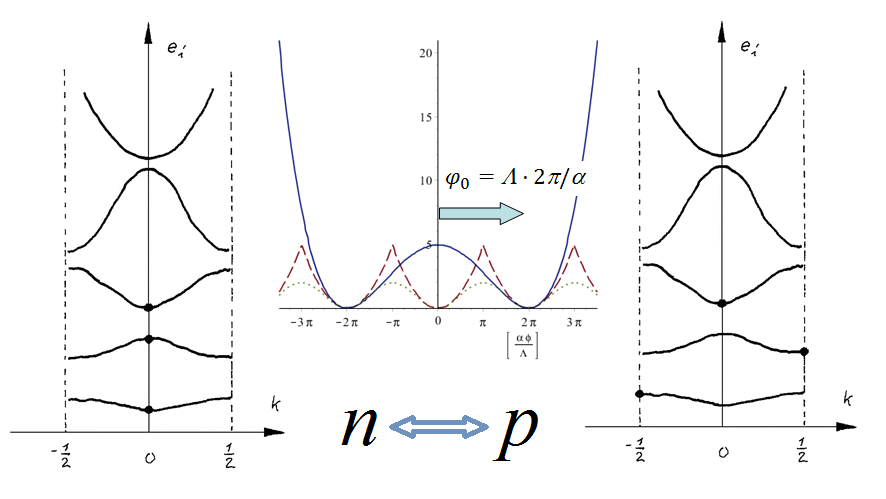}
\caption{The nucleon transformation $\rm p\rightarrow n$ in stars burning hydrogen to helium (and so forth) is manifested in the Higgs potential with a non-zero value at $\phi=0$. We interpret the non-zero value of the Higgs potential from each detained neutron in nuclei as a contribution to the dark energy content in the universe. {\it Left and right}: Reduced zone schemes, see p. 160 in \cite{AshcroftMerminBlochTheorem}, for Bloch wave numbers for the neutron state (left) and the proton state (right) in solutions of (\ref{eq:schroedingerU3}) with respectively $2\pi$ and $4\pi$ periodic wavefunctions \cite{TrinhammerBohrStibiusHiggsPreprint}. {\it Middle}: Higgs potential (solid, blue) matching the Manton-inspired potential \cite{Manton} (dashed, red) and a Wilson-inspired potential \cite{Wilson} (dotted, green). The Manton and Wilson inspired potentials yield the same value for the Higgs mass and the electroweak energy scale whereas only the Manton inspired potential gives a satisfactory reproduction of the baryon spectrum \cite{TrinhammerBohrStibiusHiggsPreprint}. Figure adapted from \cite{TrinhammerBohrStibiusEPS2015}.}
\label{fig:higgsPotentialFit}
\end{center}
\end{figure}

\end{widetext}

In \cite{TrinhammerBohrStibiusHiggsPreprint, TrinhammerBohrStibiusEPS2015} we gave a closed expression for an accurate Higgs mass
\begin{equation}
 m_{\rm H}c^2=\frac{1}{\sqrt{2}}\frac{2\pi}{\alpha_{\rm W}}\frac{\pi}{\alpha_{\rm e}}m_{\rm e}c^2.
\end{equation}
This expression is derived from fitting the Higgs potential to the intrinsic potential as expressed in eq. (\ref{eq:HiggsPotentialWithConstant}) and shown in fig. \ref{fig:higgsPotentialFit}. With corrected quark masses and updated value for $\alpha_{\rm W}^{-1}=127.984\pm 0.020$ this yields $m_{\rm H}c^2=(125.090\pm0.020)\ \rm GeV$ in accordance with the weighted average $125.14(25)\ \rm GeV$ from the CMS \cite{HiggsMassCMSrun2} and ATLAS \cite{HiggsMassATLASrun2} detectors in Run 2 at the LHC at CERN. We look forward to see if our calculation stays within the experimental determination. However we cannot expect LHC to reach an accuracy in the Higgs mass determination that would rule out our calculation. On the other hand a future International Linear Collider \cite{ILCtechReport} would give an indication with its expected $35\ \rm MeV$ accuracy in Higgs mass determination.

An international linear collider \cite{ILCtechReport} would be well suited to test our predictions of Higgs self-couplings relative to the Standard Model expectations. In \cite{TrinhammerBohrStibiusHiggsPreprint} we derived the electroweak energy scale of the Standard Model $v_{\rm SM}$
\begin{equation}	\label{eq:vSM}
 v_{\rm SM}=v\sqrt{V_{ud}}
\end{equation}
expressed in our $v/\sqrt{2}\equiv\varphi_0$ from (\ref{eq:HiggsPotentialWithConstant}) and containing the up-down quark mixing matrix element $V_{ud}$ of the CKM-matrix \cite{RPP2016}. This matrix element enters in (\ref{eq:vSM}) because we get our $v$ from scaling the Higgs potential by considering the neutron beta decay whereas $v_{\rm SM}$ is got from considering the muon decay. In other words (\ref{eq:vSM}) expresses the relation between the Fermi coupling constants $G_{\rm F\mu}$ and $G_{\rm F\beta}$ for muon decay and beta decay respectively. The presence of $V_{ud}$ is carried into our prediction for the Higgs self-couplings and for the Higgs couplings to the gauge bosons $W$ and $Z$ of the electroweak interactions. In \cite{TrinhammerJMP} we write for instance for the quartic Higgs self-coupling in (\ref{eq:HiggsPotentialWithConstant}) relative to the Standard Model
\begin{equation}	\label{eq:higgsSelfCouplingQuartic}
 \kappa_{\rm HHHH}=\frac{\lambda^2/4}{\lambda_{\rm SM}}=\frac{1/8}{m_{\rm H}^2c^4/(2v_{\rm SM}^2)}=V_{ud}.
\end{equation}
Here we used the Standard Model Higgs potential parameters as defined by \cite{GuptaRzehakWells}
\begin{equation}
 V_{\rm H}=-\frac{m_{\rm H}^2c^4}{2}\phi^2+\lambda_{\rm SM}\phi^4.
\end{equation} 
Equation (\ref{eq:higgsSelfCouplingQuartic}) yields $\kappa_{\rm HHHH}=V_{ud}=0.97417(21)$ \cite{RPP2016}. The most accurate tests of the couplings derived from the Higgs mechanism at a possible future ILC are in Higgs couplings to the gauge bosons $W$ and $Z$ where sub-percent uncertainty is expected \cite{ILCtechReport}.

In \cite{TrinhammerJMP} we argued for an electron-based anti-neutrino mass
\begin{equation}
 \frac{m_{\overline{\nu}}}{m_{\rm e}}=\frac{\alpha_{\rm Z}}{8\sin^2\theta_{\rm W}\cos^2\theta_{\rm W}}\left(\frac{\alpha_{\rm e}}{\pi}\right)^2
\end{equation}
which gives $m_{\overline{\nu}}c^2=15.152(4)\ \rm meV$. With neutrino mixing in normal hierarchy favoured by recent NO$\nu$A neutrino oscillation observations \cite{NOvAarXiv2017} this corresponds to a beta decay neutrino mass of $\sqrt{<m_\beta^2>}=17.6\pm 0.2\ \rm meV/c^2$ in accordance with the correlation between beta neutrino mass and cosmological constraints shown in figure 10 of \cite{NeutrinoMassConstraintsItalians}. This is intriguingly close to the sensitivity of the Cyclotron radiation emission spectroscopy technique prospected in the Project 8 proposal \cite{FraenkleKATRINproject8}. In \cite{TrinhammerJMP} we also gave a second scenario which yielded $m_{\overline{\nu}}c^2=0.9165(4)\ \rm eV$. This is within the reach of the upcoming KATRIN experiment in Karlsruhe, Germany \cite{ThuemmlerKATRINintro}, but in conflict with the cosmological limit on the sum-total $\Sigma\leq 0.68\ \rm eV$ of the neutrino masses, see p. 121 in \cite{RPP2016}.

In \cite{TrinhammerDarkEnergyResearchGateIQM4} we gave an interpretation of the constant term in our Higgs potential (\ref{eq:HiggsPotentialWithConstant}) as related to the dark energy content of the universe. Observation of accelerated recession of supernovae in low and high red-shift galaxies \cite{RiessEtAl} led to the acceptance of a major dark energy content in the universe \cite{Padmanabhan, PeeblesRatra}. The present observed ratio \cite{RPP2016}
\begin{equation}	\label{eq:OmegaBomegalLambdaObserved}
 \frac{\Omega_\Lambda}{\Omega_b}|_{\rm observed}=\frac{0.692(12)}{0.0484(10)}=14.3(\pm 0.4).
\end{equation}
between the dark energy content and the baryonic content remains unexplained. We suggest that the dark energy content of the universe is a manifestation of detained neutron decay, expressed as a constant term in the Higgs "potential" (\ref{eq:HiggsPotentialWithConstant}). We leave the question unanswered as to how the underlying coupling to accelerated expansion of the universe should be described. Several models have been discussed \cite{BezrukovShaposhnikov, BhattacharyaChakrabortyDasMondal, BezrukovMagninShaposhnikov2009, SimoneHertzbergWilczek, BezrukovMagninShaposhnikovSibiryakov2011}. We derived the ratio $\Omega_\Lambda/\Omega_{\rm b}=13.6(1.7)$ from an intrinsic conception of the structural changes taking place during transformations between protons and neutrons in the nuclear fusion processes inside stars. We take the Higgs mechanism to mediate the electroweak neutron to proton transformation with Higgs field $\phi=0$ for a neutronic state and $\phi=\varphi_0$ for a protonic state, see fig. \ref{fig:higgsPotentialFit}, i.e. $<n|\phi|n>=0$ and $<p|\phi|p>=\varphi_0$ where $|n>$ and $|p>$ are the neutron and proton states respectively. We thus assume that for each detained neutron there is one $\delta$-contribution to the dark energy. This yields \cite{TrinhammerDarkEnergyResearchGateIQM4}
\begin{gather}	\label{eq:OmegaBomegaLambdaSecondApprox}
 \frac{\Omega_\Lambda}{\Omega_b}|_{\rm model}=\frac{\sum_{\rm neutrons}\delta}{\sum_{\rm baryons} m_{\rm baryon}c^2}\approx\frac{n_{\rm neutron}\cdot\varphi_0/2}{n_{\rm baryon}\cdot m_{\rm n}c^2}\\ \nonumber
 \approx\frac{\varphi_0/2}{2/Y\cdot m_{\rm n}c^2}= 13.6(\pm 1.7),
\end{gather}
where the relative helium content $Y$ of the universe is determined from the neutron and proton particle densities as \cite{UnsoeldBaschek} 
\begin{equation}
 \frac{n_{\rm n}+n_{\rm p}}{n_{\rm n}}=\frac{2}{Y}.
\end{equation}
The last approximation in (\ref{eq:OmegaBomegaLambdaSecondApprox}) consists of disregarding contributions from elements heavier than helium. This is a fine approximation since the total of these, the {\it metallicity} $Z$, is of the order of one percent \footnote{For instance an ordinary main sequence star like our own Sun is modelled satisfactorily by starting out from $X:Y:Z=0.73:0.25:0.02$ with $X$ for hydrogen, see p. 282 in \cite{UnsoeldBaschek}.}. In (\ref{eq:OmegaBomegaLambdaSecondApprox}) we used for the relative primordial helium mass content $Y_{\rm p}=0.296\pm0.030$ in recent determinations from the Cosmic Microwave Background measured at the South Pole Telescope \cite{KeislerEtAl}. We look forward to improved determinations of the helium fraction of baryonic matter in the universe. The helium content $Y$ shows a slightly increasing tendency with star generation number \cite{VillanovaGeislerPiottoGratton} thus indicating an increasing value in (\ref{eq:OmegaBomegaLambdaSecondApprox}). We therefore also look forward to observations to determine whether the dark energy content of the universe is increasing with time as suggested by (\ref{eq:OmegaBomegaLambdaSecondApprox}) with increasing $Y$ as the amount of neutrons pile up inside stars from the ongoing fusion processes there. An increase in the dark energy content would be in accordance with the observed accelerated recession of supernovae \cite{RiessEtAl}.

\section{Conclusion}
We have presented the foundation and new applications of intrinsic quantum mechanics and we have given reference to a variety of other results derived from the intrinsic conception, namely predictions on neutral pentaquarks, proton radius, precise Higgs mass, Higgs self-couplings, Higgs couplings to gauge bosons, beta decay neutrino mass and dark energy to baryon matter ratio. We take intrinsic quantum mechanics to represent a step, not so much beyond the Standard Model of particle physics, but to represent a step {\it behind} the Standard Model. We have given an example of how exterior derivatives on Lie group configuration spaces relate between spacetime quantum fields and intrinsic configuration spaces expressed via the momentum form on intrinsic wavefunctions. We have hinted at a profound relation between the strong and electroweak sectors of particle physics seen in a possible origin of the electroweak mixing angle in up and down quark flavour generators. Generators that were also used to derive a proton spin structure function and the proton magnetic dipole moment. The idea of intrinsic quantum mechanics has led to a conception of mass as introtangled energy-momentum with the introtangling mediated by kinematic generators like the momentum and angular momentum generators. This conception is in accordance with the energy-mass equivalence in the theory of relativity. We have shown that left-invariance in intrinsic space corresponds to local gauge invariance in spacetime. Unlike the extra dimensions introduced in string theory, the configuration spaces of intrinsic quantum mechanics are non-spatial. This orthogonality to spacetime opens for a unified description without requiring quantum gravity. One may argue that quantum dynamics originate in excitation of intrinsic degrees of freedom as generalizations of the concept of spin degrees of freedom. To clarify such questions we need more study both at the conceptual level and at the experimental level to look for our proposed tests in baryon spectroscopy, neutrino mass determinations and  deviations from Standard Model expectations in couplings of the Higgs particle to itself and to gauge bosons.

\section{Acknowledgments}

I thank Povl Holm for solving the one-dimensional parametric Schr\"odinger equation by Maclaurin series to check against my first solution based on Sturm-Louiville theory.
I thank Hans Bruun Nielsen for pointing to the Rayleigh-Ritz method. I thank Torben Amtrup for proving the sine fraction integrals needed in the Rayleigh-Ritz method for a trigonometric base. I thank Holger Bech Nielsen for helpful discussions, in particular for mentioning the word "momentum form" for the exterior derivative. I thank Gestur Olafsson for explaining me how to derive the Laplace-Beltrami operator on $U(N)$. I thank Henrik Georg Bohr for co-work on the proton radius. I thank Henrik Georg Bohr and Mogens Stibius Jensen for interest and helpful discussions on the intrinsic viewpoint in general and for co-work on the Higgs mass in particular. I thank colleagues at the Technical University of Denmark for an inspiring working environment.

\end{small}
 
\end{document}